\documentclass{aa}  

\DeclareMathAlphabet{\altmathcal}{OMS}{cmsy}{m}{n}
\usepackage{mathptmx}
\usepackage{pifont}

\usepackage[T1]{fontenc}
\usepackage{txfonts}

\usepackage{graphicx}	
\usepackage{amsmath}	
\usepackage{amssymb}	
\usepackage{gensymb}
\usepackage{siunitx}
\usepackage{xcolor}     
\usepackage{cancel}
\usepackage{multirow}
\usepackage{mathrsfs}
\usepackage{ulem}

\usepackage[colorlinks=true,linkcolor=blue,citecolor=blue]{hyperref}

\DeclareMathOperator{\sech}{sech}

\renewcommand{\vec}[1]{\mathbf{#1}}
\renewcommand{\tens}[1]{\mathsf{#1}}
\newcommand{\vect}[1]{\mathbf{#1}}

\newcommand{\pd}[2]{\frac{\partial #1}{\partial #2} }
\newcommand{\der}[2]{\frac{\mathrm{d} #1}{\mathrm{d} #2} }

\newcommand{\DS}{\displaystyle}

\newcommand{\N}{\ding{55}}

\newcommand{\ad}{{\mathrm ad}}

\newcommand{\change}[1]{{#1}}

\begin{document} 

   \title{A robust super-time-stepping scheme for Ohmic and ambipolar diffusion}

   \subtitle{}

   \author{G. Mattia        \inst{1}\fnmsep\thanks{mattia@mpia.de} \and
           M. Flock         \inst{1}                               \and
           D. Melon Fuksman \inst{1}                               \and
           A. Tzouvanou     \inst{1,2}                             \and
           V. Berta         \inst{3,4}                             \and
           D. Crocco        \inst{5}
          }

    \institute{
    Max-Planck-Institut für Astronomie, Königstuhl 17, 69117 Heidelberg, Germany
    \and
    Fakultät für Physik und Astronomie, Universität Heidelberg, Im Neuenheimer Feld 226, 69120 Heidelberg, Germany
    \and
    Canadian Institute for Theoretical Astrophysics, 60 St. George St, Toronto, ON M5S 3H8, Canada
    \and
    Department of Astronomy, Columbia University, New York, NY 10027, USA
    \and
    Dipartimento di Fisica, Università degli Studi di Torino , Via Pietro Giuria 1, I-10125 Torino, Italy \\
    }

   \date{Received ***; accepted ***}

 
  \abstract
  {Non-ideal magnetohydrodynamics (MHD) is a key tool for modeling magnetic flux transport in astrophysical systems such as molecular clouds, protostellar cores, and protoplanetary disks. Conventional explicit methods for non-ideal MHD diffusion are severely limited by timestep constraints, while substepping approaches can be unstable due to truncation errors near boundaries and strong magnetic-field gradients.} 
  {
   Our main goal is to address these limitations by developing robust super-time-stepping methods for Ohmic and ambipolar diffusion.
  }
  {We present a super-time-stepping method based on the stability of the Gegenbauer polynomials. The method is designed to enhance robustness in the presence of strongly anisotropic resistivity and to reduce sensitivity to truncation errors near boundaries. We implement the scheme in the PLUTO code and assess its performance through dedicated Ohmic and ambipolar diffusion tests. We also compare this novel numerical scheme against two common astrophysical problems, namely magnetic reconnection and the magnetorotational instability.}
  {The novel Runge-Kutta-Gegenbauer scheme retains computational efficiency beyond purely explicit schemes while providing excellent stability compared with other traditional substepping methods. It remains stable in the presence of strongly anisotropic diffusion, enabling accurate magnetic-field evolution in regimes characteristic of protoplanetary disks and collapsing dense cores. Benchmark tests, including magnetic reconnection and magnetorotational-instability setups, confirm the method's accuracy, efficiency, and suitability for large-scale non-ideal MHD simulations.}
  {}


   \maketitle

\section{Introduction}

Magnetic fields play a key role in the dynamics and evolution of various astrophysical environments, 
including molecular clouds and star-forming regions 
\citep{mckee2007,crutcher2012}, the interstellar medium of galaxies \citep{beck2015}, protostellar envelopes, and protoplanetary disks \citep{armitage2011,turner2014,lesur2023book}, as well as the launching regions of winds and jets \citep{fendt2006,pudritz2007}.
In such environments, the gas is only partially ionized, and the degree of ionization, along with the collisional efficiency between ionized and neutral species, determines the coupling between the gas and the magnetic field.
Magnetized plasmas in these systems can be described within the framework of magnetohydrodynamics (MHD), which treats the gas as a conducting fluid coupled to the magnetic field.
In the limit of strong coupling, this description reduces to the ideal MHD approximation \citep{alfven1942,chiuderi2005}, which assumes a perfectly conducting medium.
However, in weakly ionized plasmas, where collisional coupling is inefficient, the assumptions underlying the ideal MHD approximation break down, and non-ideal MHD effects must be taken into account.
More specifically, Ohmic diffusion, associated with collisions among charge carriers that damp electric currents and break the ideal-MHD flux-freezing condition, and ambipolar diffusion, arising from ion-neutral collisions and associated with the relative drift between charged and neutral components, are responsible for the partial decoupling between magnetic field and gas motion (e.g., \citealt{spitzer1978,shu1992,tsukamoto2022}).

These non-ideal MHD effects play a central role in many of the aforementioned astrophysical contexts. For instance, in molecular clouds, ambipolar diffusion enables neutrals to slip past the magnetized ion component, allowing magnetic flux to be transported more slowly than the bulk neutral gas during collapse.
This reduces magnetic support in dense cores and facilitates gravitational contraction \citep{mouschovias1976, shu1987, marchand2016}.
At higher densities, such as during the later stages of stellar core formation, both Ohmic and ambipolar diffusion produce magnetic flux loss and favor the formation of rotationally supported disks \citep{tomida2015,wurster2021}.
Additionally, in protoplanetary disks, which are weakly ionized over most of their volume, ambipolar diffusion dominates in low-density outer regions, where ion–neutral drift strongly suppresses the magnetorotational instability (MRI) \citep{simon2013,bethune2017,cui2021,hu2023}.
Ohmic diffusion instead becomes important in dense midplane and inner-disk regions characterized by low ionization.
Together, these non-ideal effects weaken MHD turbulence and regulate angular momentum transport, thereby influencing disk accretion and evolution \citep{bai2011,lesur2014,gressel2015}.

An accurate numerical treatment of Ohmic and ambipolar diffusion poses significant challenges for robustness and computational efficiency.
These two effects enable magnetic structures to evolve independently of the bulk flow, introducing characteristic timescales that can be significantly shorter than the typical timescales of the ideal MHD regime \citep{parker1979,brandenburg2005,armitage2011}.
As a result, numerical schemes for non-ideal MHD equations must address strong scale separation, which can impose severe constraints on the maximum achievable time step.
Moreover, non-ideal processes can generate sharp local magnetic gradients and significant ion–neutral drift.
If these features are not adequately resolved or treated, they can result in prohibitively small timesteps, as well as artificial magnetic dissipation or an inaccurate description of the magnetic flux transport and magnetic braking in specific regimes \citep{wardle2007,tomida2013,masson2016,wurster2021}.

To address the timestep constraints and numerical challenges posed by Ohmic and ambipolar diffusion, a variety of numerical schemes have been developed within the framework of non-ideal MHD to preserve stability, accuracy, and the magnetic field's divergence-free condition while maintaining computational efficiency, even in strongly diffusive/ambipolar regimes.
Current approaches include fully explicit (EXPL) treatments of magnetic diffusion \citep{masson2012}, as well as accelerated methods such as Chebyshev-based super-time-stepping (STS) \citep{alexiades1996, choi2009} and Runge-Kutta-Legendre (RKL) integrators \citep{meyer2012, meyer2014, vaidya2017}.
Variants of these algorithms are implemented in a wide range of modern MHD codes, including PLUTO \citep{mignone2007}, Athena \citep{stone2020}, Idefix \citep{lesur2023}, Bifrost \citep{nobrega-siverio2020}, Arepo \citep{marinacci2018, zier2024}, PHANTOM \citep{wurster2014}, Ramses \citep{masson2012}, and Flash \citep{duffin2008}.
While STS and RKL methods have significantly improved the efficiency of explicit treatment of parabolic diffusive operators, they can still suffer from reduced robustness in the presence of strongly anisotropic resistivity and from sensitivity to truncation errors near the grid boundaries. 
These limitations motivate the development of alternative numerical approaches that go beyond purely explicit schemes and remain stable and accurate across a broad range of non-ideal MHD regimes.

In this context, Runge–Kutta–Gegenbauer (RKG) schemes \citep{osullivan2019,skaras2021,caplan2024} offer a stabilized explicit time-integration strategy characterized by substantially enlarged stability regions for diffusion-dominated operators compared to the aforementioned methods. 
In this paper, we introduce a numerical implementation of the RKG algorithm, explicitly designed for the diffusive terms arising in non-ideal MHD, with a particular focus on Ohmic and ambipolar diffusion.
The proposed approach is fully explicit, code-agnostic, and straightforwardly applicable to finite-volume MHD frameworks, while preserving the magnetic field's divergence-free constraint to machine accuracy.
We demonstrate that the RKG-based scheme provides a robust and efficient treatment of non-ideal MHD effects across a wide range of regimes, including cases characterized by strongly anisotropic resistivities, as commonly encountered in astrophysical plasmas. 
Through a set of targeted numerical benchmarks implemented within the PLUTO code \citep{mignone2007}, we show that the method remains stable in regimes where standard accelerated explicit techniques, such as STS and RKL schemes, exhibit reduced robustness or fail.
The proposed approach retains the computational efficiency typical of explicit accelerated methods, enabling substantially larger timesteps than standard explicit diffusion methods while maintaining accuracy.
We demonstrate that the RKG is applicable to simulations of astrophysical scenarios, such as resistive magnetic reconnection and ambipolar magneto-rotational instability, where it preserves the correct physical evolution while substantially relaxing the parabolic timestep constraint in strongly diffusive regimes.
These results establish the RKG scheme as a robust and portable alternative for numerically integrating non-ideal MHD effects.

The paper is structured as follows.
In Section \ref{sec::equations}, we briefly describe the non-ideal MHD equations.
In Section \ref{sec::hyperbolic}, we review the hyperbolic integration and the issues related to a purely explicit integration scheme for hyperbolic-parabolic partial differential equations.
In Section \ref{sec::method}, we introduce the mathematical foundations and the numerical implementation of the RKG algorithm.
In Section \ref{sec::benchmarks}, we test the properties of the RKG scheme through a set of numerical benchmarks.
In Section \ref{sec::applications}, we investigate different numerical methods by simulating two different astrophysical processes, i.e., magnetic reconnection and magneto-rotational instability.
Finally, in Section \ref{sec::conclusions} we draw our conclusions.
Further technical details are collected in the appendices: in Appendix \ref{app:gegenbauer} we review the Gegenbauer polynomials; in Appendix \ref{app:rkgcoefficients} we illustrate the RKG scheme; in Appendices \ref{app::field_diff} and \ref{app::barenblatt} we present multidimensional diffusion benchmarks (resistive and ambipolar), and in Appendix \ref{app::resistive_2D} we discuss the behavior of the resistivity tensor in purely two-dimensional MHD.

\section{Governing equations}
\label{sec::equations}

Here, we describe the non-ideal (i.e., resistive and ambipolar) MHD equations adopted in this paper.
We neglect the Hall-MHD term, whose dispersive nature and associated modifications to the MHD wave structure significantly complicate the Riemann problem and limit the development of numerical schemes beyond purely explicit methods.
\change{\change{Therefore, Hall-driven effects such as the generation of incompressible perturbations and modifications of the Poynting flux in partially ionized plasmas \citep{GonzalezMorales2020}, as well as polarity-dependent angular momentum transport in protoplanetary disks \citep{Zhao2021}, are not captured by the numerical scheme described in this paper.
Since the Hall term is dispersive and significantly alters the characteristic structure of the MHD system, implementing it beyond purely explicit schemes remains challenging and will be investigated in future work.}}

Electrical currents in plasmas are carried exclusively by the charged component (ions and electrons).
In weakly ionized environments, however, neutral species dominate the mass and momentum budget, i.e.

\begin{equation}
    \rho_i\ll\rho_n\simeq\rho,
\end{equation}

where $\rho_i$, $\rho_n$, and $\rho$ are, respectively, the ion, the neutral, and the total mass densities.
Collisions between charged particles and neutrals limit the plasma conductivity and give rise to Ohmic diffusion. In contrast, the unbalanced coupling between the magnetized charged fluid and the neutral background produces a drift between ions and neutrals that manifests as ambipolar diffusion in the induction equation. Under the single-fluid approximation, the system can be described by the standard MHD equations supplemented with non-ideal terms that account for Ohmic and ambipolar diffusion.

\subsection{Non-ideal MHD Equations}

The MHD formulation consists of a set of partial differential equations (here provided in Heaviside-Lorentz units) in the form
\begin{equation}
\label{eq::MHD_pde}
\pd{\vect{U}}{t} + \nabla\cdot\tens{F} + \vect{S} = 0,
\end{equation}
with $\vect{U}$, $\tens{F}$ and $\vect{S}$ the conserved variables, fluxes and source terms.
These equations describe the conservation of mass
\begin{equation}
\pd{\rho}{t} + \nabla\cdot(\rho\vect{v}) = 0,
\end{equation}
momentum
\begin{equation}
\pd{\rho\vect{v}}{t} + \nabla\cdot\left[\rho\vect{v}\vect{v} + \left(p + \DS\frac{B^2}{2}\right)\tens{I} - \vect{B}\vect{B}\right]^T - \rho\vect{g} = 0,
\end{equation}
total energy
\begin{equation}
\pd{E_t}{t} + \nabla\cdot\left[\left(\DS\frac{\rho v^2}{2} + \rho e + p\right)\vect{v} + \vect{E}\times\vect{B}\right] - \rho\vect{v}\cdot\vect{g} = 0,
\end{equation}
and magnetic field $\vec{B}$
\begin{equation}
\pd{\vect{B}}{t} + \nabla\times\vect{E} = 0,
\end{equation}
where $\vec{v}$ and $\vec{g}$ represent, respectively, the fluid velocity and the external gravity vector.
The total energy density is expressed as
\begin{equation}
E_t = \rho e + \rho\DS\frac{v^2}{2} + \DS\frac{B^2}{2},
\end{equation}
where a suitable equation of state provides the closure between internal energy $e$, density $\rho$, and pressure $p$ (see the next subsection).
The electric field $\vec{E}$ is composed of the advective hyperbolic term $\vect{E}_\mathrm{hyp}$ and a non-ideal parabolic term $\vect{E}_\mathrm{par}$:
\begin{equation}
\vect{E} = \vect{E}_\mathrm{hyp} + \vect{E}_\mathrm{par} = -\vect{v}\times\vect{B} + \eta_\Omega\vect{J} - \eta_\ad(\vect{J}\times\vect{B})\times\vect{B},
\end{equation}
where $\vect{J} = \nabla\times\vect{B}$ represents the electric current density and $\eta_\Omega$ and $\eta_\ad$ correspond, respectively, to the resistive and ambipolar resistivity coefficients.
Finally, the absence of magnetic monopoles is provided by the non-evolutionary Maxwell equation
\begin{equation}
\nabla\cdot\vect{B} = 0.
\end{equation}
To ensure the correct dimensionality, the ambipolar diffusion coefficient is written as
\begin{equation}
\label{eq::eta_ad_expr}
\eta_\ad = \DS\frac{1}{\gamma_\ad\rho_i\rho},
\end{equation}
where $(\gamma_\ad\rho_i)^{-1}$ and $\gamma$ represent, respectively, the mean collisional time and the collisional coupling constant between ions and neutral.
Note that the ambipolar diffusion term is sometimes written with an explicit factor $1/B^2$ (see, e.g., \citealt{nobrega-siverio2020}).
In that case, the magnetic field dependence is absorbed into the definition of the ambipolar diffusivity.

\subsection{Equation of state (EoS)}

To solve the MHD equations, a proper closure relating thermodynamic quantities is required.
For the sake of simplicity, if not specified, we will assume an isothermal gas, i.e.
\begin{equation}
p = c_{\mathrm iso}^2\rho,
\end{equation}
where the sound speed $c_{\mathrm iso}$ can be dependent on time and position, but must always be provided through a suitable algebraic expression.
As a consequence, the energy equation is not evolved in this regime.

Another alternative also used in this paper is to prescribe a calorically ideal gas, where the internal energy can be written as
\begin{equation}
\rho e = \DS\frac{p}{\Gamma - 1},
\end{equation}
with $\Gamma = 5/3$ being the (constant in time and space) ratio of specific heats.
More refined EoS would go beyond the scope of this paper and are therefore not considered in this work.

\section{Hyperbolic and parabolic integration}
\label{sec::hyperbolic}


The set of non-ideal MHD equations of Eq. \ref{eq::MHD_pde} can be rewritten as
\begin{equation}
\pd{U}{t} = \mathcal{H}(U) + \mathcal{P}(U) + \vec{S},
\end{equation}
where $\mathcal{H}$ denotes the hyperbolic flux divergence terms and  $\mathcal{P}$ represents the parabolic contributions arising from Ohmic and ambipolar diffusion.
The hyperbolic operator is characterized by finite signal speeds set by the eigenvalues of the MHD system, whereas the parabolic operator introduces diffusive behavior through second-order spatial derivatives.

\subsection{Explicit Runge--Kutta integration}

In a fully explicit framework, the hyperbolic and parabolic contributions are combined into a single right-hand side $\mathcal{L} = \mathcal{H}(U) + \mathcal{P}(U) + \vec{S}$ and advanced in time simultaneously.
Both operators are evaluated at each intermediate stage of the time integrator without further splitting.
The overall accuracy is therefore determined solely by the spatial reconstruction (when employed) and the temporal integrator.

Time advancement is then performed using a fully explicit time integrator, such as an s-stage Runge–Kutta (RK) scheme.
Given the solution $U^n$ at time $t^n$, an explicit RK method computes a sequence of intermediate stage values $U^i$ according to
\begin{equation}
    \vec{U}^i = \vec{U}^n + \Delta t\sum_{j = 1}^{i-1}a_{ij}\mathcal{L}(\vec{U}^j) \qquad\text{for }i = 1,\dots,s,
\end{equation}
and updates the solution as,
\begin{equation}
    \vec{U}^{n+1} = \vec{U}^n + \Delta t\sum_{i  = 1}^sb_i\mathcal{L}(\vec{U}^i).
\end{equation}
The coefficients $a_{ij}$ and $b_i$ that define the method are typically summarized in a Butcher's tableau \citep{butcher1987}.

\subsection{Hyperbolic and parabolic timestep}

When both hyperbolic and parabolic operators are treated explicitly, the timestep must satisfy the stability limits imposed by each contribution.
The system is therefore advanced using the Courant-Friedrichs-Levy (CFL)
\begin{equation}
\Delta t = \min(\Delta t_{\mathrm h}, \Delta t_{\mathrm p}),
\end{equation}
where
\begin{equation}
\label{eq::deltat_hp}
\begin{array}{lcl}
\Delta t_{\mathrm h} & = & C_h\DS\min_{ijk}\left[\DS\min_d\left(\DS\frac{\Delta x_d}{|\lambda_{d,\max}|} \right)\right],\\ \noalign{\bigskip}
\Delta t_{\mathrm p} & = & \DS\frac{C_p}{2}\DS\min_{ijk}\left[\DS\min_d\left(\DS\frac{\Delta x^2_d}{\chi_{d,\max}} \right)\right].
\end{array}
\end{equation}

$C_h$ and $C_p$ are, respectively, the hyperbolic and parabolic Courant numbers \citep{courant1928}, $\lambda_d$ is the fast magnetosonic speed, $\Delta x_d$ is the mesh spacing in the three directions, and $\chi_d$ is the maximum between the Ohmic ($\chi_\Omega = \eta_{\Omega}$) and the ambipolar diffusion ($\chi_\ad = \eta_\ad B^2$) coefficient in each direction (note that the ambipolar diffusion coefficient here is the same in all directions by construction).
Because $\Delta t_p$ scales quadratically with the grid spacing, it rapidly becomes more restrictive than the hyperbolic CFL condition at high resolution, severely limiting the efficiency of standard explicit schemes.

\section{The Runge-Kutta-Gegenbauer scheme}
\label{sec::method}

In this section, we introduce the Runge-Kutta-Gegenbauer scheme and discuss its properties and applicability.
The hyperbolic and parabolic operators are then advanced in a split fashion according to the above decomposition, with the diffusive part integrated using the present method.

\subsection{The numerical scheme}

Super-time-stepping methods are designed to solve ordinary differential equations of the form
\begin{equation}
    \der{\vect{U}}{t} = \mathcal{M}(\vect{U}(t)),
\end{equation}
where $\mathcal{M}$ is a spatially discretized parabolic operator.
An explicit $s-$stage time-integration scheme with superstep $dt$ yields the update \citep{vanderhouwen1980}:
\begin{equation}
\label{eq::stabilitypol}
\vect{U}(t+dt) = R_s(dt \mathcal{M})\vect{U}(t),
\end{equation}
where $R_s$ is a polynomial of degree $s$, also referred to as the stability polynomial \citep{butcher1987,hairer2010}.
In practice, this update is realized through an explicit stage recursion,  with the operator $\mathcal{M}(\vect{U})$ evaluated explicitly at each stage \citep{osullivan2019}.
Here we adopt the stability polynomial proposed by \citet{verwer1990,skaras2021}, i.e.:
\begin{equation}
\label{eq::stabilitypolynomial}
 R_s(z) = a_s + b_sC_s^{(\alpha)}(w_0 + w_1z),
\end{equation}
where in our case $C_s^{(\alpha)}$ denotes the Gegenbauer polynomial of degree $s$ (see Appendix \ref{app:gegenbauer} for a brief description of the Gegenbauer polynomials and their properties).

The coefficients $w_0$ and $w_1$ define a linear rescaling of the polynomial argument, while $a_s$ and $b_s$ are normalization constants determined by imposing the stability polynomial (and its derivatives) to be unity at $z = 0$ \citep{osullivan2019}. Once the coefficients $a_s$, $b_s$, $w_0$, and $w_1$ are computed, it is possible to derive a general numerical scheme by solving Eq. \ref{eq::stabilitypol}.
The resulting compact RKG form (valid for both $1-\mathrm{st}$ and $2-\mathrm{nd}$ order) is
\begin{equation}
\label{eq::RKGcompactform}
\begin{array}{rcl}
Y_0 & = & \vec{U}(t), \\ \noalign{\medskip}
Y_1 & = & Y_0 + b_1w_1\tau\mathcal{M}Y_0, \\ \noalign{\medskip}
Y_j & = & \mu_jY_{j-1} + \nu_jY_{j-2} + (1-\mu_j-\nu_j)Y_0+ \\ \noalign{\smallskip}
&&\tilde{\mu}_j\tau\mathcal{M}Y_{j-1}+\tilde{\gamma}_j\tau\mathcal{M}Y_0, \qquad 2\leq j\leq s, \\ \noalign{\medskip}
\vec{U}(t + dt) & = & Y_s,
\end{array}
\end{equation}
where the explicit form of the coefficients at each stage depends on $\alpha$ and $s$ and is given in Appendix~\ref{app:rkgcoefficients}.
This formulation naturally recovers the RKL \citep{meyer2012,meyer2014,vaidya2017} by setting $\alpha = 1/2$.
On the other hand, the RKG coefficients of \citet{skaras2021,caplan2024} differ from those reported here due to the different normalization of the Gegenbauer polynomials (see Appendix \ref{app:gegenbauer}).

\subsection{Time-stepping substages}

\begin{figure}
\centering
\includegraphics[width=0.49\textwidth]{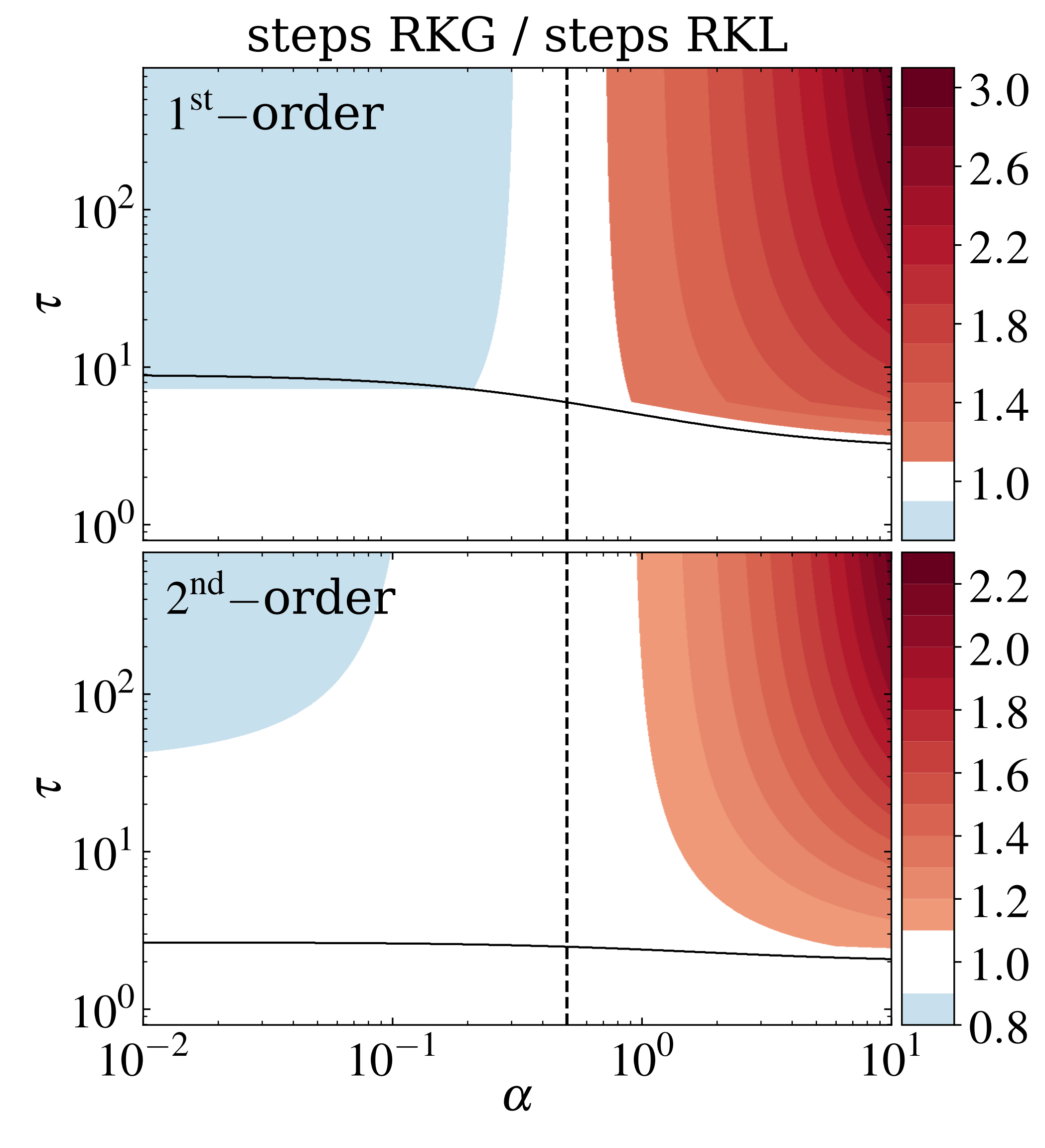}

\caption{Ratio between the RKG steps and the RKL steps (i.e., RKG with $\alpha = 0.5$, represented by the vertical dashed line) for different values of $\alpha$ and $\tau$. In the top (bottom) panel, the ratio is shown for the $1^{\rm st}$- order ($2^{\rm nd}$-order) algorithms. The solid black lines indicate the $\tau$ values below which the RKG method would require fewer than 3 steps.}
\label{fig::nsteps_rkg}%
\end{figure}

The number of substages $s$ required for a given super-step is (see Appendix \ref{app:rkgcoefficients} for the derivation):
\begin{equation}
\label{eq::substeps}
s = \left\{
\begin{array}{lcl}
-\alpha + \DS\sqrt{\alpha^2 + \tau(1 + 2\alpha)} & \qquad& 1\mathrm{st}\text{ order}, \\ \noalign{\medskip}
-\alpha + \DS\sqrt{(\alpha + 1)^2 + \tau(3 + 2\alpha)} & \qquad& 2\mathrm{nd}\text{ order},
\end{array}\right.
\end{equation}
where $\tau$ is the ratio between the hyperbolic and parabolic time steps (see Eq. \ref{eq::deltat_hp}),
\begin{equation}
\tau = \DS\frac{\Delta t_h}{\Delta t_p} = \DS\frac{1}{w_1}.
\end{equation}

Following \citet{caplan2024}, we enforce a minimum of three substages to preserve stability.
Fig. \ref{fig::nsteps_rkg} shows the ratio between the number of RKG substages and the corresponding RKL substages (i.e., $\alpha = 0.5$, indicated by the vertical dashed line) as a function of $\tau$ for different values of $\alpha$.
The top and bottom panels refer to the first- and second-order schemes, respectively.

For increasing $\alpha$ coefficient, the RKG scheme requires more substages than RKL (for a given $\tau$), thus demanding a higher computational cost in exchange for enhanced stability (as shown in the results of Section \ref{sec::benchmarks}).
Note that this ratio is only related to the evolution of the parabolic flux, while the advective flux requires the same computational time regardless of the parabolic scheme (provided that it is not fully explicit).
As a consequence, the ratios presented here do not necessarily reflect the overall simulation's computing time.

\section{Numerical benchmarks}
\label{sec::benchmarks}

\begin{figure*}[!t]
\centering
\includegraphics[width=0.97\textwidth]{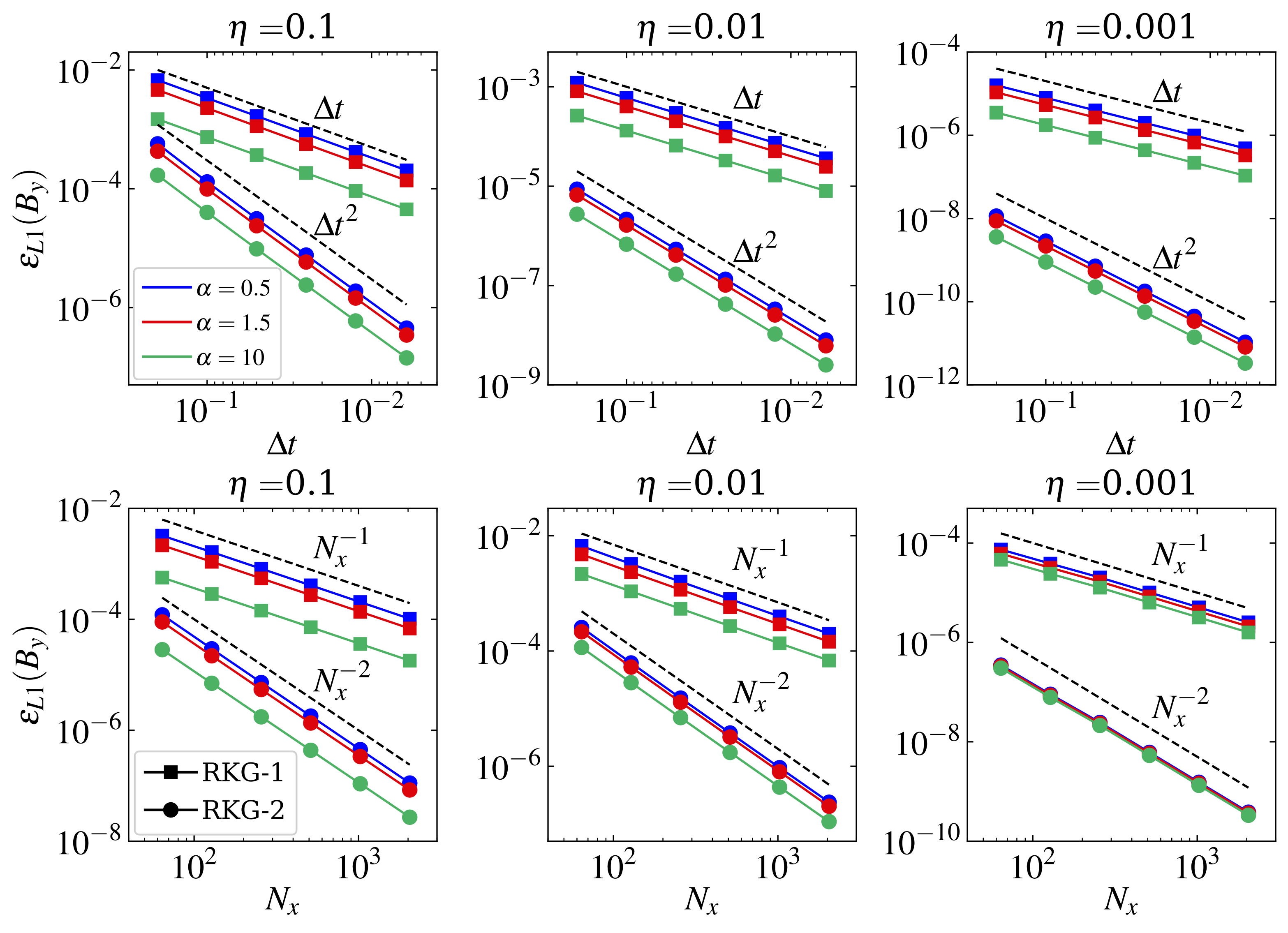}
\caption{$L_1$ errors for RKG-1 and RKG-2 on a decaying sine mode of the one-dimensional scalar decay, versus timestep (top row) and grid resolution under adaptive CFL stepping (bottom row), for different values of the resistivity $\eta$ and the Gegenbauer parameter $\alpha$.}
\label{Fig::scalar}%
\end{figure*}

To test the performance and applicability of our numerical scheme, this section presents an extensive suite of tests.
All tests reported in this section are performed with our implementation within the PLUTO code \citep{mignone2007}, linear spatial reconstruction \citep{toro2009}, $2^{\text nd}$ order Runge-Kutta time integrator \citep{gottlieb2001}, and the UCT-HLL upwind constrained transport method \citep{mignone2021}.
A comparison between RKG and STS \citep{alexiades1996} and between RKG and RKL \citep{vaidya2017} is also provided.
If not specified, the RKG and RKL schemes are assumed to be $2^{\text nd}$ order.
Unless otherwise stated, the numerical experiments therefore employ $\alpha=10$, which provides improved stability properties for the time-integration method considered.
All quantities (e.g., timestep, resistivity, ambipolar diffusion coefficient, and conserved variables, etc.) are expressed in code units.
Further validation is provided through additional numerical benchmarks presented in Appendices \ref{app::field_diff} and \ref{app::barenblatt}.

\subsection{One-dimensional scalar decay}

To test the convergence properties of the RKG schemes, we set up a series of simple one-dimensional experiments (in Cartesian geometry) in which only the diffusion operator is evolved. 
This test is the only one that was not implemented in the PLUTO code but instead run separately to isolate the behavior of the diffusion operator without the influence of the spatial reconstruction, advection, and magnetic monopole cleaning/suppression.
We start with a 1D parabolic differential equation written in the form
\begin{equation}
\pd{B_y}{t} = \eta\pd{^2B_y}{x^2},
\end{equation}
and we set the initial condition starting from the analytical solution
\begin{equation}
B_y(x,t) = \sin(mx)\exp(-\eta m^2 t),
\end{equation}
by setting $t = 0$ and $m = 4$ over a domain $x\in[0,2\pi]$ with 256 grid points.
We solve the equation numerically until the final time $t = 2$.
We report in Fig. \ref{Fig::scalar} the error and function of the timestep for different values of $\alpha$ and both the $1^{\text st}$- and $2^{\text nd}$-order RKG schemes (RKG-1 and RKG-2, respectively).

The top panel was produced by fixing the number of parabolic substeps and halving the global timestep $\Delta t$ successively from 0.2 to 0.006125.
The bottom panel was produced by computing both the hyperbolic and parabolic timesteps as functions of the resolution by fixing the fastest wave speed to $10^{-3}$, $10^{-2}$, and $10^{-1}$ for the $\eta = 0.1, 0.01$, and 0.001 cases, respectively, for different grid resolutions spanning from 64 to 2048 points.
As a consequence of the chosen wave speed, the ratio between the hyperbolic and parabolic timesteps is fixed to 900, 9, and 0.09 for the high-, intermediate-, and low-resistivity cases, respectively.
To properly link the increase in resolution with the decrease in timestep, we adopt the procedure presented in Section 4.1 of \citet{vaidya2017}.
Both algorithms show the expected convergence order, with a decreasing error for higher values of $\alpha$.

\begin{figure*}[t!]
\centering
\includegraphics[width=0.97\textwidth]{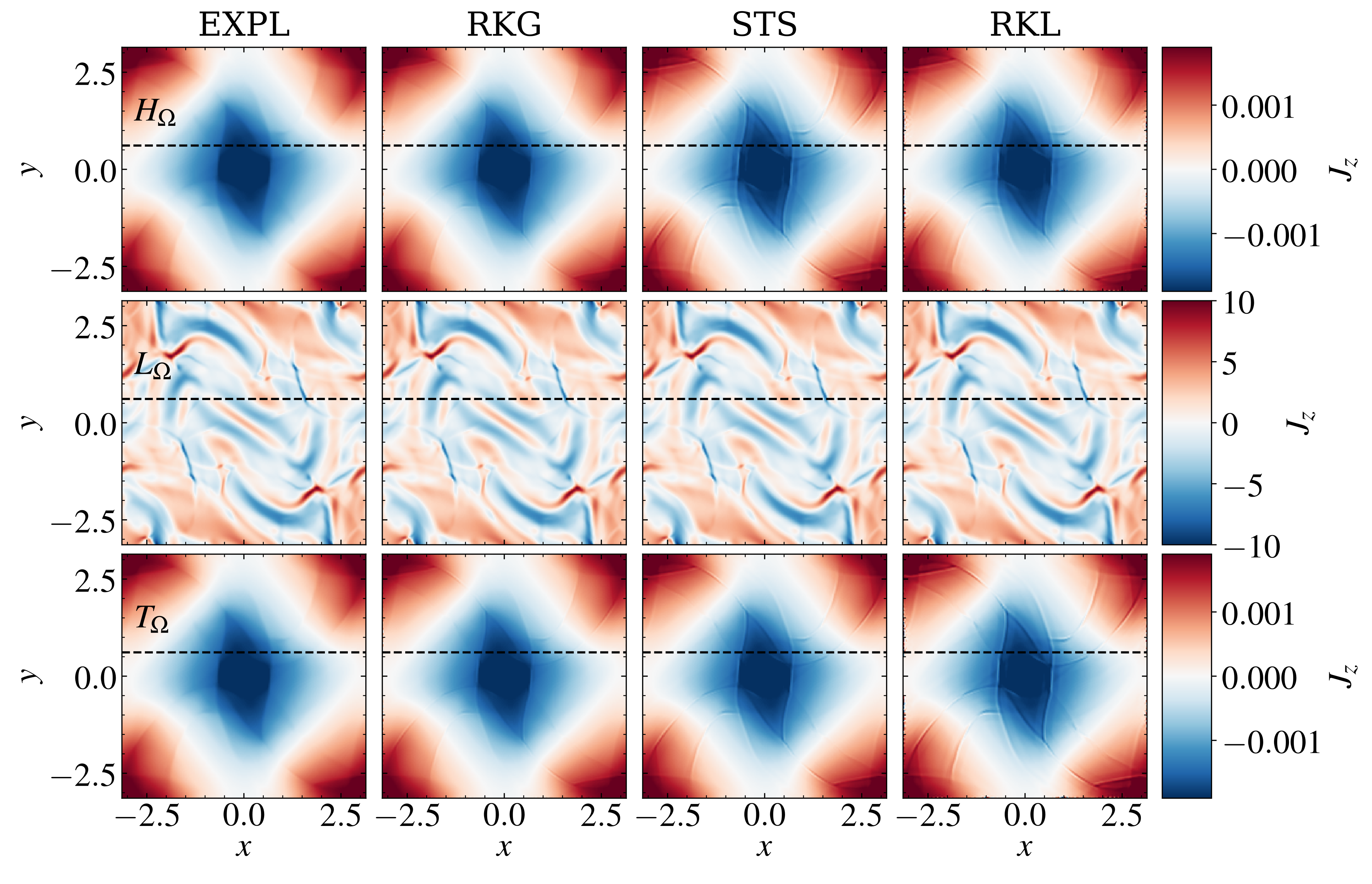}
\caption{Electric current for the resistive Orszag-Tang vortex cases. The top, middle, and bottom panels show, respectively, the cases $H_\Omega$, $L_\Omega$, and $ T_\Omega$. In the different columns, from left to right, the EXPl, RKG, STS, and RKL cases are reported, respectively. The black dashed lines represent the slice shown in Fig. \ref{Fig::ot_res_slice}.}
\label{Fig::ot_res_J}%
\end{figure*}

We also observe that the separation between the error curves corresponding to different values of $\alpha$ becomes more pronounced for larger values of $\eta$, and, more generally, for larger hyperbolic-to-parabolic timestep ratios.
Such differences can be directly linked to the temporal discretization of the two methods, since a higher $\alpha$ results in more substeps.
More specifically, the RKG-2 algorithm needs $\sim60$ substeps with $\alpha = 0.5$ (i.e., RKL scheme) and $\sim134$ steps with $\alpha = 10$ in the high-resistivity case, while both algorithms require only the minimum number of substeps (i.e., 3) in the low-resistivity regime.
As a consequence, in the high-resistivity regime ($\eta = 0.1$), the $\alpha = 10$ curve is shifted by approximately a factor of two in $N_x$ relative to the $\alpha = 0.5$ and $\alpha = 1.5$ cases, meaning that the same error requires about twice the resolution.
This is a clear confirmation that, especially when the parabolic timestep drops below the hyperbolic one, robust numerical schemes that minimize computational costs while preserving accuracy become indispensable.

\subsection{Orszag-Tang (OT) vortex}

To investigate the stability and accuracy of the RKG numerical scheme, we investigated its impact on the solution of the Orszag-Tang vortex in the presence of non-negligible Ohmic and ambipolar diffusion \citep{kayanikhoo2024}.

We assume an initial constant density and pressure distributions $\rho = 25/9$ and $p = 5/3$, while the velocity and magnetic field are set, respectively, to $\vect{v} = (-\sin y, \sin x, 0)$ and $\vect{B} = (-\sin y, \sin 2x, 0)$.
In all simulations, we employ a resolution of $200\times 200$ over the computational domain $[-\pi,\pi] \times [-\pi,\pi]$.
To effectively capture the differences between the numerical methods, we ran the simulations until $t = 6.28$ in code units.
We employ an HLL Riemann solver \citep{harten1983}, ideal equation of state, and we set the CFL number to 0.35.

\begin{figure}
\centering
\includegraphics[width=0.47\textwidth]{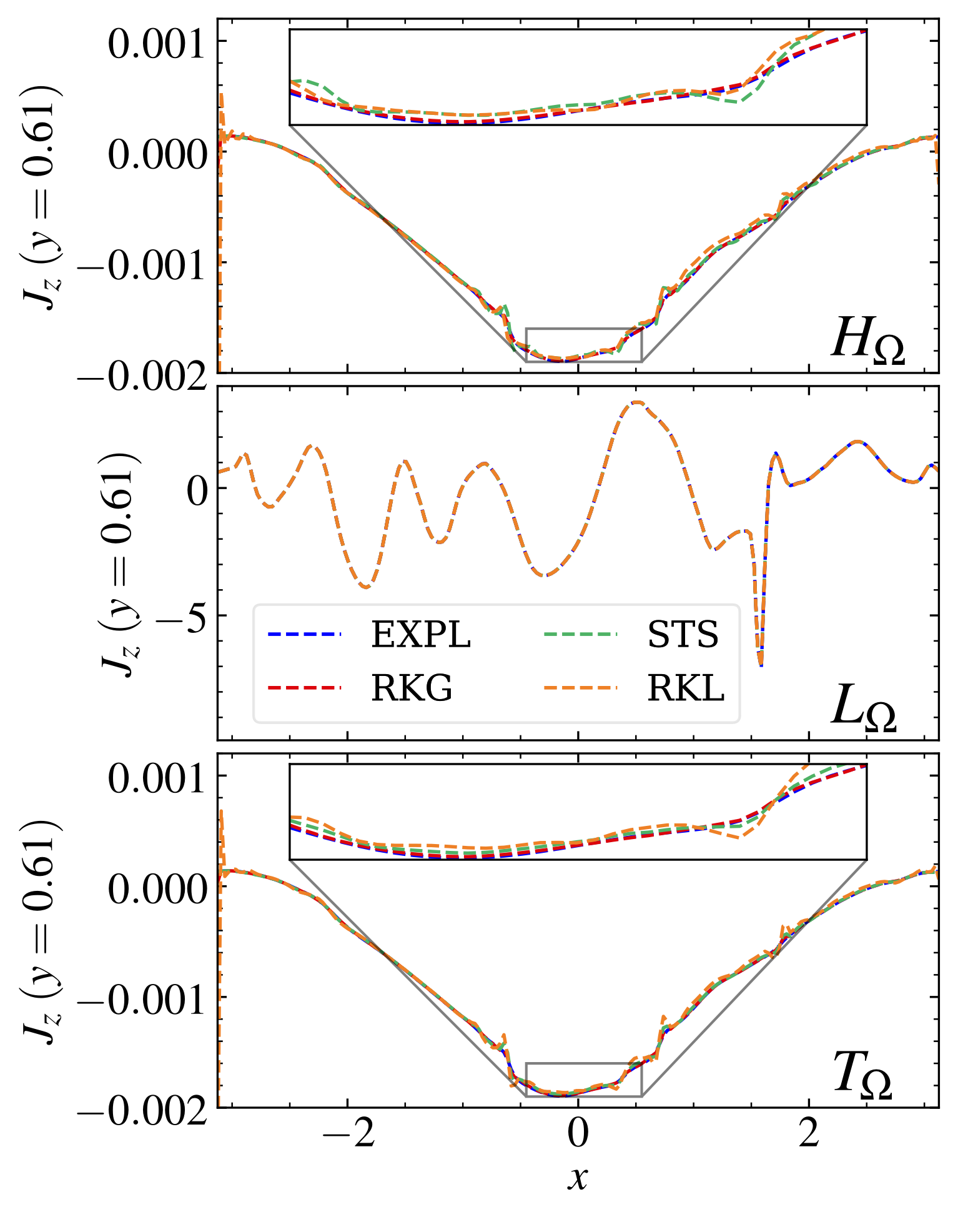}
\caption{Slices of the electric current for the resistive Orszag-Tang vortex cases. The top, middle, and bottom panels show, respectively, the cases $H_\Omega$, $L_\Omega$, and $T_\Omega$. The cuts, shown also as dashed lines in Fig. \ref{Fig::ot_res_J}, are placed at $y = \pi/5$.}
\label{Fig::ot_res_slice}%
\end{figure}

\setlength{\tabcolsep}{4pt}
\begin{table}
    \centering
    \caption{List of different OT cases and their required computing time in seconds. }
    \label{tab:OT}
\begin{tabular}{c|cccc|cccc}
 Case & $\eta_x$ & $\eta_y$ & $\eta_z$ & $\gamma_\ad$ & 
 EXPL & STS & RKL & RKG \\ 
 \hline
 L$_\Omega$ \rule{0pt}{2.5ex} & 0.01 & 0.01  & 0.01 & 0   & 117  & 88  & 105 & 106 \\ 
 H$_\Omega$ \rule{0pt}{2.5ex} & 1.0  & 1.0   & 1.0  & 0   & 2757 & 125 & 186 & 242 \\
 T$_\Omega$ \rule{0pt}{2.5ex} & 2.0  & 2.0   & 1.0  & 0   & 4630 & 149 & 202 & 312 \\
 A$_\Omega$ \rule{0pt}{2.5ex} & 0.1  & 0.1   & 1.0  & 0   & \N   & \N  & 155 & 187 \\
 L$_\ad$    \rule{0pt}{2.5ex} & 0.0  & 0.0   & 0.0  & 10  & 1372 & 125 & 166 & 206 \\ 
 H$_\ad$    \rule{0pt}{2.5ex} & 0.0  & 0.0   & 0.0  & 0.5 & 389  & 91  & 128 & 136 \\ 
\end{tabular}
\end{table}

\begin{figure}
\centering
\includegraphics[width=0.49\textwidth]{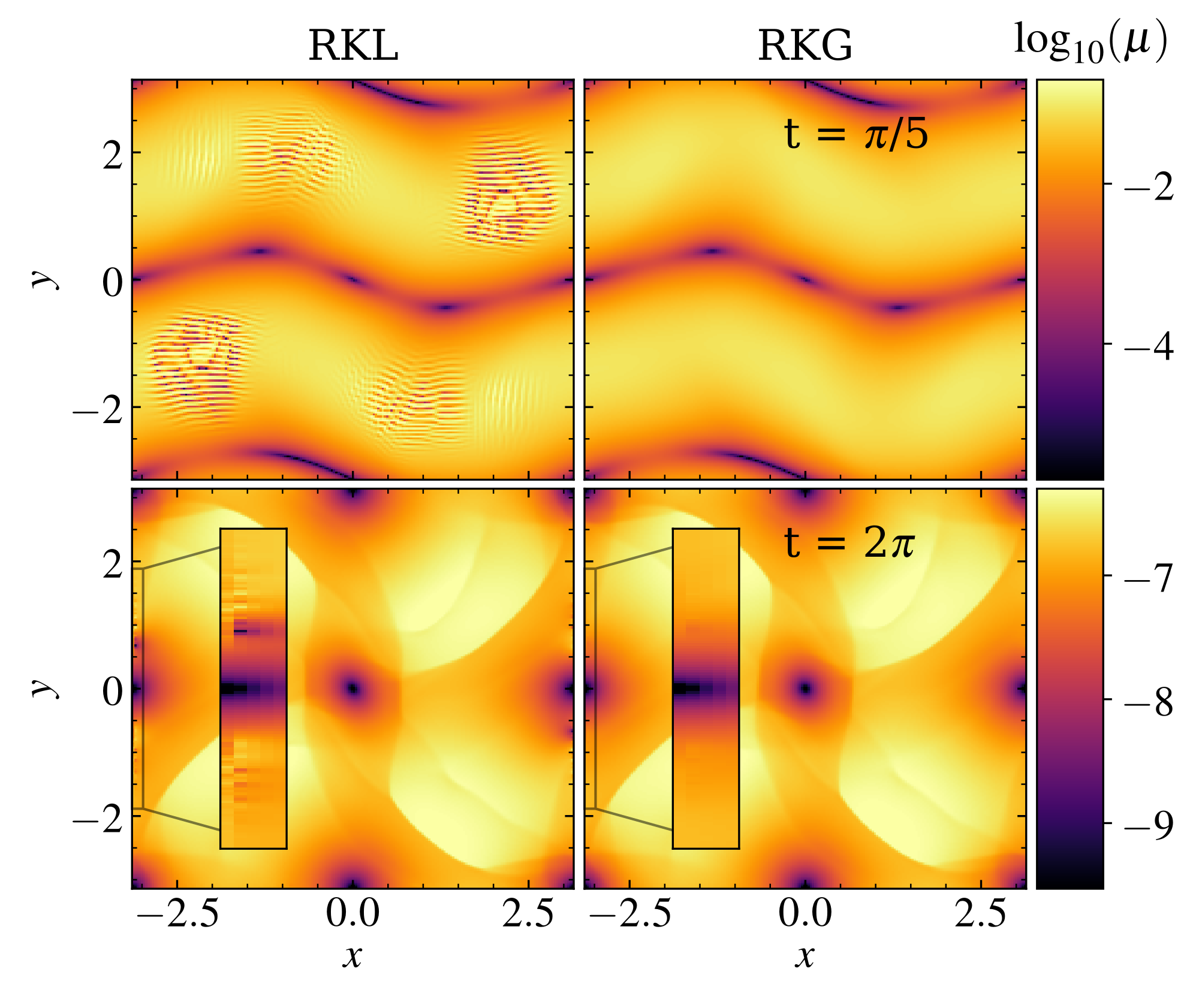}
\caption{Plasma magnetization for the case $A_\Omega$ of the Orszag-Tang vortex test at times $t = \pi/2$ (top panels) and $t = 2\pi$ (bottom panels) for the RKL (left panel) and RKG (right panels) methods.}
\label{Fig::ot_res_han}%
\end{figure}

\begin{figure}
\centering
\includegraphics[width=0.47\textwidth]{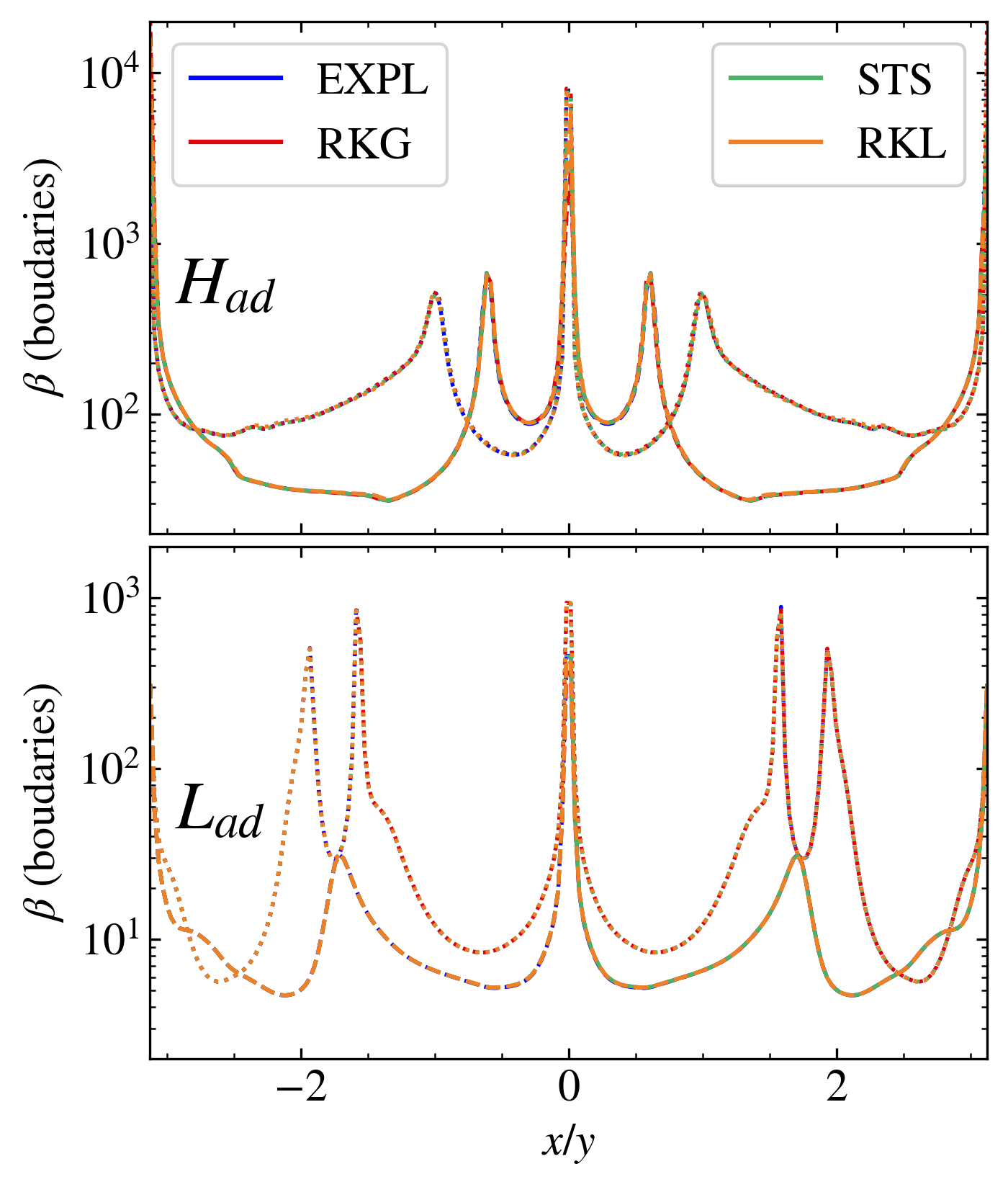}
\caption{Cuts of the plasma-$\beta$ for the ambipolar Orszag-Tang vortex tests. Dashed and dotted lines represent, respectively, $x$- and $y$-boundaries. The top and bottom panels show, respectively, the results obtained from the cases $H_{ad}$ and $L_{ad}$.}
\label{Fig::ot_amb_beta}%
\end{figure}

Note that, since the $z-$component of the magnetic field is set to zero and the problem is two-dimensional, ideally, the $x-$ and $y-$ resistivity components should not play any role in the temporal evolution; therefore, any contribution from that side must be purely numerical (see Appendix \ref{app::resistive_2D} for the algebraic details).
As a first point, the runtime (reported in Table \ref{tab:OT}) varies widely depending on the numerical scheme: while the explicit scheme is competitive in the weakly diffusive case ($L_\Omega$), it becomes up to $\sim30$ times slower than the fastest stable substepping method in the strongly diffusive runs. On the other hand, STS consistently delivers the shortest runtime among the substepping schemes, with RKL and RKG typically within a factor $\lesssim2$ of STS.

Figure \ref{Fig::ot_res_J} compares the 2D structure of the diffusive contributions (represented by the electric current $\vec{J} = \nabla\times\vec{B}$) at the final time. All schemes reproduce the same large–scale morphology, and none show signs of spurious instability or qualitative drift in the electric current. The RKG solution is qualitatively indistinguishable from the purely explicit run in both the smooth components in the high-resistivity regimes (top and bottom rows) and the sharper, filamentary structures in the low-resistivity case (middle row). By contrast, STS and especially RKL exhibit faint, arc-like or streaky features in the $H_\Omega$ and $T_\Omega$ runs, suggesting a slightly larger truncation error/dispersion in regimes where the solution is dominated by smooth gradients. Importantly, these differences remain small and do not alter the current's location or overall amplitude.

The 1D cuts in Fig. \ref{Fig::ot_res_slice} confirm this impression more quantitatively. Along the horizontal slice (dashed line in the 2D panels of Fig. \ref{Fig::ot_res_J}), the four schemes overlap over most of the domain, with only minor separations near the extrema; these are most visible in the insets, where STS/RKL show a slightly larger deviation from the explicit profile, while RKG tracks it essentially point by point. The middle diagnostic is virtually identical for all integrators, indicating that the time-stepping choice does not affect the sharp, shock-like features dominated by the advective component of the induction equation. Overall, the Orszag–Tang test supports the conclusion that RKG preserves the accuracy of explicit solutions in a strongly nonlinear, mixed hyperbolic–parabolic setting, while STS/RKL introduces only small, localized, geometry-independent artifacts in the smoother diffusive channels.

Figure \ref{Fig::ot_res_han} shows the evolution of the plasma magnetization of case $A_\Omega$, for which both the purely explicit integrator and STS become unstable and crash before reaching the final time. In this more restrictive regime caused by the strong anisotropy of the Ohmic resistivity, the stability margin of STS is not sufficient to prevent runaway oscillations. By contrast, both stabilized Runge–Kutta super-stepping methods (i.e., RKL and RKG) remain more stable and complete the run.
A qualitative difference between the RKL and the RKG approaches is already visible at early times ($t = \pi/5$, top panels of Fig.\ref{Fig::ot_res_han}): RKL develops small-scale, grid-aligned ripples in several regions, indicative of dispersive overshoots near steep gradients, whereas RKG produces a smooth field with no comparable high-frequency contamination. At late times ($t=2\pi$, bottom panels of Fig.\ref{Fig::ot_res_han}), the two stable schemes yield consistent large-scale morphology, but the residual striping in RKL persists inside the highlighted slice region at the boundaries, while RKG remains essentially free of such artifacts. 

For the ambipolar Orszag–Tang configurations, Fig. \ref{Fig::ot_amb_beta} shows plasma-$\beta$ profiles extracted along two representative boundary cuts (dashed and dotted lines for the $x$- and $y$-boundaries, respectively), used to probe the evolution of the most magnetically dominated (and potentially unstable) regions. At moderate ambipolar strength (top panel), all numerical schemes coincide for a given cut. The explicit, RKG, STS, and RKL curves are essentially indistinguishable, indicating that substepping schemes do not alter either the amplitude or the placement of low-$\beta$ structures in this regime.
The same conclusion holds in the weak ambipolar case (bottom panel). Although the dynamic range of $\beta$ increases and the profiles become sharper, the four methods again overlap for both boundary directions, reproducing the same extrema and gradients.

\subsection{Alfv\'en waves}

As a final benchmark, we consider the propagation of linear Alfvén waves in a weakly ionized plasma, i.e., including either Ohmic resistivity or ambipolar diffusion.
The initial conditions are adapted from \citet{choi2009}.
We start with a background plasma density $\rho = 1$ and magnetic fields $\vec{B} = B_0\hat{x}$, with $B_0 = 1$, yielding a characteristic Alfv\'en speed $v_A = 1/\sqrt{2}$.
A standing wave is initialized along the diagonal on the $x-y$ plane through a transverse velocity perturbation of the form
\begin{equation}
    \vec{v} = v_pv_A\sin(k_xx + k_yy)\hat{z},
 \end{equation}
where the initial peak amplitude has been set to $v_p = 0.1$.
The domain consists of a three-dimensional box with $x, y, z \in [0, 1]$ and a resolution of $128$ cells per direction.
We employ a Roe Riemann solver \citep{roe1981}, isothermal equation of state with $c_s = 1$, and we set the CFL number to 0.3.
However, we noticed that the RKL scheme performs significantly worse than the other algorithms in the strongly resistive and ambipolar regimes, failing to correctly reproduce the damping rate even after a few oscillation periods.
For this reason, we reduced the CFL number of the RKL scheme to 0.2 in the strongly resistive and ambipolar cases to limit the large numerical errors observed at higher CFL values and improve accuracy, at the expense of increased computational cost.
Periodic boundary conditions are applied in all directions.

In the strong coupling approximation \citep{balsara1996,lesaffre2007}, and for the background field and perturbation considered here, the linearized non-ideal MHD equations yield the same dispersion relation for Alfvén waves damped by Ohmic resistivity or ambipolar diffusion, differing only in the effective diffusivity coefficient.
Assuming that the perturbation is $\propto\exp(ikx - i\omega t)$, the dispersion relation can be written as
\begin{equation}
\label{eq::disp_rel_alfven}
    \omega + ik^2\eta\omega - v_Ak^2 = 0,
\end{equation}
where $k$ is the component of the wavevector parallel to the background magnetic field and $\omega = \omega_R + i\omega_I$ is the complex angular frequency of the wave.
In the resistive case, $\eta$ corresponds to the Ohmic resistivity $\eta_\Omega$, while for the ambipolar diffusion case, it is given by
\begin{equation}
    \eta = \DS\frac{v_A^2}{\gamma_\ad\rho_i}.
\end{equation}

Solving the dispersion relation yields a damped oscillatory solution provided that $(v_Ak)^2$ is larger than the square of the damping rate $\gamma_d$. The latter is given by
\begin{equation}
    \gamma_d = \DS\frac{1}{2}\eta k^2.
\end{equation}
In this regime, the amplitude of the transverse magnetic perturbation evolves as
\begin{equation}
    h(t) = h_0\sin(\omega_Rt)\exp(\omega_it),
\end{equation}
where $\omega_R$ and $\omega_i$ are obtained by solving the dispersion relation reported in Eq. \ref{eq::disp_rel_alfven}.
The initial velocity perturbation fixes the factor $h_0$:
\begin{equation}
    h_0 = \DS\frac{kv_0}{\omega_R\sqrt{2}},
\end{equation}
\begin{figure}
\centering
\includegraphics[width=0.47\textwidth]{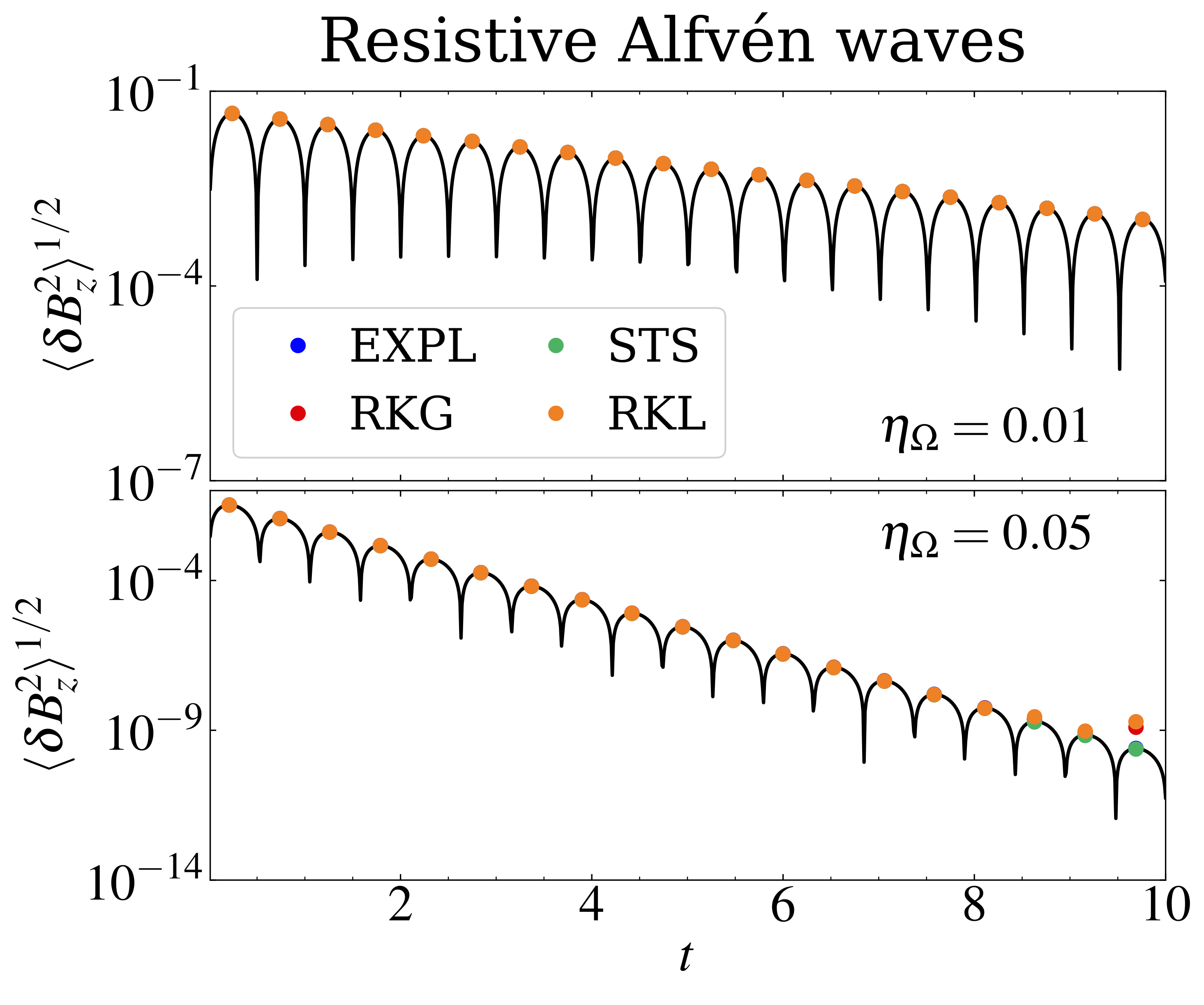}
\caption{Temporal evolution of the root-mean-square transverse magnetic perturbation 
$\langle B_z^2 \rangle^{1/2}$ for resistive Alfv\'en waves with 
$\eta_\Omega = 0.01$ (top) and $\eta_\Omega = 0.05$ (bottom). 
Solid lines show the analytical solution, while symbols denote numerical results 
(EXPL, RKG, STS, and RKL) sampled at successive oscillation maxima.}
\label{Fig::res_alfven}%
\end{figure}
\begin{figure}
\centering
\includegraphics[width=0.47\textwidth]{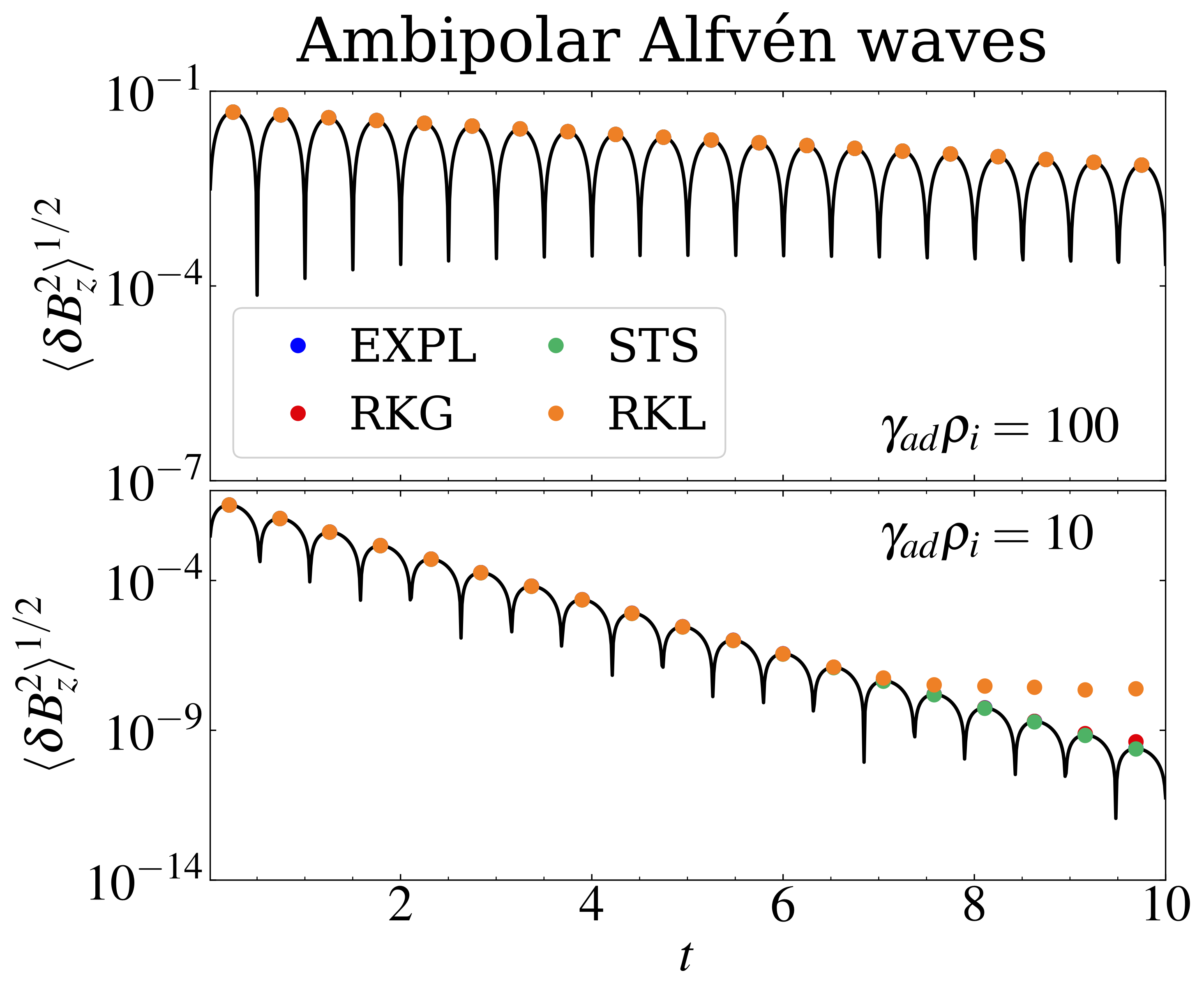}
\caption{Same as Fig. \ref{Fig::res_alfven} but for the ambipolar case.}
\label{Fig::amb_alfven}%
\end{figure}
where $v_0 = v_pv_A$.
Note that if one sets $h_0 = v_0$ directly, the analytical solution slightly underestimates the numerical results at early times due to small deviations between the numerical initialization and the exact linear eigenmode.
However, this mismatch does not affect the measured damping rate.

\setlength{\tabcolsep}{4pt}
\begin{table}
    \centering
    \caption{\change{List of different Alfv\'en waves test cases and their required computing time in minutes. Note that, due to the reduced maximum CFL, the RKL scheme yields a lower maximum timestep.}}
    \label{tab:AW}
\begin{tabular}{c|cccc|cccc}
 Case & $\eta_x$ & $\eta_y$ & $\eta_z$ & $\gamma_\ad\rho_i$ & 
 EXPL & STS & RKL & RKG \\ 
 \hline
 L$_\Omega$ \rule{0pt}{2.5ex} & 0.01 & 0.01  & 0.01 & 0   & 129  & 58  & 67 & 67 \\ 
 H$_\Omega$ \rule{0pt}{2.5ex} & 0.5  & 0.5   & 0.5  & 0   & 417 & 64 & 115 & 107 \\
 L$_\ad$    \rule{0pt}{2.5ex} & 0.0  & 0.0   & 0.0  & 100 & 172 & 61 & 71 & 70 \\ 
 H$_\ad$    \rule{0pt}{2.5ex} & 0.0  & 0.0   & 0.0  & 10  & 1185  & 89  & 171 & 182 \\ 
\end{tabular}
\end{table}

\change{
A summary of the test cases and their corresponding numerical runtimes is reported in Table \ref{tab:AW}.
As a first consideration, the EXPL scheme is consistently the most expensive across all test cases, with computing times ranging from roughly 2 to 20 times longer than those of the other numerical schemes.
The STS scheme is the fastest in all cases, while RKL and RKG show comparable performance.
It is worth noting that the RKL scheme incurs additional computational overhead in the strongly resistive and ambipolar cases, as a reduced CFL number of 0.2 was required to ensure accurate results, limiting the maximum allowed timestep and thereby increasing the total number of numerical steps required to complete the simulation.
}

Figures \ref{Fig::res_alfven} and \ref{Fig::amb_alfven} show, respectively, the temporal evolution of the root-mean-square transverse magnetic-field component, $\langle B_z^2\rangle^{1/2}$.
The markers denote numerical results sampled at successive oscillation maxima, while solid lines indicate the analytical solution.
Sampling at the maxima removes phase shifts between the numerical and analytical solutions, thereby isolating the exponential decay and enabling a direct comparison of the damping rates.
In the resistive Alfv\'en wave test, all schemes accurately reproduce the analytical decay in the weakly diffusive regime ($\eta_\Omega = 0.01$, case $L_\Omega$).
Differences between schemes are negligible over the full integration time. At higher resistivity ($\eta_\Omega = 0.05$, case $H_\Omega$), the role of the numerical scheme becomes significant.
The explicit scheme continues to match the analytical solution, but the CFL conditions strongly restrict the timestep.
The STS scheme performs best among the accelerated methods, remaining in close agreement with the analytical solution over most of the evolution.
This behavior is consistent with the fact that STS advances the electromotive force through successive updates without accumulating polynomial sums, which can degrade the accuracy of the numerical schemes in the presence of very weak magnetic fields.
RKG remains in close agreement with the explicit solution over most of the evolution, with deviations appearing only at very low magnetic field amplitudes.
The RKL scheme, despite the lower CFL, also exhibits a departure from the analytical envelope in the strongly diffusive regime.

The same trends are observed in the ambipolar diffusion test.
In the weakly diffusive regime ($\gamma_\ad \rho_i = 100$, case $L_\ad$), all schemes reproduce the analytical decay with comparable accuracy.
For stronger ambipolar diffusion ($\gamma_\ad\rho_i = 10$, case $H_\ad$), the explicit scheme and STS are still able to closely follow the analytical solution over several decades in amplitude.
RKG also remains in good agreement, with deviations limited to the lowest magnetic field values.
By contrast, RKL shows significant departures from the expected exponential decay at late times, remaining the least accurate of the four schemes and exhibiting peak amplitudes that systematically exceed the analytical solution.

Overall, these results show that while all schemes perform well in weakly diffusive regimes, differences emerge in strongly diffusive regimes at low magnetic fields.
STS provides the most accurate behavior among the accelerated schemes, while RKG reproduces the correct damping over most of the evolution with minor deviations at low amplitudes.
The RKL scheme, on the other hand, exhibits larger systematic errors in the strongly damped regime, even with a reduced CFL (and thus a smaller timestep).

\section{Astrophysical applications}
\label{sec::applications}

In this section, we test the RKG numerical scheme in benchmarks relevant to astrophysics simulations to assess its robustness and accuracy in more complex scenarios.

\subsection{Resistive Magnetic reconnection}

Magnetic reconnection has long been proposed as one of the most relevant physical processes responsible for the rapid release of magnetic energy and the efficient acceleration of particles in a broad range of astrophysical environments \citep{zweibel2009}.
In the non-relativistic regime, it is associated with phenomena such as solar flares and coronal mass ejections \citep{wang2023}, and plays a key role in the dynamics of magnetized turbulence and current sheet formation in accretion disk coronae and the interstellar medium \citep{loureiro2016}.

\begin{figure}
\centering
\includegraphics[width=0.49\textwidth]{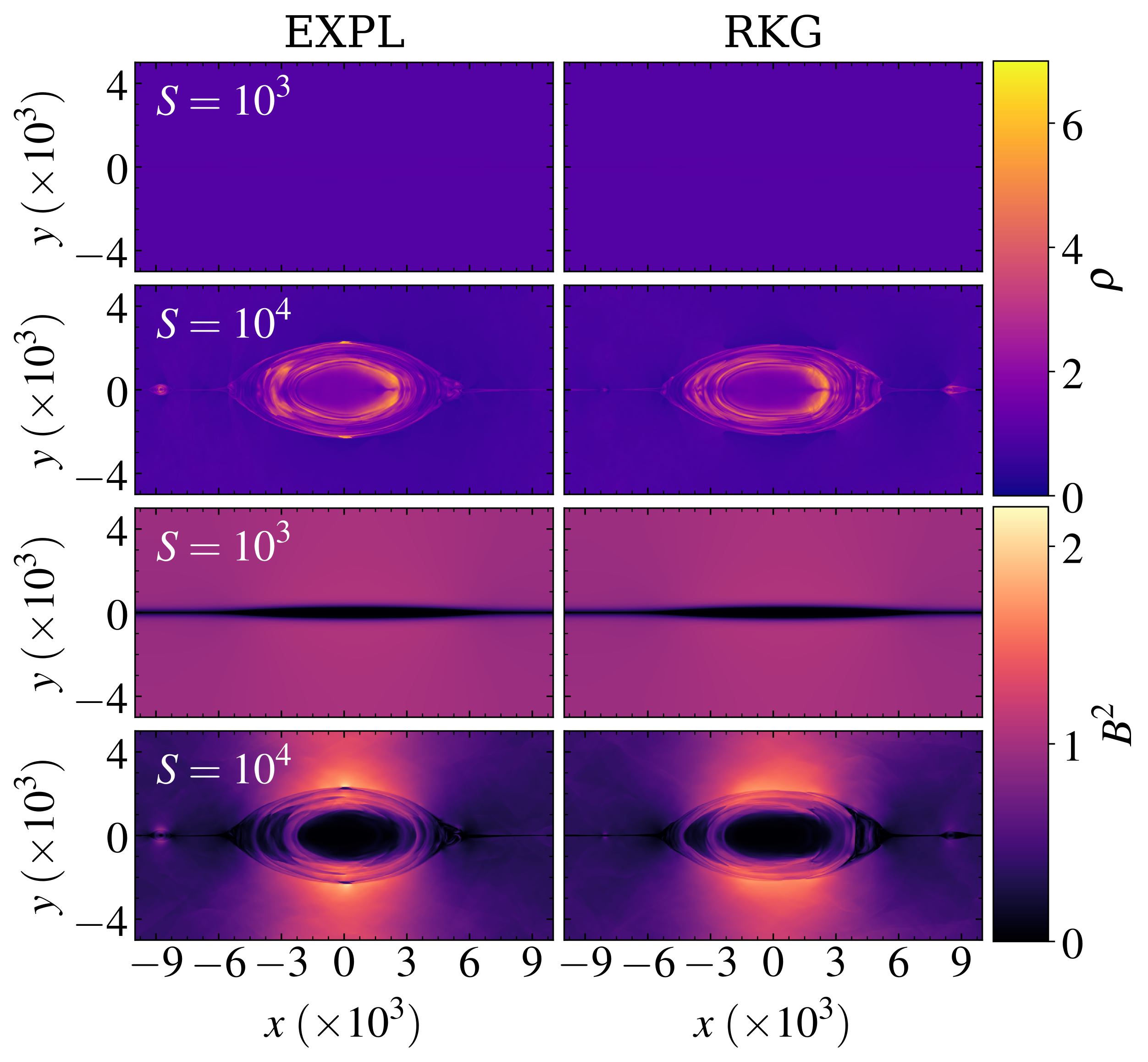}

\caption{Comparison between the EXPL (left) and RKG (right) schemes for magnetic reconnection at Lundquist numbers $S=10^3$ (first and third rows) and $S=10^4$ (second and fourth rows). The top two rows show the density $\rho$, while the bottom two rows display the magnetic energy density $B^2$. Snapshots are shown at time $t = 3\times10^5$.}
\label{Fig::rec_display10}%
\end{figure}

\begin{figure}
\centering
\includegraphics[width=0.49\textwidth]{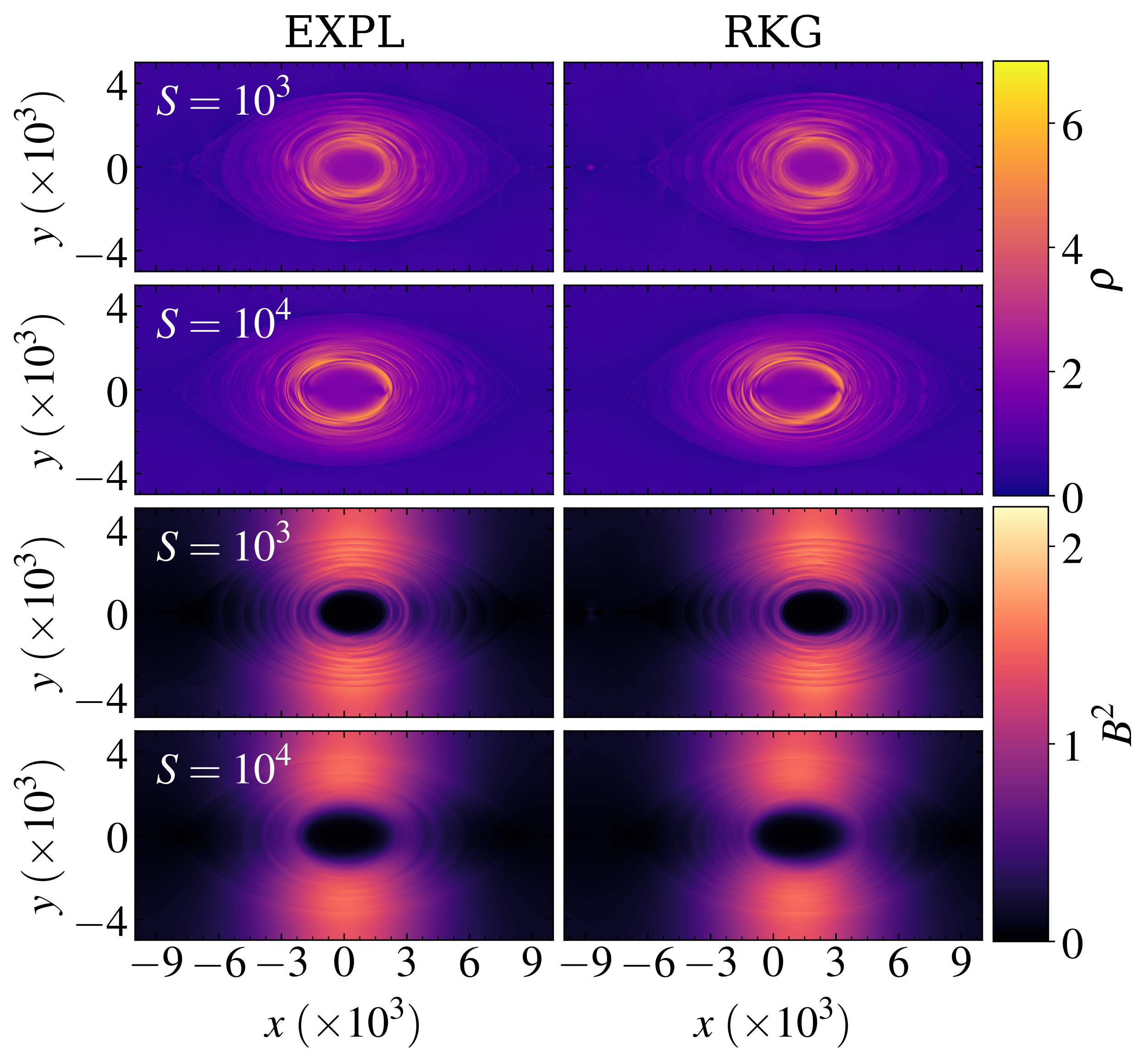}

\caption{Same as Fig. \ref{Fig::rec_display10} but at the final time $t = 6\times10^5$.}
\label{Fig::rec_display20}%
\end{figure}

\begin{figure}
\centering
\includegraphics[width=0.49\textwidth]{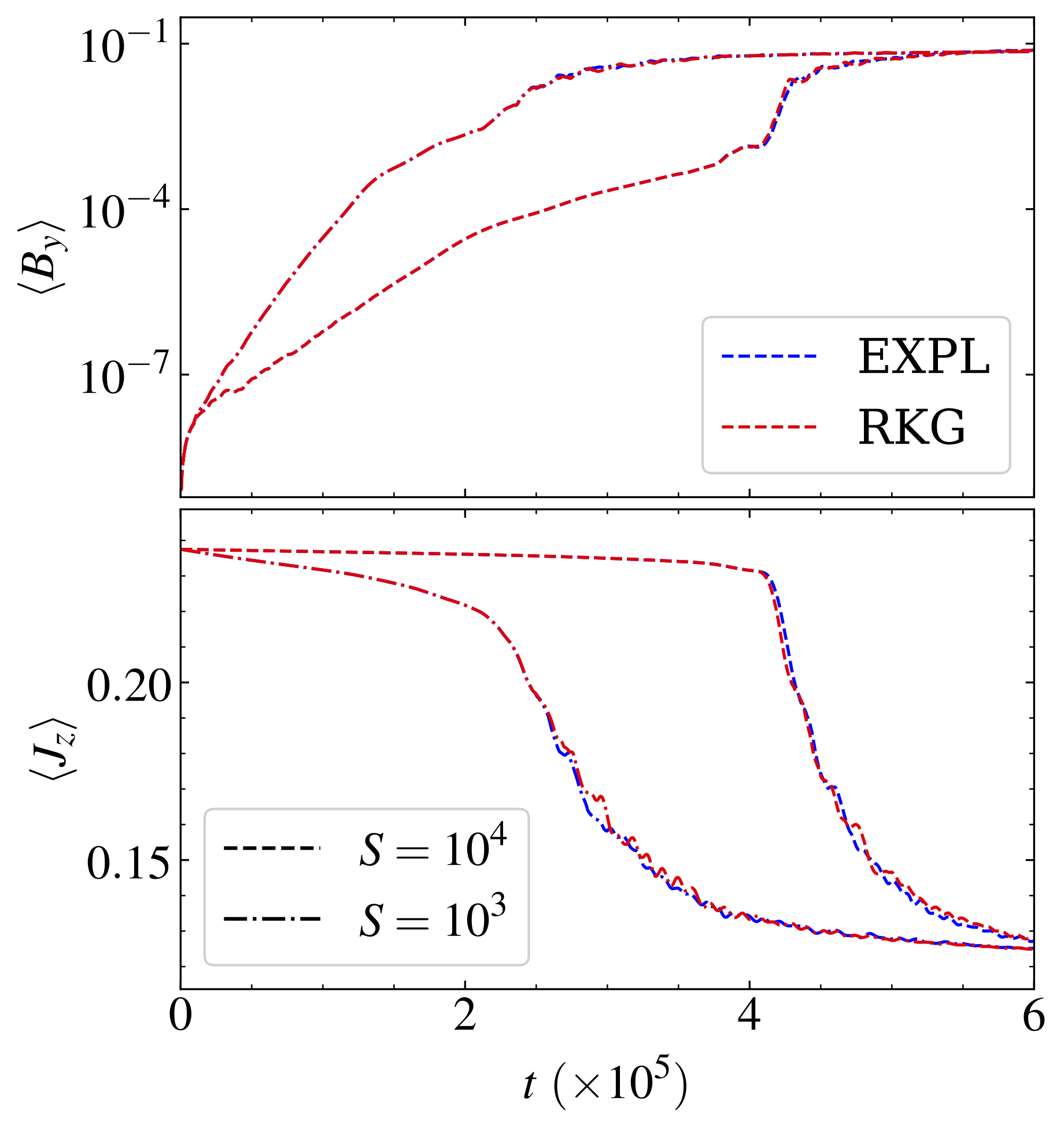}

\caption{Temporal evolution of the domain-averaged transverse magnetic field $\langle B_y \rangle$ (top) and out-of-plane current density $\langle J_z \rangle$ (bottom) for magnetic reconnection at $S=10^3$ and $S=10^4$. Blue dashed lines denote the EXPL scheme, while red dashed lines correspond to RKG. The dashed and dotted-dashed line styles identify, respectively, the high and low values of the Lundquist number.}
\label{Fig::rec_gr}%
\end{figure}

Our setup follows closely the one described in \citet{puzzoni2021,puzzoni2022}.
For this test, we adopt WENOZ reconstruction \citep{borges2008}, a third-order Runge–Kutta integrator \citep{gottlieb2001}, an ideal equation of state, the UCT-HLLD scheme \citep{mignone2021}, a CFL number of 0.4, and the HLLD Riemann solver \citep{miyoshi2005}.
We perform 2-dimensional simulations on a Cartesian mesh with a resolution of $N_x\times N_y = 1536\times768$, where $x\in[-12.8,12.8]$ and $y\in[-6.4,6.4]$, with periodic boundary conditions along $x$ and reflective boundaries along $y$.
The initial equilibrium consists of a Harris-type current sheet, with uniform density $\rho = 1$ and vanishing velocity $\vec{v} = 0$.
The magnetic field reverses across the midplane following
\begin{equation}
\label{eq::rec_Beq}
    B_x = B_0\tanh(y/a),
\end{equation}
where $a = 250$ is the half-thickness of the current sheet.
No guide field $B_z$ is applied.
The gas pressure is then chosen to enforce transverse total-pressure balance:
\begin{equation}
p(y) = \frac{B_0^2}{2}(\beta + 1) - \frac{B_x(y)^2}{2},
\end{equation}
where $\beta = 0.1$ is the input plasma beta parameter.
Reconnection is triggered by introducing 20 small-amplitude modes along $x$, with different wavenumbers $k$.
To ensure that the initial magnetic field is divergence-free, we reconstruct it from a vector potential
\begin{equation}
    A_z(x,y) = A_0(y) + \delta A_z(x,y),
\end{equation}
where $A_0 = aB_0\log([\cosh(y/a)]$ corresponds to the equilibrium field set in Eq. \ref{eq::rec_Beq} and 
\begin{equation}
    \delta A_z(x,y) = \DS\frac{\epsilon B_0}{N_m}\DS\sum_{m = 0}^{N_m}\DS\frac{1}{k}\sin(kx + \phi_m)\sech\left(\DS\frac{y}{a}\right).
\end{equation}
Here $N_m = 20$ is the number of modes, $\epsilon = 10^{-3}$ is the perturbation amplitude, $\phi_m\in[0,2\pi]$ is the random phase (different for every mode), and $k = 2\pi(m+1)/L$ is the value of each wavenumber with $L = 25.6$ being the domain length along the $x-$axis.
The resistivity is defined by fixing the Lundquist number
\begin{equation}
S = \DS\frac{v_Aa}{\eta},
\end{equation}
where we set $v_A = 1$.
We choose 4 configurations: 2 with $S = 10^3$ \change{(case $S3$)} and 2 with $S = 10^4$ \change{(case $S4$)}.
For each chosen value of the Lundquist number, we performed one simulation with the EXPL scheme and one with the novel RKG scheme (fixing $\alpha = 10$).
Note that the reconnection process strongly depends on the magnitude of the resistivity. If diffusion processes dominate over advection, the current sheet will broaden without triggering any instability.
For this reason, this test is not well-suited for assessing the computational speed-up provided by the different time integration schemes, as shown in Table \ref{tab:reconnection}.
Instead, it is designed to examine their numerical behavior in configurations that are very close to the ideal MHD regime.

\setlength{\tabcolsep}{4pt}
\begin{table}
    \centering
    \caption{\change{List of different magnetic reconnection cases and their required computing time in minutes.}}
    \label{tab:reconnection}
\begin{tabular}{c|cc|cc}
 Case & $S$ & $\eta_\Omega$ & EXPL  & RKG \\ 
 \hline
 $S3$ \rule{0pt}{2.5ex} & $10^3$ & 0.4  & 1476 & 1588 \\ 
 $S4$ \rule{0pt}{2.5ex} & $10^4$ & 0.04 & 1153 & 1224 \\
\end{tabular}
\end{table}

Figures \ref{Fig::rec_display10} and \ref{Fig::rec_display20} display the density (top panels) and magnetic energy (bottom panels) at $t = 3\times10^5$ and $t = 6\times10^5$, respectively. For each snapshot, the solutions computed with the EXPL (left) and RKG (right) time integrators are compared for different Lundquist numbers.
The comparison of the resolved fields shows a very close agreement between the EXPL and RKG schemes.
For both Lundquist numbers and at both output times, the overall structure of the reconnecting region is essentially the same.
In particular, at $t = 6\times10^5$, when the nonlinear stage is fully developed and plasmoid-like, concentric structures are clearly visible, and the two integrators still produce nearly identical configurations.
In both regimes, switching to the RKG scheme does not introduce any noticeable smoothing of gradients, additional small-scale features, or distortions of the current sheet.
The magnetic energy distribution and density layering, which are sensitive to current-sheet evolution and magnetic topology changes, display the same spatial organization and comparable extrema in both methods.

To reinforce these conclusions, we show in Fig. \ref{Fig::rec_gr} the evolution of the average magnetic field component $B_y$ (top panel) and the average electric current component $J_z$ (bottom panel), calculated in the same fashion as \citet{berta2026}.
The temporal evolution of the averaged $B_y$ shows identical growth rates, onset times of the nonlinear phase, and saturation levels for EXPL and RKG, for both $S = 10^3$ and $S = 10^4$. Likewise, the evolution of the averaged current density $J_z$ captures the same transition from the initial quasi-equilibrium to the reconnection-driven drop and final relaxed state.
These results indicate that, in an advection-dominated reconnection regime, the RKG integrator serves as a non-intrusive temporal discretization, preserving the effective dissipation rate and the nonlinear energy conversion dynamics of the explicit method.
The dominant differences in the simulations arise from the Lundquist number rather than from the choice of time integrator. The separation in transition times and relaxation behavior between the $S = 10^3$ and the $S = 10^4$ cases is clearly resolved, while EXPL and RKG remain superposed within each case.

\subsection{Ambipolar Magneto-rotational instability}

\begin{figure}
\centering
\includegraphics[width=0.47\textwidth]{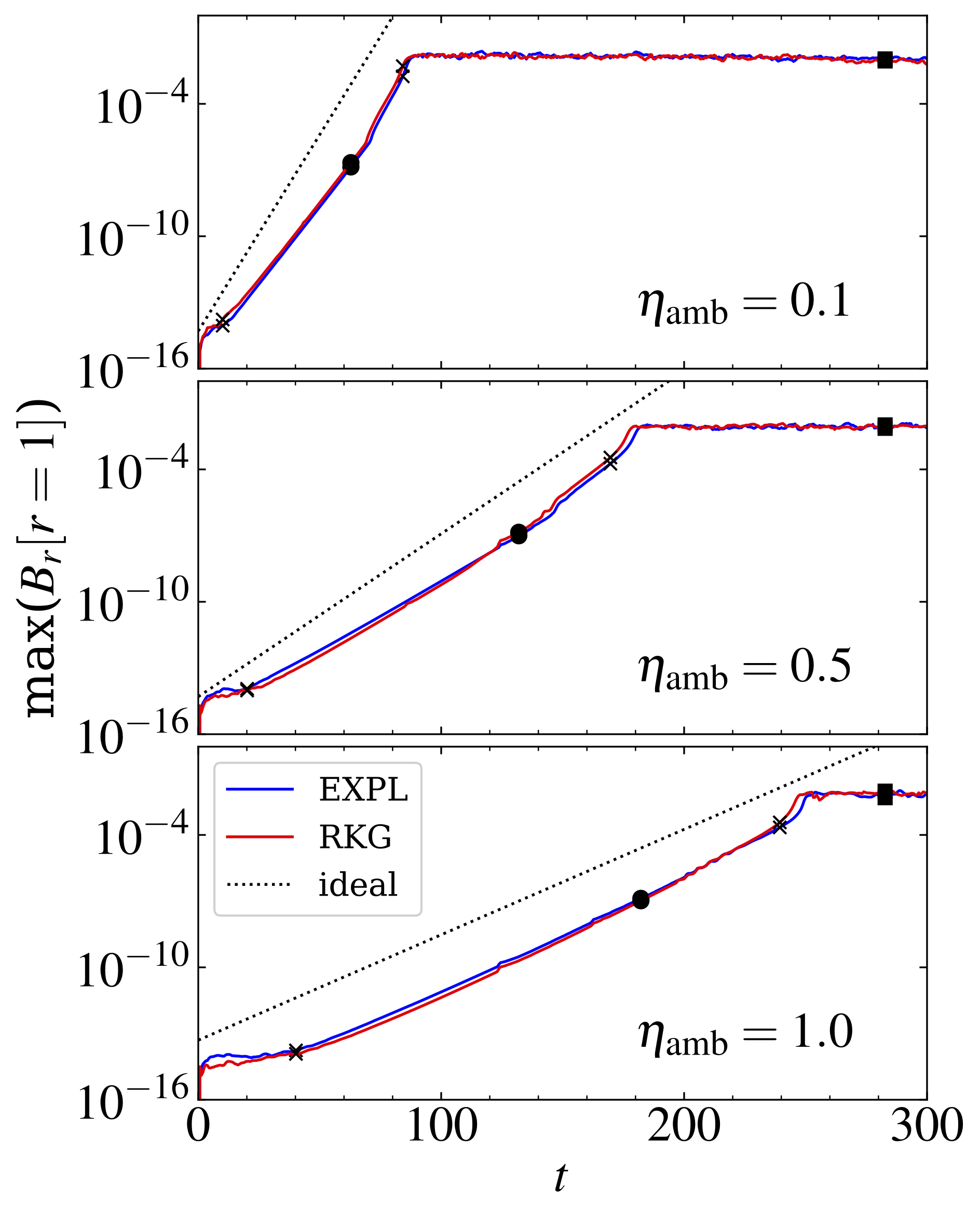}
\caption{Temporal evolution of $\max(B_r[r = 1])$ for $\eta_\ad=0.1$, $0.5$, and $1.0$ (top to bottom), comparing the EXPL (blue) and RKG (red) schemes for ambipolar MRI. The dotted line indicates the theoretical growth rate. Black crosses mark the time interval used to fit the linear growth rate; filled circles and squares denote the time corresponding to Fig.~\ref{Fig::mri_display1} and Fig.~\ref{Fig::mri_display2}, respectively.}
\label{Fig::mri_gr}%
\end{figure}

\begin{figure}
\centering
\includegraphics[width=0.49\textwidth]{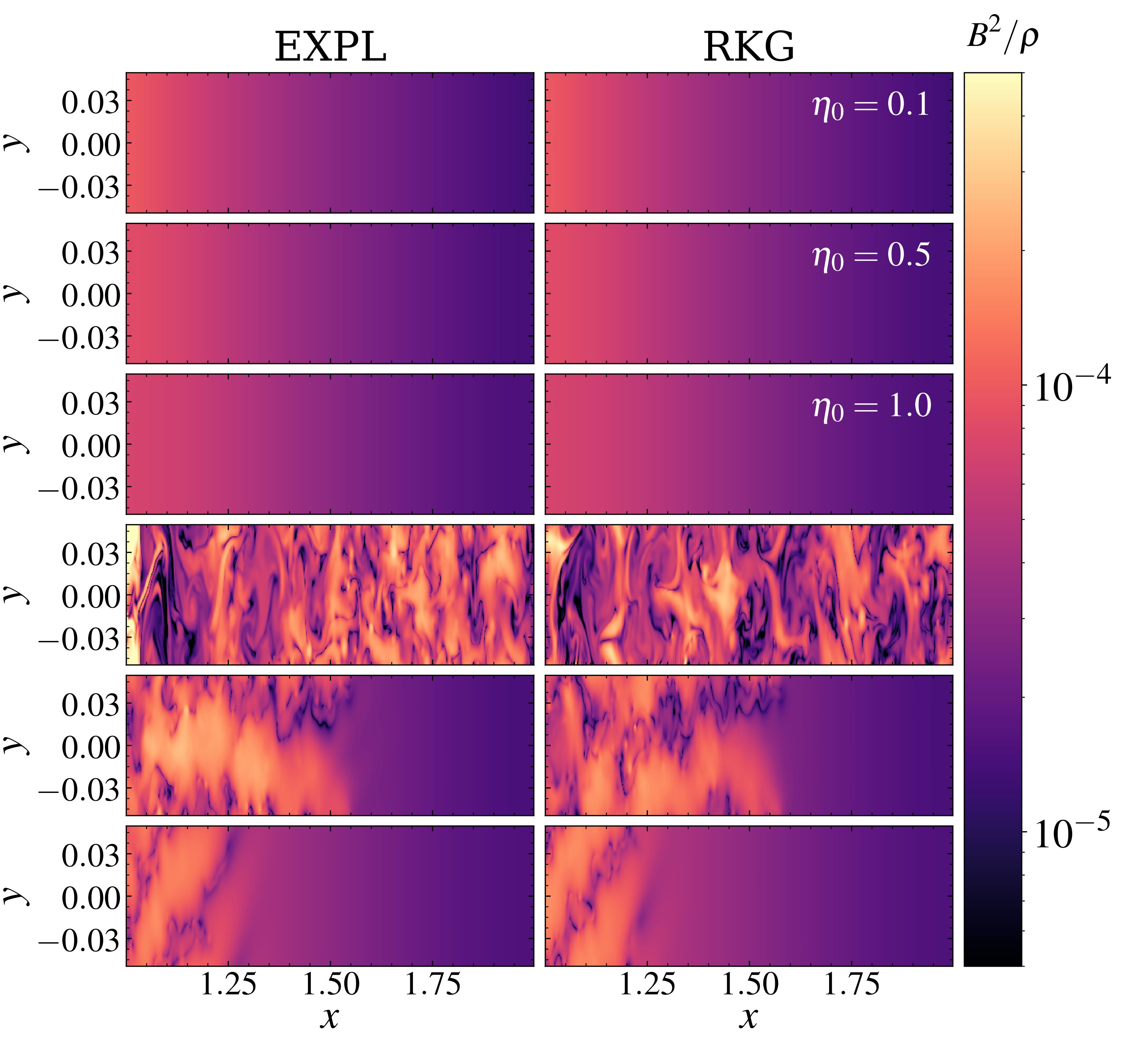}
\caption{Spatial distribution of the local Alfv\'en speed squared, $B^2/\rho$, for ambipolar MRI, shown for $\eta_0 = 0.1$, $0.5$, and $1.0$. The left and right columns compare the EXPL and RKG schemes, respectively. Top and bottom panels correspond to the times marked by the filled circles and filled squares in Fig.~\ref{Fig::mri_gr}, respectively.}
\label{Fig::mri_display1}%
\end{figure}

\begin{figure}
\centering
\includegraphics[width=0.49\textwidth]{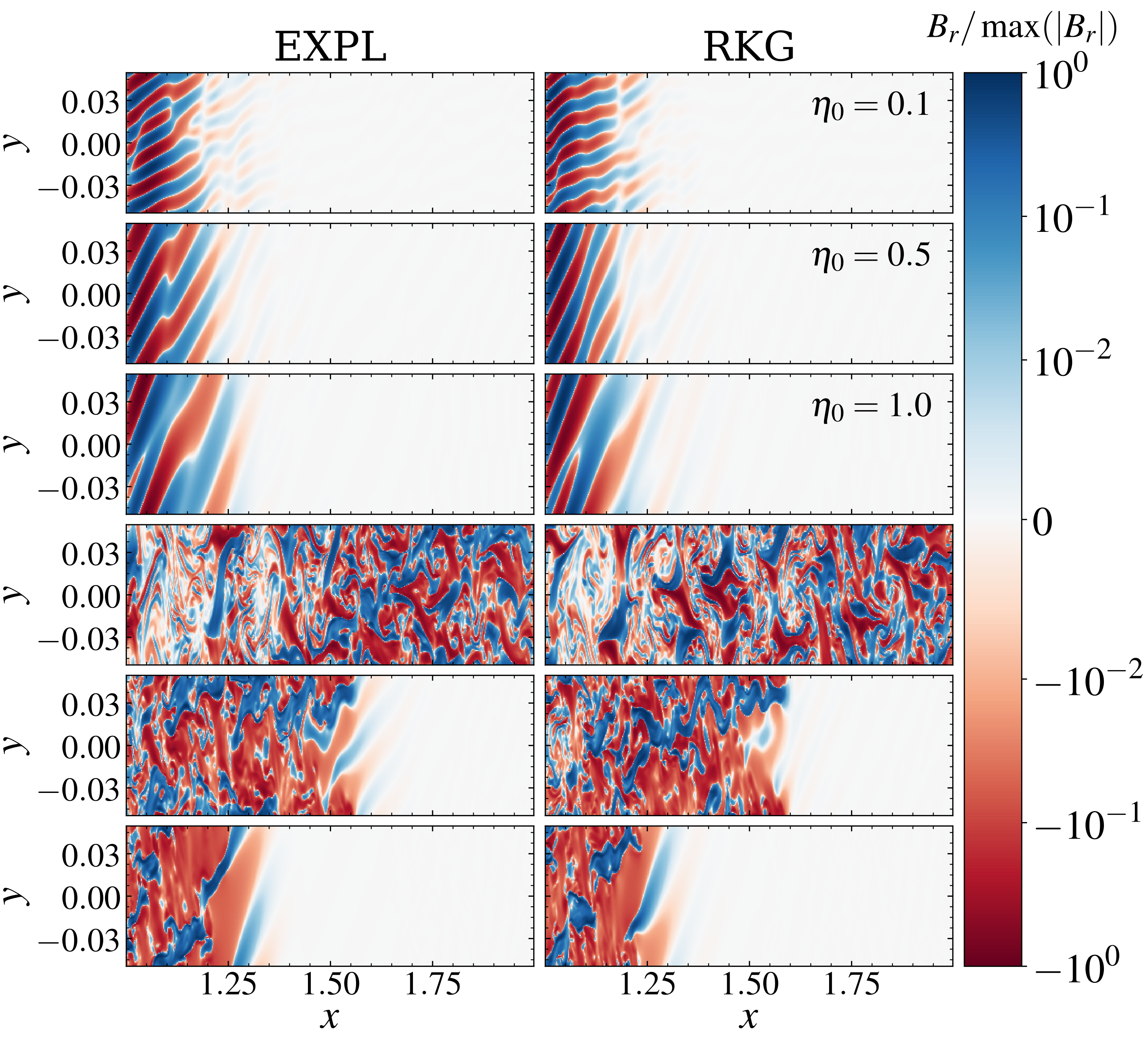}
\caption{Same as Fig.~\ref{Fig::mri_display1} but showing the normalized radial magnetic field $B_r/\max(|B_r|)$.}
\label{Fig::mri_display2}%
\end{figure}

The magneto-rotational instability is a key driver to generate turbulence and transport angular momentum \citep{balbus1991,hawley1991,fromang2007}, especially in the thermally ionized protoplanetary disk regions close to the star \citep{flock2017,lesur2023book,zhu2024}.
For our final test, we use a unstratified global disk model on a 3D polar grid.
The setup combines the global geometry of the linear MRI test by \cite{flock2010} with the ambipolar and magnetic field configuration of \cite{kunz2004} to determine the growth rate of the linear MRI.
The advantage of this setup is that it allows careful testing of the geometry terms for non-Cartesian coordinates alongside the EMF reconstruction and magnetic field update, which directly control the MRI's linear growth and nonlinear saturation.
By extracting and comparing the growth rate with the predictions of \cite{kunz2004}, we aim to assess the accuracy and robustness of the new method in reproducing the ambipolar-diffusion–modified MRI in a global disk geometry. 

For the setup, we use an axisymmetric (i.e., no dependence on the azimuthal direction) 3D polar grid, with $R_{beg}=1$, $R_{end}=2$, $Z_{beg}=-0.05$, $Z_{end}=0.05$, and a resolution of $512 \times 128$ in $R$ and $z$ respectively.
For this test, we adopt parabolic reconstruction \citep{mignone2014}, a third-order Runge–Kutta integrator \citep{gottlieb2001}, the UCT-contact scheme \citep{mignone2021}, a CFL number of 0.3, and the HLLD Riemann solver \citep{miyoshi2005}.
The disk setup is in radial equilibrium (and constant along the vertical direction) with uniform density $\rho=1$, sound speed $c_s=H\Omega$ with $H=0.1$ and $\Omega = 1/\sqrt{R^3}$, and azimuthal velocity 
\begin{equation}
v_\phi=R\Omega(\sqrt{1.0-3(H/R)^2}.
\end{equation}
The magnetic field is set so that the Alfv\'{e}n speeds are $v_A^\phi=0.1$ and $v_A^Z=0.02$.
We set the ambipolar diffusion coefficient to be:
\begin{equation}
    \eta_\ad = \eta_0R^{3/2},
\end{equation}
with $\eta_0 = 0.1, 0.5, 1.0$ \change{(cases $L_\ad$, $M_\ad$, and $H_\ad$, respectively)}.
The setup is performed over 50 inner orbits (corresponding to $t = 314$ in code units) until the MRI saturates.
This allows direct comparison with the local linear analysis of \citet{kunz2004}, for which the expected theoretical growth rates are approximately $\gamma = 0.41 \Omega^{-1}, 0.17 \Omega^{-1}, 0.11 \Omega^{-1}$, respectively, for $\eta_0 = 0.1, 0.5, 1.0$ for the three diffusion values.

Figure \ref{Fig::mri_gr} shows the temporal evolution of the maximum radial magnetic field at the innermost radius for three representative cases.
For reference, the ideal MRI growth scaling is indicated by the dotted line.
The temporal evolution of the maximum radial magnetic field shows a clear exponential growth phase followed by saturation in all cases.
The growth rate decreases systematically with increasing $\eta_0$, consistent with the ambipolar-modified MRI dispersion relation.
Both EXPL and RKG schemes exhibit comparable linear behavior, with measured growth rates of $0.34 \Omega^{-1}$ in the weak-ambipolar regime and $0.12\Omega^{-1}$ in the high-ambipolar regime.
In the intermediate-ambipolar regime, the EXPL scheme yields a growth rate of $0.15 \Omega^{-1}$, while the RKG scheme yields a growth rate of $0.16 \Omega^{-1}$.
The extracted values show a good agreement with the theoretical prediction of \citet{kunz2004}, confirming that the ambipolar diffusion implementation captures the correct linear physics in global geometry.
The small deviations observed in the low- and high-ambipolar regimes may be related to the highly nonlinear nature of ambipolar diffusion and the radial dependence of the coefficient $\eta_\ad$ and go beyond the scope of this paper.

In Figure \ref{Fig::mri_display1} and \ref{Fig::mri_display2}, we report, respectively, the spatial distribution of the Alfv\'en speed and the radial magnetic field (the latter normalized to its maximum at the selected time) at different simulation stages (represented by the circular and square symbols in Fig \ref{Fig::mri_gr}) to assess the impact of the numerical schemes during the linear regime and the nonlinear saturation stage.
The MRI initially develops at small radii and propagates outward. This is consistent with the radial dependence of the orbital frequency $\Omega\propto R^{-3/2}$, which leads to faster linear growth at smaller radii.
As expected, the spatial morphology and growth pattern are nearly identical between EXPL and RKG.
After the saturation reported in Fig. \ref{Fig::mri_gr}, the instability has entered the nonlinear regime.
At this stage, the magnetic and velocity fields become fully turbulent in the active region.
The turbulence spreads radially from the inner disk, eventually filling the computational domain.

\begin{figure}
\centering
\includegraphics[width=0.49\textwidth]{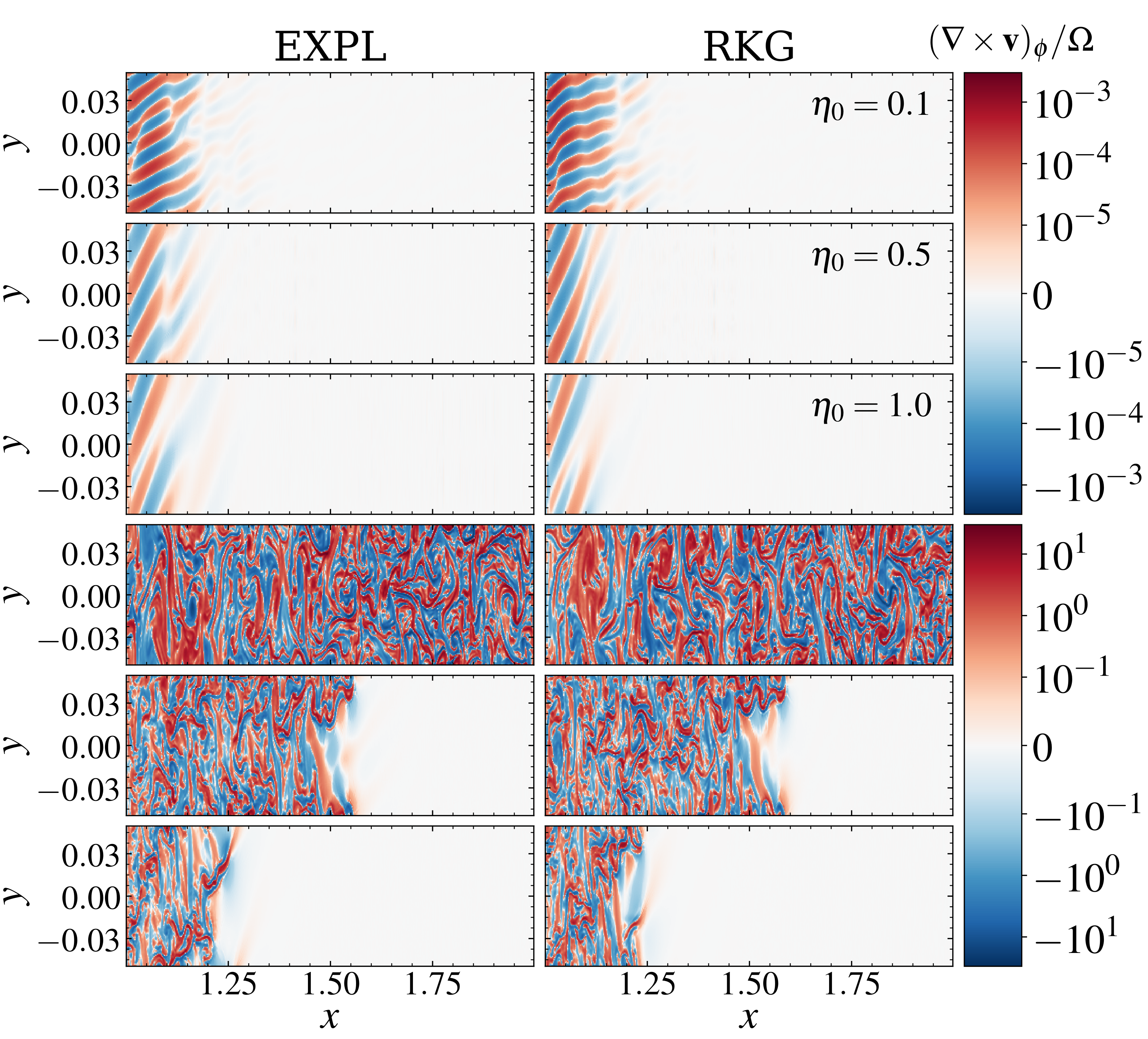}
\caption{Same as Fig.~\ref{Fig::mri_display1} but showing the vorticity $\varpi_\phi = (\nabla \times \mathbf{v})_\phi/\Omega$.}
\label{Fig::mri_display3}%
\end{figure}

\change{
Finally, in Figure \ref{Fig::mri_display3}, we compare the evolution of the toroidal vorticity, $\varpi_\phi = (\nabla \times \mathbf{v})_\phi/\Omega$, obtained with the EXPL and RKG schemes.
In the linear regime, both methods produce nearly identical vorticity patterns, indicating that the RKG integrator preserves the propagation and dissipation of smooth structures without introducing spurious oscillations.
In the nonlinear regime, the two schemes also show very similar distributions of small-scale vortical structures, suggesting that the accelerated integration does not significantly alter the effective numerical diffusivity or the generation of turbulent-like features.}
The saturation amplitude and turbulent structure are consistent across the two schemes, indicating that both methods robustly reproduce the nonlinear ambipolar MRI dynamics.
In addition to physical accuracy, the RKG integrator provides a substantial performance advantage.
For this test, RKG is approximately 2–2.5 times faster than the explicit scheme \change{(see Table \ref{tab:mri})}, while maintaining comparable accuracy in both the linear growth phase and nonlinear saturation.

\setlength{\tabcolsep}{4pt}
\begin{table}
    \centering
    \caption{\change{List of different magneto-rotational instability cases and their required computing time in seconds.}}
    \label{tab:mri}
\begin{tabular}{c|c|cc}
 Case &  $\eta_0$ & EXPL  & RKG \\ 
 \hline
 $L_\ad$ \rule{0pt}{2.5ex} & 0.1   & 3649 & 1905 \\ 
 $M_\ad$ \rule{0pt}{2.5ex} & 0.5  & 3929 & 1794 \\
 $H_\ad$ \rule{0pt}{2.5ex} & 1.0  & 4543 & 1793 \\
\end{tabular}
\end{table}

\section{Conclusions}
\label{sec::conclusions}

In this paper, we have presented a robust super–time–stepping scheme based on the stability of the Gegenbauer polynomials for the explicit integration of Ohmic and ambipolar diffusion in non-ideal MHD.
The method is fully explicit, straightforward to implement within finite-volume frameworks, and preserves the divergence-free condition of the magnetic field to machine precision when coupled to constrained transport.
Across an extensive suite of benchmarks, the RKG integrator consistently demonstrated excellent stability and accuracy.
In diffusion-dominated regimes, where classical explicit methods become prohibitively expensive and traditional STS variants suffer from robustness issues, RKG maintains stable evolution without introducing spurious oscillations or grid-aligned artifacts.
At the same time, it preserves the correct linear damping rates and nonlinear dynamics over many decades in amplitude.

By testing the numerical scheme against two representative astrophysical scenarios, i.e., ambipolar MRI in a global disk and magnetic reconnection, we demonstrate that the proposed RKG integrator retains the robustness and accuracy of conventional explicit substepping approaches while extending their practical efficiency.
In the ambipolar MRI problem, the scheme reproduces the ambipolar-modified theoretical linear growth rates and captures the subsequent nonlinear saturation without altering the instability dynamics.
In the reconnection setup, which probes thin current sheets and strongly diffusive regions, the magnetic field evolution and dissipation remain consistent with the fully explicit reference solution.
Across both tests, the enlarged stability region of the RKG method does not introduce artificial damping or spurious structures.

The advantage of RKG becomes increasingly pronounced in regimes where the parabolic timestep constraint dominates the evolution.
In strongly diffusive regions or at high resolution, where the explicit timestep would otherwise be severely restricted, the RKG method allows substantially larger effective timesteps than purely explicit schemes, thereby offering a significant computational benefit without compromising the physical solution.

By combining the simplicity and locality of explicit schemes with enlarged stability regions (compared to other substepping schemes) and enhanced robustness in anisotropic or strongly diffusive regimes, the RKG scheme is particularly well-suited for large-scale simulations of weakly ionized astrophysical plasmas, including protoplanetary disks, collapsing dense cores, and magnetized turbulence in partially ionized environments.
Extensions to higher-order methods \citep{berta2024} and GPU-accelerated implementations of the PLUTO code \citep{rossazza2025} will be considered in future work.

\begin{acknowledgements}
We thank D. Ostertag, A. Mignone, and B. Vaidya for the valuable discussions.
We thank F. Beckenbauer (not involved directly in this work) for indirect inspiration.
The data analysis and visualization were carried out using the PyPLUTO package \citep{mattia2025} and the Python libraries NumPy \citep{harris2020}, Matplotlib \citep{hunter2007}, and SciPy \citep{virtanen2020}.
The 3D simulations and the ones presented in Section \ref{sec::applications} were performed on the ASTRONODES and the VERA cluster of the Max Planck Institute for Astronomy and Max Planck Society.
We used ChatGPT (OpenAI) version 5.2 for language editing and clarity improvements. The scientific content, analysis, and conclusions are solely those of the authors.
\end{acknowledgements}

\bibliographystyle{aa.bst}
\bibliography{main}

@BOOK{abramowitz1972,
       author = {{Abramowitz}, M. and {Stegun}, I.~A.},
        title = "{Handbook of Mathematical Functions}",
         year = 1972,
       adsurl = {https://ui.adsabs.harvard.edu/abs/1972hmfw.book.....A}
}

@article{alexiades1996,
author = {Alexiades, V. and Amiez, Genevgve and Gremaud, Pierre},
year = {1996},
month = {01},
pages = {31-42},
title = {Super-time-stepping acceleration of explicit schemes for parabolic problems},
volume = {12},
journal = {Communications in Numerical Methods in Engineering},
doi = {10.1002/(SICI)1099-0887(199601)12:13.0.CO;2-5}
}

@ARTICLE{alfven1942,
       author = {{Alfv{\'e}n}, H.},
        title = "{Existence of Electromagnetic-Hydrodynamic Waves}",
      journal = {\nat},
         year = 1942,
        month = oct,
       volume = {150},
       number = {3805},
        pages = {405-406},
          doi = {10.1038/150405d0},
       adsurl = {https://ui.adsabs.harvard.edu/abs/1942Natur.150..405A}
}

@ARTICLE{armitage2011,
       author = {{Armitage}, Philip J.},
        title = "{Dynamics of Protoplanetary Disks}",
      journal = {\araa},
         year = 2011,
        month = sep,
       volume = {49},
       number = {1},
        pages = {195-236},
          doi = {10.1146/annurev-astro-081710-102521},
archivePrefix = {arXiv},
       eprint = {1011.1496},
 primaryClass = {astro-ph.SR},
       adsurl = {https://ui.adsabs.harvard.edu/abs/2011ARA&A..49..195A}
}

@ARTICLE{bai2011,
       author = {{Bai}, Xue-Ning and {Stone}, James M.},
        title = "{Effect of Ambipolar Diffusion on the Nonlinear Evolution of Magnetorotational Instability in Weakly Ionized Disks}",
      journal = {\apj},
         year = 2011,
        month = aug,
       volume = {736},
       number = {2},
          eid = {144},
        pages = {144},
          doi = {10.1088/0004-637X/736/2/144},
archivePrefix = {arXiv},
       eprint = {1103.1380},
 primaryClass = {astro-ph.EP},
       adsurl = {https://ui.adsabs.harvard.edu/abs/2011ApJ...736..144B}
}

@ARTICLE{balsara1996,
       author = {{Balsara}, Dinshaw S.},
        title = "{Wave Propagation in Molecular Clouds}",
      journal = {\apj},
         year = 1996,
        month = jul,
       volume = {465},
        pages = {775},
          doi = {10.1086/177462},
       adsurl = {https://ui.adsabs.harvard.edu/abs/1996ApJ...465..775B}
}

@article{barenblatt1952,
  author       = {Barenblatt, G. I.},
  title        = {On some unsteady motions of a liquid and a gas in a porous medium},
  journal      = {Prikl.\ Mat.\ Mekh.},
  volume       = {16},
  number       = {1},
  pages        = {67--78},
  year         = {1952},
}

@ARTICLE{beck2015,
       author = {{Beck}, Rainer},
        title = "{Magnetic fields in spiral galaxies}",
      journal = {\aapr},
         year = 2015,
        month = dec,
       volume = {24},
          eid = {4},
        pages = {4},
          doi = {10.1007/s00159-015-0084-4},
archivePrefix = {arXiv},
       eprint = {1509.04522},
 primaryClass = {astro-ph.GA},
       adsurl = {https://ui.adsabs.harvard.edu/abs/2015A&ARv..24....4B}
}

@ARTICLE{berta2024,
       author = {{Berta}, V. and {Mignone}, A. and {Bugli}, M. and {Mattia}, G.},
        title = "{A 4$^{th}$-order accurate finite volume method for ideal classical and special relativistic MHD based on pointwise reconstructions}",
      journal = {Journal of Computational Physics},
         year = 2024,
        month = feb,
       volume = {499},
          eid = {112701},
        pages = {112701},
          doi = {10.1016/j.jcp.2023.112701},
archivePrefix = {arXiv},
       eprint = {2310.11831},
 primaryClass = {astro-ph.HE},
       adsurl = {https://ui.adsabs.harvard.edu/abs/2024JCoPh.49912701B}
}

@ARTICLE{berta2026,
       author = {{Berta}, V. and {Bugli}, M. and {Mignone}, A. and {Mattia}, G. and {Del Zanna}, L. and {Truzzi}, S.},
        title = "{2D or not 2D? Exploring 3D relativistic magnetic reconnection dynamics with highly accurate numerical simulations}",
      journal = {\mnras},
         year = 2026,
        month = mar,
       volume = {546},
       number = {4},
          eid = {stag286},
        pages = {stag286},
          doi = {10.1093/mnras/stag286},
archivePrefix = {arXiv},
       eprint = {2512.15954},
 primaryClass = {astro-ph.HE},
       adsurl = {https://ui.adsabs.harvard.edu/abs/2026MNRAS.546ag286B}
}

@ARTICLE{borges2008,
       author = {{Borges}, Rafael and {Carmona}, Monique and {Costa}, Bruno and {Don}, Wai Sun},
        title = "{An improved weighted essentially non-oscillatory scheme for hyperbolic conservation laws}",
      journal = {Journal of Computational Physics},
         year = 2008,
        month = mar,
       volume = {227},
       number = {6},
        pages = {3191-3211},
          doi = {10.1016/j.jcp.2007.11.038},
       adsurl = {https://ui.adsabs.harvard.edu/abs/2008JCoPh.227.3191B}
}

@ARTICLE{brandenburg2005,
       author = {{Brandenburg}, Axel and {Subramanian}, Kandaswamy},
        title = "{Astrophysical magnetic fields and nonlinear dynamo theory}",
      journal = {\physrep},
         year = 2005,
        month = oct,
       volume = {417},
       number = {1-4},
        pages = {1-209},
          doi = {10.1016/j.physrep.2005.06.005},
archivePrefix = {arXiv},
       eprint = {astro-ph/0405052},
 primaryClass = {astro-ph},
       adsurl = {https://ui.adsabs.harvard.edu/abs/2005PhR...417....1B}
}

@book{butcher1987,
  address = {Chichester},
  author = {Butcher, J. C.},
  edition = {First},
  isbn = {0471910465 9780471910466},
  publisher = {Wiley},
  refid = {802658434},
  title = {The Numerical analysis of ordinary differential equations : Runge-Kutta and general linear methods},
  url = {https://www.worldcat.org/title/numerical-analysis-of-ordinary-differential-equations-runge-kutta-and-general-linear-methods/oclc/802658434&referer=brief_results},
  year = 1987
}

@article{caplan2024,
doi = {10.1088/1742-6596/2742/1/012020},
url = {https://dx.doi.org/10.1088/1742-6596/2742/1/012020},
year = {2024},
month = {apr},
publisher = {IOP Publishing},
volume = {2742},
number = {1},
pages = {012020},
author = {Caplan, Ronald M. and Johnston, Craig D. and Daldoff, Lars K. S. and Linker, Jon A.},
title = {Advancing parabolic operators in thermodynamic MHD models II: Evaluating a Practical Time Step Limit for Unconditionally Stable Methods},
journal = {Journal of Physics: Conference Series},
}

@BOOK{chiuderi2005,
       author = {{Chiuderi}, C. and {Velli}, M.},
        title = "{Basics of Plasma Astrophysics}",
         year = 2015,
          doi = {10.1007/978-88-470-5280-2},
       adsurl = {https://ui.adsabs.harvard.edu/abs/2015bps..book.....C}
}

@ARTICLE{choi2009,
       author = {{Choi}, Eunwoo and {Kim}, Jongsoo and {Wiita}, Paul J.},
        title = "{An Explicit Scheme for Incorporating Ambipolar Diffusion in a Magnetohydrodynamics Code}",
      journal = {\apjs},
         year = 2009,
        month = apr,
       volume = {181},
       number = {2},
        pages = {413-420},
          doi = {10.1088/0067-0049/181/2/413},
archivePrefix = {arXiv},
       eprint = {0812.3748},
 primaryClass = {astro-ph},
       adsurl = {https://ui.adsabs.harvard.edu/abs/2009ApJS..181..413C}
}

@ARTICLE{courant1928,
       author = {{Courant}, R. and {Friedrichs}, K. and {Lewy}, H.},
        title = "{{\"U}ber die partiellen Differenzengleichungen der mathematischen Physik}",
      journal = {Mathematische Annalen},
         year = 1928,
        month = jan,
       volume = {100},
        pages = {32-74},
          doi = {10.1007/BF01448839},
       adsurl = {https://ui.adsabs.harvard.edu/abs/1928MatAn.100...32C}
}

@ARTICLE{crutcher2012,
       author = {{Crutcher}, Richard M.},
        title = "{Magnetic Fields in Molecular Clouds}",
      journal = {\araa},
         year = 2012,
        month = sep,
       volume = {50},
        pages = {29-63},
          doi = {10.1146/annurev-astro-081811-125514},
       adsurl = {https://ui.adsabs.harvard.edu/abs/2012ARA&A..50...29C}
}

@ARTICLE{dedner2002,
   author = {{Dedner}, A. and {Kemm}, F. and {Kr{\"o}ner}, D. and {Munz}, C.-D. and 
	{Schnitzer}, T. and {Wesenberg}, M.},
    title = "{Hyperbolic Divergence Cleaning for the MHD Equations}",
  journal = {Journal of Computational Physics},
     year = 2002,
    month = jan,
   volume = 175,
    pages = {645-673},
      doi = {10.1006/jcph.2001.6961},
   adsurl = {http://adsabs.harvard.edu/abs/2002JCoPh.175..645D}
}

@article{doha1991,
title = {The coefficients of differentiated expansions and derivatives of ultraspherical polynomials},
journal = {Computers \& Mathematics with Applications},
volume = {21},
number = {2},
pages = {115-122},
year = {1991},
issn = {0898-1221},
doi = {https://doi.org/10.1016/0898-1221(91)90089-M},
url = {https://www.sciencedirect.com/science/article/pii/089812219190089M},
author = {E.H. Doha},
}

@ARTICLE{duffin2008,
       author = {{Duffin}, D.~F. and {Pudritz}, R.~E.},
        title = "{Simulating hydromagnetic processes in star formation: introducing ambipolar diffusion into an adaptive mesh refinement code}",
      journal = {\mnras},
         year = 2008,
        month = dec,
       volume = {391},
       number = {4},
        pages = {1659-1673},
          doi = {10.1111/j.1365-2966.2008.14026.x},
archivePrefix = {arXiv},
       eprint = {0810.0299},
 primaryClass = {astro-ph},
       adsurl = {https://ui.adsabs.harvard.edu/abs/2008MNRAS.391.1659D}
}

@INCOLLECTION{erdelyi1953,
       author = {{Erdelyi}, A.},
        title = "{Higher Transcendental Functions}",
    booktitle = {Higher Transcendental Functions},
         year = 1953,
        pages = {59},
       adsurl = {https://ui.adsabs.harvard.edu/abs/1953hft1.book...59E}
}

@ARTICLE{fendt2006,
       author = {{Fendt}, Christian},
        title = "{Collimation of Astrophysical Jets: The Role of the Accretion Disk Magnetic Field Distribution}",
      journal = {\apj},
         year = 2006,
        month = nov,
       volume = {651},
       number = {1},
        pages = {272-287},
          doi = {10.1086/507976},
archivePrefix = {arXiv},
       eprint = {astro-ph/0511611},
 primaryClass = {astro-ph},
       adsurl = {https://ui.adsabs.harvard.edu/abs/2006ApJ...651..272F}
}

@ARTICLE{zhu2024,
       author = {{Zhu}, Zhaohuan and {Stone}, James M. and {Calvet}, Nuria},
        title = "{A global 3D simulation of magnetospheric accretion - I. Magnetically disrupted discs and surface accretion}",
      journal = {\mnras},
         year = 2024,
        month = feb,
       volume = {528},
       number = {2},
        pages = {2883-2911},
          doi = {10.1093/mnras/stad3712},
archivePrefix = {arXiv},
       eprint = {2309.15318},
 primaryClass = {astro-ph.SR},
       adsurl = {https://ui.adsabs.harvard.edu/abs/2024MNRAS.528.2883Z}
}

@ARTICLE{fromang2007,
       author = {{Fromang}, S. and {Papaloizou}, J.},
        title = "{MHD simulations of the magnetorotational instability in a shearing box with zero net flux. I. The issue of convergence}",
      journal = {\aap},
         year = 2007,
        month = dec,
       volume = {476},
       number = {3},
        pages = {1113-1122},
          doi = {10.1051/0004-6361:20077942},
archivePrefix = {arXiv},
       eprint = {0705.3621},
 primaryClass = {astro-ph},
       adsurl = {https://ui.adsabs.harvard.edu/abs/2007A&A...476.1113F}
}

@ARTICLE{harris2020,
       author = {{Harris}, Charles R. and {Millman}, K. Jarrod and {van der Walt}, St{\'e}fan J. and {Gommers}, Ralf and {Virtanen}, Pauli and {Cournapeau}, David and {Wieser}, Eric and {Taylor}, Julian and {Berg}, Sebastian and {Smith}, Nathaniel J. and {Kern}, Robert and {Picus}, Matti and {Hoyer}, Stephan and {van Kerkwijk}, Marten H. and {Brett}, Matthew and {Haldane}, Allan and {del R{\'\i}o}, Jaime Fern{\'a}ndez and {Wiebe}, Mark and {Peterson}, Pearu and {G{\'e}rard-Marchant}, Pierre and {Sheppard}, Kevin and {Reddy}, Tyler and {Weckesser}, Warren and {Abbasi}, Hameer and {Gohlke}, Christoph and {Oliphant}, Travis E.},
        title = "{Array programming with NumPy}",
      journal = {\nat},
         year = 2020,
        month = sep,
       volume = {585},
       number = {7825},
        pages = {357-362},
          doi = {10.1038/s41586-020-2649-2},
archivePrefix = {arXiv},
       eprint = {2006.10256},
 primaryClass = {cs.MS},
       adsurl = {https://ui.adsabs.harvard.edu/abs/2020Natur.585..357H}
}

@ARTICLE{hawley1991,
       author = {{Hawley}, John F. and {Balbus}, Steven A.},
        title = "{A Powerful Local Shear Instability in Weakly Magnetized Disks. II. Nonlinear Evolution}",
      journal = {\apj},
         year = 1991,
        month = jul,
       volume = {376},
        pages = {223},
          doi = {10.1086/170271},
       adsurl = {https://ui.adsabs.harvard.edu/abs/1991ApJ...376..223H}
}

@ARTICLE{hunter2007,
       author = {{Hunter}, John D.},
        title = "{Matplotlib: A 2D Graphics Environment}",
      journal = {Computing in Science and Engineering},
         year = 2007,
        month = jan,
       volume = {9},
       number = {3},
        pages = {90-95},
          doi = {10.1109/MCSE.2007.55},
       adsurl = {https://ui.adsabs.harvard.edu/abs/2007CSE.....9...90H}
}

@ARTICLE{balbus1991,
       author = {{Balbus}, Steven A. and {Hawley}, John F.},
        title = "{A Powerful Local Shear Instability in Weakly Magnetized Disks. I. Linear Analysis}",
      journal = {\apj},
         year = 1991,
        month = jul,
       volume = {376},
        pages = {214},
          doi = {10.1086/170270},
       adsurl = {https://ui.adsabs.harvard.edu/abs/1991ApJ...376..214B}
}

@ARTICLE{flock2017,
       author = {{Flock}, M. and {Fromang}, S. and {Turner}, N.~J. and {Benisty}, M.},
        title = "{3D Radiation Nonideal Magnetohydrodynamical Simulations of the Inner Rim in Protoplanetary Disks}",
      journal = {\apj},
         year = 2017,
        month = feb,
       volume = {835},
       number = {2},
          eid = {230},
        pages = {230},
          doi = {10.3847/1538-4357/835/2/230},
archivePrefix = {arXiv},
       eprint = {1612.02740},
 primaryClass = {astro-ph.EP},
       adsurl = {https://ui.adsabs.harvard.edu/abs/2017ApJ...835..230F}
}

@ARTICLE{gottlieb2001,
       author = {{Gottlieb}, Sigal and {Shu}, Chi-Wang and {Tadmor}, Eitan},
        title = "{Strong Stability-Preserving High-Order Time Discretization Methods}",
      journal = {SIAM Review},
         year = 2001,
        month = jan,
       volume = {43},
       number = {1},
        pages = {89-112},
          doi = {10.1137/S003614450036757X},
       adsurl = {https://ui.adsabs.harvard.edu/abs/2001SIAMR..43...89G}
}

@ARTICLE{gressel2015,
       author = {{Gressel}, Oliver and {Turner}, Neal J. and {Nelson}, Richard P. and {McNally}, Colin P.},
        title = "{Global Simulations of Protoplanetary Disks With Ohmic Resistivity and Ambipolar Diffusion}",
      journal = {\apj},
         year = 2015,
        month = mar,
       volume = {801},
       number = {2},
          eid = {84},
        pages = {84},
          doi = {10.1088/0004-637X/801/2/84},
archivePrefix = {arXiv},
       eprint = {1501.05431},
 primaryClass = {astro-ph.EP},
       adsurl = {https://ui.adsabs.harvard.edu/abs/2015ApJ...801...84G}
}

@ARTICLE{grundy1982,
       author = {{Grundy}, R.~E. and {McLaughlin}, R.},
        title = "{Eigenvalues of the Barenblatt-Pattle Similarity Solution in Nonlinear Diffusion}",
      journal = {Proceedings of the Royal Society of London Series A},
         year = 1982,
        month = sep,
       volume = {383},
       number = {1784},
        pages = {89-100},
          doi = {10.1098/rspa.1982.0122},
       adsurl = {https://ui.adsabs.harvard.edu/abs/1982RSPSA.383...89G}
}

@book{hairer2010,
  address = {Berlin},
  author = {Hairer, E. and N{\o}rsett, S.P. and Wanner, G.},
  edition = {Second},
  publisher = {Springer},
  title = {Solving Ordinary Differential 
 Equations {I} Nonstiff problems},
  year = 2000
}

@article{harten1983,
  author = {Harten, A. and Lax, P. and Leer, B.},
  title = {On Upstream Differencing and Godunov-Type Schemes for Hyperbolic Conservation Laws},
  journal = {SIAM Review},
  volume = {25},
  number = {1},
  pages = {35-61},
  year = {1983},
  doi = {10.1137/1025002},
  URL = { https://doi.org/10.1137/1025002},
  eprint = { https://doi.org/10.1137/1025002}
}

@article{karageorghis1992,
title = {On the coefficients of differentiated expansions of ultraspherical polynomials},
journal = {Applied Numerical Mathematics},
volume = {9},
number = {2},
pages = {133-141},
year = {1992},
issn = {0168-9274},
doi = {https://doi.org/10.1016/0168-9274(92)90010-B},
url = {https://www.sciencedirect.com/science/article/pii/016892749290010B},
author = {Andreas Karageorghis and Timothy N. Phillips},
}

@ARTICLE{kayanikhoo2024,
       author = {{Kayanikhoo}, Fatemeh and {{\v{C}}emelji{\'c}}, Miljenko and {Wielgus}, Maciek and {Klu{\'z}niak}, W{\l}odek},
        title = "{Energy distribution and substructure formation in astrophysical MHD simulations}",
      journal = {\mnras},
         year = 2024,
        month = feb,
       volume = {527},
       number = {4},
        pages = {10151-10167},
          doi = {10.1093/mnras/stad3807},
archivePrefix = {arXiv},
       eprint = {2308.16062},
 primaryClass = {astro-ph.HE},
       adsurl = {https://ui.adsabs.harvard.edu/abs/2024MNRAS.52710151K}
}

@ARTICLE{kunz2004,
       author = {{Kunz}, Matthew W. and {Balbus}, Steven A.},
        title = "{Ambipolar diffusion in the magnetorotational instability}",
      journal = {\mnras},
         year = 2004,
        month = feb,
       volume = {348},
       number = {1},
        pages = {355-360},
          doi = {10.1111/j.1365-2966.2004.07383.x},
archivePrefix = {arXiv},
       eprint = {astro-ph/0309707},
 primaryClass = {astro-ph},
       adsurl = {https://ui.adsabs.harvard.edu/abs/2004MNRAS.348..355K}
}

@ARTICLE{flock2010,
       author = {{Flock}, M. and {Dzyurkevich}, N. and {Klahr}, H. and {Mignone}, A.},
        title = "{High-order Godunov schemes for global 3D MHD simulations of accretion disks. I. Testing the linear growth of the magneto-rotational instability}",
      journal = {\aap},
         year = 2010,
        month = jun,
       volume = {516},
          eid = {A26},
        pages = {A26},
          doi = {10.1051/0004-6361/200912443},
archivePrefix = {arXiv},
       eprint = {0906.5516},
 primaryClass = {astro-ph.EP},
       adsurl = {https://ui.adsabs.harvard.edu/abs/2010A&A...516A..26F}
}

@ARTICLE{lesur2014,
       author = {{Lesur}, Geoffroy and {Kunz}, Matthew W. and {Fromang}, S{\'e}bastien},
        title = "{Thanatology in protoplanetary discs. The combined influence of Ohmic, Hall, and ambipolar diffusion on dead zones}",
      journal = {\aap},
         year = 2014,
        month = jun,
       volume = {566},
          eid = {A56},
        pages = {A56},
          doi = {10.1051/0004-6361/201423660},
archivePrefix = {arXiv},
       eprint = {1402.4133},
 primaryClass = {astro-ph.SR},
       adsurl = {https://ui.adsabs.harvard.edu/abs/2014A&A...566A..56L}
}

@ARTICLE{hu2023,
       author = {{Hu}, Xiao and {Li}, Zhi-Yun and {Wang}, Lile and {Zhu}, Zhaohuan and {Bae}, Jaehan},
        title = "{Gap opening in protoplanetary discs: gas dynamics from global axisymmetric non-ideal MHD simulations with consistent thermochemistry}",
      journal = {\mnras},
         year = 2023,
        month = aug,
       volume = {523},
       number = {4},
        pages = {4883-4894},
          doi = {10.1093/mnras/stad1632},
archivePrefix = {arXiv},
       eprint = {2304.05972},
 primaryClass = {astro-ph.EP},
       adsurl = {https://ui.adsabs.harvard.edu/abs/2023MNRAS.523.4883H}
}

@ARTICLE{cui2021,
       author = {{Cui}, Can and {Bai}, Xue-Ning},
        title = "{Global three-dimensional simulations of outer protoplanetary discs with ambipolar diffusion}",
      journal = {\mnras},
         year = 2021,
        month = oct,
       volume = {507},
       number = {1},
        pages = {1106-1126},
          doi = {10.1093/mnras/stab2220},
archivePrefix = {arXiv},
       eprint = {2106.10167},
 primaryClass = {astro-ph.EP},
       adsurl = {https://ui.adsabs.harvard.edu/abs/2021MNRAS.507.1106C}
}

@ARTICLE{bethune2017,
       author = {{B{\'e}thune}, William and {Lesur}, Geoffroy and {Ferreira}, Jonathan},
        title = "{Global simulations of protoplanetary disks with net magnetic flux. I. Non-ideal MHD case}",
      journal = {\aap},
         year = 2017,
        month = apr,
       volume = {600},
          eid = {A75},
        pages = {A75},
          doi = {10.1051/0004-6361/201630056},
archivePrefix = {arXiv},
       eprint = {1612.00883},
 primaryClass = {astro-ph.EP},
       adsurl = {https://ui.adsabs.harvard.edu/abs/2017A&A...600A..75B}
}

@ARTICLE{simon2013,
       author = {{Simon}, Jacob B. and {Bai}, Xue-Ning and {Stone}, James M. and {Armitage}, Philip J. and {Beckwith}, Kris},
        title = "{Turbulence in the Outer Regions of Protoplanetary Disks. I. Weak Accretion with No Vertical Magnetic Flux}",
      journal = {\apj},
         year = 2013,
        month = feb,
       volume = {764},
       number = {1},
          eid = {66},
        pages = {66},
          doi = {10.1088/0004-637X/764/1/66},
archivePrefix = {arXiv},
       eprint = {1210.4164},
 primaryClass = {astro-ph.SR},
       adsurl = {https://ui.adsabs.harvard.edu/abs/2013ApJ...764...66S}
}

@ARTICLE{lesaffre2007,
       author = {{Lesaffre}, Pierre and {Balbus}, Steven A.},
        title = "{Exact shearing box solutions of magnetohydrodynamic flows with resistivity, viscosity and cooling}",
      journal = {\mnras},
         year = 2007,
        month = oct,
       volume = {381},
       number = {1},
        pages = {319-333},
          doi = {10.1111/j.1365-2966.2007.12270.x},
archivePrefix = {arXiv},
       eprint = {0709.1388},
 primaryClass = {astro-ph},
       adsurl = {https://ui.adsabs.harvard.edu/abs/2007MNRAS.381..319L}
}

@INPROCEEDINGS{lesur2023book,
       author = {{Lesur}, G. and {Flock}, M. and {Ercolano}, B. and {Lin}, M.-K. and {Yang}, C. and {Barranco}, J.~A. and {Benitez-Llambay}, P. and {Goodman}, J. and {Johansen}, A. and {Klahr}, H. and {Laibe}, G. and {Lyra}, W. and {Marcus}, P.~S. and {Nelson}, R.~P. and {Squire}, J. and {Simon}, J.~B. and {Turner}, N.~J. and {Umurhan}, O.~M. and {Youdin}, A.~N.},
        title = "{Hydro-, Magnetohydro-, and Dust-Gas Dynamics of Protoplanetary Disks}",
    booktitle = {Protostars and Planets VII},
         year = 2023,
       editor = {{Inutsuka}, S. and {Aikawa}, Y. and {Muto}, T. and {Tomida}, K. and {Tamura}, M.},
       series = {Astronomical Society of the Pacific Conference Series},
       volume = {534},
        month = jul,
        pages = {465},
          doi = {10.48550/arXiv.2203.09821},
archivePrefix = {arXiv},
       eprint = {2203.09821},
 primaryClass = {astro-ph.EP},
       adsurl = {https://ui.adsabs.harvard.edu/abs/2023ASPC..534..465L}
}

@ARTICLE{lesur2023,
       author = {{Lesur}, G.~R.~J. and {Baghdadi}, S. and {Wafflard-Fernandez}, G. and {Mauxion}, J. and {Robert}, C.~M.~T. and {Van den Bossche}, M.},
        title = "{IDEFIX: A versatile performance-portable Godunov code for astrophysical flows}",
      journal = {\aap},
         year = 2023,
        month = sep,
       volume = {677},
          eid = {A9},
        pages = {A9},
          doi = {10.1051/0004-6361/202346005},
archivePrefix = {arXiv},
       eprint = {2304.13746},
 primaryClass = {astro-ph.IM},
       adsurl = {https://ui.adsabs.harvard.edu/abs/2023A&A...677A...9L}
}

@ARTICLE{marchand2016,
       author = {{Marchand}, P. and {Masson}, J. and {Chabrier}, G. and {Hennebelle}, P. and {Commer{\c{c}}on}, B. and {Vaytet}, N.},
        title = "{Chemical solver to compute molecule and grain abundances and non-ideal MHD resistivities in prestellar core-collapse calculations}",
      journal = {\aap},
         year = 2016,
        month = jul,
       volume = {592},
          eid = {A18},
        pages = {A18},
          doi = {10.1051/0004-6361/201526780},
archivePrefix = {arXiv},
       eprint = {1604.05613},
 primaryClass = {astro-ph.GA},
       adsurl = {https://ui.adsabs.harvard.edu/abs/2016A&A...592A..18M}
}

@ARTICLE{loureiro2016,
       author = {{Loureiro}, N.~F. and {Uzdensky}, D.~A.},
        title = "{Magnetic reconnection: from the Sweet-Parker model to stochastic plasmoid chains}",
      journal = {Plasma Physics and Controlled Fusion},
         year = 2016,
        month = jan,
       volume = {58},
       number = {1},
          eid = {014021},
        pages = {014021},
          doi = {10.1088/0741-3335/58/1/014021},
archivePrefix = {arXiv},
       eprint = {1507.07756},
 primaryClass = {physics.plasm-ph},
       adsurl = {https://ui.adsabs.harvard.edu/abs/2016PPCF...58a4021L}
}

@ARTICLE{marinacci2018,
       author = {{Marinacci}, Federico and {Vogelsberger}, Mark and {Kannan}, Rahul and {Mocz}, Philip and {Pakmor}, R{\"u}diger and {Springel}, Volker},
        title = "{Non-ideal magnetohydrodynamics on a moving mesh}",
      journal = {\mnras},
         year = 2018,
        month = may,
       volume = {476},
       number = {2},
        pages = {2476-2492},
          doi = {10.1093/mnras/sty397},
archivePrefix = {arXiv},
       eprint = {1710.10265},
 primaryClass = {astro-ph.IM},
       adsurl = {https://ui.adsabs.harvard.edu/abs/2018MNRAS.476.2476M}
}

@ARTICLE{masson2012,
       author = {{Masson}, J. and {Teyssier}, R. and {Mulet-Marquis}, C. and {Hennebelle}, P. and {Chabrier}, G.},
        title = "{Incorporating Ambipolar and Ohmic Diffusion in the AMR MHD Code RAMSES}",
      journal = {\apjs},
         year = 2012,
        month = aug,
       volume = {201},
       number = {2},
          eid = {24},
        pages = {24},
          doi = {10.1088/0067-0049/201/2/24},
archivePrefix = {arXiv},
       eprint = {1206.2476},
 primaryClass = {astro-ph.SR},
       adsurl = {https://ui.adsabs.harvard.edu/abs/2012ApJS..201...24M}
}

@ARTICLE{masson2016,
       author = {{Masson}, J. and {Chabrier}, G. and {Hennebelle}, P. and {Vaytet}, N. and {Commer{\c{c}}on}, B.},
        title = "{Ambipolar diffusion in low-mass star formation. I. General comparison with the ideal magnetohydrodynamic case}",
      journal = {\aap},
         year = 2016,
        month = mar,
       volume = {587},
          eid = {A32},
        pages = {A32},
          doi = {10.1051/0004-6361/201526371},
archivePrefix = {arXiv},
       eprint = {1509.05630},
 primaryClass = {astro-ph.SR},
       adsurl = {https://ui.adsabs.harvard.edu/abs/2016A&A...587A..32M}
}

@ARTICLE{mattia2025,
       author = {{Mattia}, Giancarlo and {Crocco}, Daniele and {Melon Fuksman}, David and {Bugli}, Matteo and {Berta}, Vittoria and {Puzzoni}, Eleonora and {Mignone}, Andrea and {Vaidya}, Bhargav},
        title = "{PyPLUTO: a data analysis Python package for the PLUTO code}",
      journal = {The Journal of Open Source Software},
         year = 2025,
        month = sep,
       volume = {10},
       number = {113},
          eid = {8448},
        pages = {8448},
          doi = {10.21105/joss.08448},
archivePrefix = {arXiv},
       eprint = {2501.09748},
 primaryClass = {astro-ph.IM},
       adsurl = {https://ui.adsabs.harvard.edu/abs/2025JOSS...10.8448M}
}

@ARTICLE{mckee2007,
       author = {{McKee}, Christopher F. and {Ostriker}, Eve C.},
        title = "{Theory of Star Formation}",
      journal = {\araa},
         year = 2007,
        month = sep,
       volume = {45},
       number = {1},
        pages = {565-687},
          doi = {10.1146/annurev.astro.45.051806.110602},
archivePrefix = {arXiv},
       eprint = {0707.3514},
 primaryClass = {astro-ph},
       adsurl = {https://ui.adsabs.harvard.edu/abs/2007ARA&A..45..565M}
}

@ARTICLE{meyer2012,
       author = {{Meyer}, Chad D. and {Balsara}, Dinshaw S. and {Aslam}, Tariq D.},
        title = "{A second-order accurate Super TimeStepping formulation for anisotropic thermal conduction}",
      journal = {\mnras},
         year = 2012,
        month = may,
       volume = {422},
       number = {3},
        pages = {2102-2115},
          doi = {10.1111/j.1365-2966.2012.20744.x},
       adsurl = {https://ui.adsabs.harvard.edu/abs/2012MNRAS.422.2102M}
}

@ARTICLE{meyer2014,
       author = {{Meyer}, Chad D. and {Balsara}, Dinshaw S. and {Aslam}, Tariq D.},
        title = "{A stabilized Runge-Kutta-Legendre method for explicit super-time-stepping of parabolic and mixed equations}",
      journal = {Journal of Computational Physics},
         year = 2014,
        month = jan,
       volume = {257},
        pages = {594-626},
          doi = {10.1016/j.jcp.2013.08.021},
       adsurl = {https://ui.adsabs.harvard.edu/abs/2014JCoPh.257..594M}
}

@ARTICLE{mignone2007,
  title = {{{PLUTO}}: {{A Numerical Code}} for {{Computational Astrophysics}}},
  shorttitle = {{{PLUTO}}},
  author = {Mignone, A. and Bodo, G. and Massaglia, S. and Matsakos, T. and Tesileanu, O. and Zanni, C. and Ferrari, A.},
  year = {2007},
  month = may,
  journal = {The Astrophysical Journal Supplement Series},
  volume = {170},
  pages = {228--242},
  issn = {0067-0049},
  doi = {10.1086/513316},
  urldate = {2020-12-04}
}

@ARTICLE{mignone2014,
       author = {{Mignone}, A.},
        title = "{High-order conservative reconstruction schemes for finite volume methods in cylindrical and spherical coordinates}",
      journal = {Journal of Computational Physics},
         year = 2014,
        month = aug,
       volume = {270},
        pages = {784-814},
          doi = {10.1016/j.jcp.2014.04.001},
archivePrefix = {arXiv},
       eprint = {1404.0537},
 primaryClass = {physics.comp-ph},
       adsurl = {https://ui.adsabs.harvard.edu/abs/2014JCoPh.270..784M}
}

@ARTICLE{mignone2021,
       author = {{Mignone}, A. and {Del Zanna}, L.},
        title = "{Systematic construction of upwind constrained transport schemes for MHD}",
      journal = {Journal of Computational Physics},
         year = 2021,
        month = jan,
       volume = {424},
          eid = {109748},
        pages = {109748},
          doi = {10.1016/j.jcp.2020.109748},
archivePrefix = {arXiv},
       eprint = {2004.10542},
 primaryClass = {physics.comp-ph},
       adsurl = {https://ui.adsabs.harvard.edu/abs/2021JCoPh.42409748M}
}

@ARTICLE{miyoshi2005,
       author = {{Miyoshi}, Takahiro and {Kusano}, Kanya},
        title = "{A multi-state HLL approximate Riemann solver for ideal magnetohydrodynamics}",
      journal = {Journal of Computational Physics},
         year = "2005",
        month = "Sep",
       volume = {208},
       number = {1},
        pages = {315-344},
          doi = {10.1016/j.jcp.2005.02.017},
       adsurl = {https://ui.adsabs.harvard.edu/abs/2005JCoPh.208..315M}
}

@ARTICLE{mouschovias1976,
       author = {{Mouschovias}, T.~C.},
        title = "{Nonhomologous contraction and equilibria of self-gravitating, magnetic interstellar clouds embedded in an intercloud medium: star formation. II. Results.}",
      journal = {\apj},
         year = 1976,
        month = jul,
       volume = {207},
        pages = {141-158},
          doi = {10.1086/154478},
       adsurl = {https://ui.adsabs.harvard.edu/abs/1976ApJ...207..141M}
}

@ARTICLE{nobrega-siverio2020,
       author = {{N{\'o}brega-Siverio}, D. and {Mart{\'\i}nez-Sykora}, J. and {Moreno-Insertis}, F. and {Carlsson}, M.},
        title = "{Ambipolar diffusion in the Bifrost code}",
      journal = {\aap},
         year = 2020,
        month = jun,
       volume = {638},
          eid = {A79},
        pages = {A79},
          doi = {10.1051/0004-6361/202037809},
archivePrefix = {arXiv},
       eprint = {2004.11927},
 primaryClass = {astro-ph.SR},
       adsurl = {https://ui.adsabs.harvard.edu/abs/2020A&A...638A..79N}
}

@ARTICLE{osullivan2019,
       author = {{O'Sullivan}, Stephen},
        title = "{Runge-Kutta-Gegenbauer explicit methods for advection-diffusion problems}",
      journal = {Journal of Computational Physics},
         year = 2019,
        month = jul,
       volume = {388},
        pages = {209-223},
          doi = {10.1016/j.jcp.2019.03.001},
       adsurl = {https://ui.adsabs.harvard.edu/abs/2019JCoPh.388..209O}
}

@BOOK{parker1979,
       author = {{Parker}, E.~N.},
        title = "{Cosmical magnetic fields. Their origin and their activity}",
         year = 1979,
       adsurl = {https://ui.adsabs.harvard.edu/abs/1979cmft.book.....P}
}

@INPROCEEDINGS{pudritz2007,
       author = {{Pudritz}, R.~E. and {Ouyed}, R. and {Fendt}, Ch. and {Brandenburg}, A.},
        title = "{Disk Winds, Jets, and Outflows: Theoretical and Computational Foundations}",
    booktitle = {Protostars and Planets V},
         year = 2007,
       editor = {{Reipurth}, Bo and {Jewitt}, David and {Keil}, Klaus},
        month = jan,
        pages = {277},
          doi = {10.48550/arXiv.astro-ph/0603592},
archivePrefix = {arXiv},
       eprint = {astro-ph/0603592},
 primaryClass = {astro-ph},
       adsurl = {https://ui.adsabs.harvard.edu/abs/2007prpl.conf..277P}
}

@ARTICLE{puzzoni2021,
       author = {{Puzzoni}, E. and {Mignone}, A. and {Bodo}, G.},
        title = "{On the impact of the numerical method on magnetic reconnection and particle acceleration - I. The MHD case}",
      journal = {\mnras},
         year = 2021,
        month = dec,
       volume = {508},
       number = {2},
        pages = {2771-2783},
          doi = {10.1093/mnras/stab2813},
archivePrefix = {arXiv},
       eprint = {2109.12858},
 primaryClass = {physics.plasm-ph},
       adsurl = {https://ui.adsabs.harvard.edu/abs/2021MNRAS.508.2771P}
}

@ARTICLE{puzzoni2022,
       author = {{Puzzoni}, E. and {Mignone}, A. and {Bodo}, G.},
        title = "{The impact of resistive electric fields on particle acceleration in reconnection layers}",
      journal = {\mnras},
         year = 2022,
        month = nov,
       volume = {517},
       number = {1},
        pages = {1452-1459},
          doi = {10.1093/mnras/stac2807},
archivePrefix = {arXiv},
       eprint = {2210.01113},
 primaryClass = {astro-ph.HE},
       adsurl = {https://ui.adsabs.harvard.edu/abs/2022MNRAS.517.1452P}
}

@ARTICLE{roe1981,
       author = {{Roe}, P.~L.},
        title = "{Approximate Riemann Solvers, Parameter Vectors, and Difference Schemes}",
      journal = {Journal of Computational Physics},
         year = 1981,
        month = oct,
       volume = {43},
       number = {2},
        pages = {357-372},
          doi = {10.1016/0021-9991(81)90128-5},
       adsurl = {https://ui.adsabs.harvard.edu/abs/1981JCoPh..43..357R}
}

@article{rossazza2025,
title = {The PLUTO code on GPUs: A first look at Eulerian MHD methods},
journal = {Astronomy and Computing},
pages = {101076},
year = {2026},
issn = {2213-1337},
doi = {https://doi.org/10.1016/j.ascom.2026.101076},
url = {https://www.sciencedirect.com/science/article/pii/S2213133726000181},
author = {M. Rossazza and A. Mignone and M. Bugli and S. Truzzi and L. Riha and T. Panoc and O. Vysocky and N. Shukla and A. Romeo and V. Berta}
}

@ARTICLE{shu1987,
       author = {{Shu}, Frank H. and {Adams}, Fred C. and {Lizano}, Susana},
        title = "{Star formation in molecular clouds: observation and theory.}",
      journal = {\araa},
         year = 1987,
        month = jan,
       volume = {25},
        pages = {23-81},
          doi = {10.1146/annurev.aa.25.090187.000323},
       adsurl = {https://ui.adsabs.harvard.edu/abs/1987ARA&A..25...23S}
}

@ARTICLE{shu1992,
       author = {{Shu}, F.~H.},
        title = "{Books-Received - the Physics of Astrophysics - V.2 - Gas Dynamics}",
      journal = {Journal of the British Astronomical Association},
         year = 1992,
        month = aug,
       volume = {102},
        pages = {230},
       adsurl = {https://ui.adsabs.harvard.edu/abs/1992JBAA..102..230S}
}

@ARTICLE{skaras2021,
       author = {{Skaras}, Timothy and {Saxton}, Torrey and {Meyer}, Chad and {Aslam}, Tariq D.},
        title = "{Super-time-stepping schemes for parabolic equations with boundary conditions}",
      journal = {Journal of Computational Physics},
         year = 2021,
        month = jan,
       volume = {425},
          eid = {109879},
        pages = {109879},
          doi = {10.1016/j.jcp.2020.109879},
       adsurl = {https://ui.adsabs.harvard.edu/abs/2021JCoPh.42509879S}
}

@BOOK{spitzer1978,
       author = {{Spitzer}, Lyman},
        title = "{Physical processes in the interstellar medium}",
         year = 1978,
          doi = {10.1002/9783527617722},
       adsurl = {https://ui.adsabs.harvard.edu/abs/1978ppim.book.....S}
}

@ARTICLE{stone2020,
       author = {{Stone}, James M. and {Tomida}, Kengo and {White}, Christopher J. and {Felker}, Kyle G.},
        title = "{The Athena++ Adaptive Mesh Refinement Framework: Design and Magnetohydrodynamic Solvers}",
      journal = {\apjs},
         year = 2020,
        month = jul,
       volume = {249},
       number = {1},
          eid = {4},
        pages = {4},
          doi = {10.3847/1538-4365/ab929b},
archivePrefix = {arXiv},
       eprint = {2005.06651},
 primaryClass = {astro-ph.IM},
       adsurl = {https://ui.adsabs.harvard.edu/abs/2020ApJS..249....4S}
}

@ARTICLE{tomida2013,
       author = {{Tomida}, Kengo and {Tomisaka}, Kohji and {Matsumoto}, Tomoaki and {Hori}, Yasunori and {Okuzumi}, Satoshi and {Machida}, Masahiro N. and {Saigo}, Kazuya},
        title = "{Radiation Magnetohydrodynamic Simulations of Protostellar Collapse: Protostellar Core Formation}",
      journal = {\apj},
         year = 2013,
        month = jan,
       volume = {763},
       number = {1},
          eid = {6},
        pages = {6},
          doi = {10.1088/0004-637X/763/1/6},
archivePrefix = {arXiv},
       eprint = {1206.3567},
 primaryClass = {astro-ph.SR},
       adsurl = {https://ui.adsabs.harvard.edu/abs/2013ApJ...763....6T}
}

@ARTICLE{tomida2015,
       author = {{Tomida}, Kengo and {Okuzumi}, Satoshi and {Machida}, Masahiro N.},
        title = "{Radiation Magnetohydrodynamic Simulations of Protostellar Collapse: Nonideal Magnetohydrodynamic Effects and Early Formation of Circumstellar Disks}",
      journal = {\apj},
         year = 2015,
        month = mar,
       volume = {801},
       number = {2},
          eid = {117},
        pages = {117},
          doi = {10.1088/0004-637X/801/2/117},
archivePrefix = {arXiv},
       eprint = {1501.04102},
 primaryClass = {astro-ph.SR},
       adsurl = {https://ui.adsabs.harvard.edu/abs/2015ApJ...801..117T}
}

@inbook{toro2009,
author = {Toro, Eleuterio},
year = {2009},
month = {01},
pages = {},
title = {Riemann Solvers and Numerical Methods for Fluid Dynamics: A Practical Introduction},
journal = {Riemann Solvers and Numerical Methods for Fluid Dynamics},
doi = {10.1007/b79761}
}

@ARTICLE{toth2000,
       author = {{T{\'o}th}, G{\'a}bor},
        title = "{The {\ensuremath{\nabla}}{\textperiodcentered} B=0 Constraint in Shock-Capturing Magnetohydrodynamics Codes}",
      journal = {Journal of Computational Physics},
         year = 2000,
        month = jul,
       volume = {161},
       number = {2},
        pages = {605-652},
          doi = {10.1006/jcph.2000.6519},
       adsurl = {https://ui.adsabs.harvard.edu/abs/2000JCoPh.161..605T}
}

@ARTICLE{tsukamoto2022,
       author = {{Tsukamoto}, Yusuke and {Okuzumi}, Satoshi},
        title = "{Impact of Dust Size Distribution Including Large Dust Grains on Magnetic Resistivity: An Analytical Approach}",
      journal = {\apj},
         year = 2022,
        month = jul,
       volume = {934},
       number = {1},
          eid = {88},
        pages = {88},
          doi = {10.3847/1538-4357/ac7b7b},
archivePrefix = {arXiv},
       eprint = {2208.00601},
 primaryClass = {astro-ph.SR},
       adsurl = {https://ui.adsabs.harvard.edu/abs/2022ApJ...934...88T}
}

@ARTICLE{turner2014,
       author = {{Turner}, N.~J. and {Lee}, Man Hoi and {Sano}, T.},
        title = "{Magnetic Coupling in the Disks around Young Gas Giant Planets}",
      journal = {\apj},
         year = 2014,
        month = mar,
       volume = {783},
       number = {1},
          eid = {14},
        pages = {14},
          doi = {10.1088/0004-637X/783/1/14},
archivePrefix = {arXiv},
       eprint = {1306.2276},
 primaryClass = {astro-ph.EP},
       adsurl = {https://ui.adsabs.harvard.edu/abs/2014ApJ...783...14T}
}

@ARTICLE{vanderhouwen1980,
       author = {{van Der Houwen}, P.~J. and {Sommeijer}, B.~P.},
        title = "{On the Internal Stability of Explicit,m-Stage Runge-Kutta Methods for Largem-Values}",
      journal = {Zeitschrift Angewandte Mathematik und Mechanik},
         year = 1980,
        month = jan,
       volume = {60},
       number = {10},
        pages = {479-485},
          doi = {10.1002/zamm.19800601005},
       adsurl = {https://ui.adsabs.harvard.edu/abs/1980ZaMM...60..479V}
}

@ARTICLE{vaidya2017,
       author = {{Vaidya}, Bhargav and {Prasad}, Deovrat and {Mignone}, Andrea and {Sharma}, Prateek and {Rickler}, Luca},
        title = "{Scalable explicit implementation of anisotropic diffusion with Runge-Kutta-Legendre super-time stepping}",
      journal = {\mnras},
         year = 2017,
        month = dec,
       volume = {472},
       number = {3},
        pages = {3147-3160},
          doi = {10.1093/mnras/stx2176},
archivePrefix = {arXiv},
       eprint = {1702.05487},
 primaryClass = {astro-ph.IM},
       adsurl = {https://ui.adsabs.harvard.edu/abs/2017MNRAS.472.3147V}
}

@article{verwer1990,
  author = {Verwer, J.G. and Hundsdorfer, W.H. and Sommeijer, B.P.},
  title = {Convergence Properties of the Runge-Kutta-Chebyshev Method},
  journal = {Numerische Mathematik},
  volume = {57},
  pages = {157--178},
  year = {1990},
  doi = {10.1007/BF01386405},
  url = {http://eudml.org/doc/133443}
}

@ARTICLE{virtanen2020,
       author = {{Virtanen}, Pauli and {Gommers}, Ralf and {Oliphant}, Travis E. and {Haberland}, Matt and {Reddy}, Tyler and {Cournapeau}, David and {Burovski}, Evgeni and {Peterson}, Pearu and {Weckesser}, Warren and {Bright}, Jonathan and {van der Walt}, St{\'e}fan J. and {Brett}, Matthew and {Wilson}, Joshua and {Millman}, K. Jarrod and {Mayorov}, Nikolay and {Nelson}, Andrew R.~J. and {Jones}, Eric and {Kern}, Robert and {Larson}, Eric and {Carey}, C.~J. and {Polat}, {\.I}lhan and {Feng}, Yu and {Moore}, Eric W. and {VanderPlas}, Jake and {Laxalde}, Denis and {Perktold}, Josef and {Cimrman}, Robert and {Henriksen}, Ian and {Quintero}, E.~A. and {Harris}, Charles R. and {Archibald}, Anne M. and {Ribeiro}, Ant{\^o}nio H. and {Pedregosa}, Fabian and {van Mulbregt}, Paul and {SciPy 1.  0 Contributors}},
        title = "{SciPy 1.0: fundamental algorithms for scientific computing in Python}",
      journal = {Nature Medicine},
         year = 2020,
        month = feb,
       volume = {17},
        pages = {261-272},
          doi = {10.1038/s41592-019-0686-2},
archivePrefix = {arXiv},
       eprint = {1907.10121},
 primaryClass = {cs.MS},
       adsurl = {https://ui.adsabs.harvard.edu/abs/2020NatMe..17..261V}
}

@ARTICLE{wang2023,
       author = {{Wang}, Yulei and {Cheng}, Xin and {Ding}, Mingde and {Liu}, Zhaoyuan and {Liu}, Jian and {Zhu}, Xiaojue},
        title = "{Three-dimensional Turbulent Reconnection within the Solar Flare Current Sheet}",
      journal = {\apjl},
         year = 2023,
        month = sep,
       volume = {954},
       number = {2},
          eid = {L36},
        pages = {L36},
          doi = {10.3847/2041-8213/acf19d},
archivePrefix = {arXiv},
       eprint = {2308.10494},
 primaryClass = {astro-ph.SR},
       adsurl = {https://ui.adsabs.harvard.edu/abs/2023ApJ...954L..36W}
}

@ARTICLE{wardle2007,
       author = {{Wardle}, Mark},
        title = "{Magnetic fields in protoplanetary disks}",
      journal = {\apss},
         year = 2007,
        month = oct,
       volume = {311},
       number = {1-3},
        pages = {35-45},
          doi = {10.1007/s10509-007-9575-8},
archivePrefix = {arXiv},
       eprint = {0704.0970},
 primaryClass = {astro-ph},
       adsurl = {https://ui.adsabs.harvard.edu/abs/2007Ap&SS.311...35W}
}

@ARTICLE{wurster2014,
       author = {{Wurster}, James and {Price}, Daniel and {Ayliffe}, Ben},
        title = "{Ambipolar diffusion in smoothed particle magnetohydrodynamics}",
      journal = {\mnras},
         year = 2014,
        month = oct,
       volume = {444},
       number = {2},
        pages = {1104-1112},
          doi = {10.1093/mnras/stu1524},
archivePrefix = {arXiv},
       eprint = {1408.1807},
 primaryClass = {astro-ph.SR},
       adsurl = {https://ui.adsabs.harvard.edu/abs/2014MNRAS.444.1104W}
}

@ARTICLE{wurster2021,
       author = {{Wurster}, James},
        title = "{Do we need non-ideal magnetohydrodynamic to model protostellar discs?}",
      journal = {\mnras},
         year = 2021,
        month = mar,
       volume = {501},
       number = {4},
        pages = {5873-5891},
          doi = {10.1093/mnras/staa3943},
archivePrefix = {arXiv},
       eprint = {2101.04129},
 primaryClass = {astro-ph.SR},
       adsurl = {https://ui.adsabs.harvard.edu/abs/2021MNRAS.501.5873W}
}

@ARTICLE{zier2024,
       author = {{Zier}, Oliver and {Springel}, Volker and {Mayer}, Alexander C.},
        title = "{Non-ideal magnetohydrodynamics on a moving mesh I: ohmic and ambipolar diffusion}",
      journal = {\mnras},
         year = 2024,
        month = jan,
       volume = {527},
       number = {1},
        pages = {1563-1579},
          doi = {10.1093/mnras/stad3200},
archivePrefix = {arXiv},
       eprint = {2307.11814},
 primaryClass = {astro-ph.EP},
       adsurl = {https://ui.adsabs.harvard.edu/abs/2024MNRAS.527.1563Z}
}

@ARTICLE{zweibel2009,
       author = {{Zweibel}, Ellen G. and {Yamada}, Masaaki},
        title = "{Magnetic Reconnection in Astrophysical and Laboratory Plasmas}",
      journal = {\araa},
         year = 2009,
        month = sep,
       volume = {47},
       number = {1},
        pages = {291-332},
          doi = {10.1146/annurev-astro-082708-101726},
       adsurl = {https://ui.adsabs.harvard.edu/abs/2009ARA&A..47..291Z}
}

@ARTICLE{GonzalezMorales2020,
       author = {{Gonz{\'a}lez-Morales}, P.~A. and {Khomenko}, E. and {Vitas}, N. and {Collados}, M.},
        title = "{Joint action of Hall and ambipolar effects in 3D magneto-convection simulations of the quiet Sun. I. Dissipation and generation of waves}",
      journal = {\aap},
         year = 2020,
        month = oct,
       volume = {642},
          eid = {A220},
        pages = {A220},
          doi = {10.1051/0004-6361/202037938},
archivePrefix = {arXiv},
       eprint = {2008.10429},
 primaryClass = {astro-ph.SR},
       adsurl = {https://ui.adsabs.harvard.edu/abs/2020A&A...642A.220G}
}

@ARTICLE{Zhao2021,
       author = {{Zhao}, Bo and {Caselli}, Paola and {Li}, Zhi-Yun and {Krasnopolsky}, Ruben and {Shang}, Hsien and {Lam}, Ka Ho},
        title = "{The interplay between ambipolar diffusion and Hall effect on magnetic field decoupling and protostellar disc formation}",
      journal = {\mnras},
         year = 2021,
        month = aug,
       volume = {505},
       number = {4},
        pages = {5142-5163},
          doi = {10.1093/mnras/stab1295},
archivePrefix = {arXiv},
       eprint = {2009.07820},
 primaryClass = {astro-ph.SR},
       adsurl = {https://ui.adsabs.harvard.edu/abs/2021MNRAS.505.5142Z}
}

\appendix

\section{The Gegenbauer Polynomials}
\label{app:gegenbauer}

The Gegenbauer polynomials $C_s^{(\alpha)}(x)$ form a family of classical orthogonal polynomials that arise as solutions of the Gegenbauer differential equation \citep{abramowitz1972,karageorghis1992}
\begin{equation}
\label{eq:GegenbauerODE}
(1 - x^2)C_s^{\prime\prime(\alpha)} - (2\alpha + 1)xC^{\prime(\alpha)}_s + s(s + 2\alpha)C_s^{(\alpha)} = 0,
\end{equation}
where $s\in \mathbb{N}_0$ and $\alpha\in(-1/2,\infty)$ represent, respectively, the polynomial degree and the Gegenbauer parameter.
For each integer $s \ge 0$, this equation admits a polynomial solution of degree $s$, which defines $C_s^{(\alpha)}(x)$ up to normalization.

\begin{figure}

\centering
\includegraphics[width=0.49\textwidth]{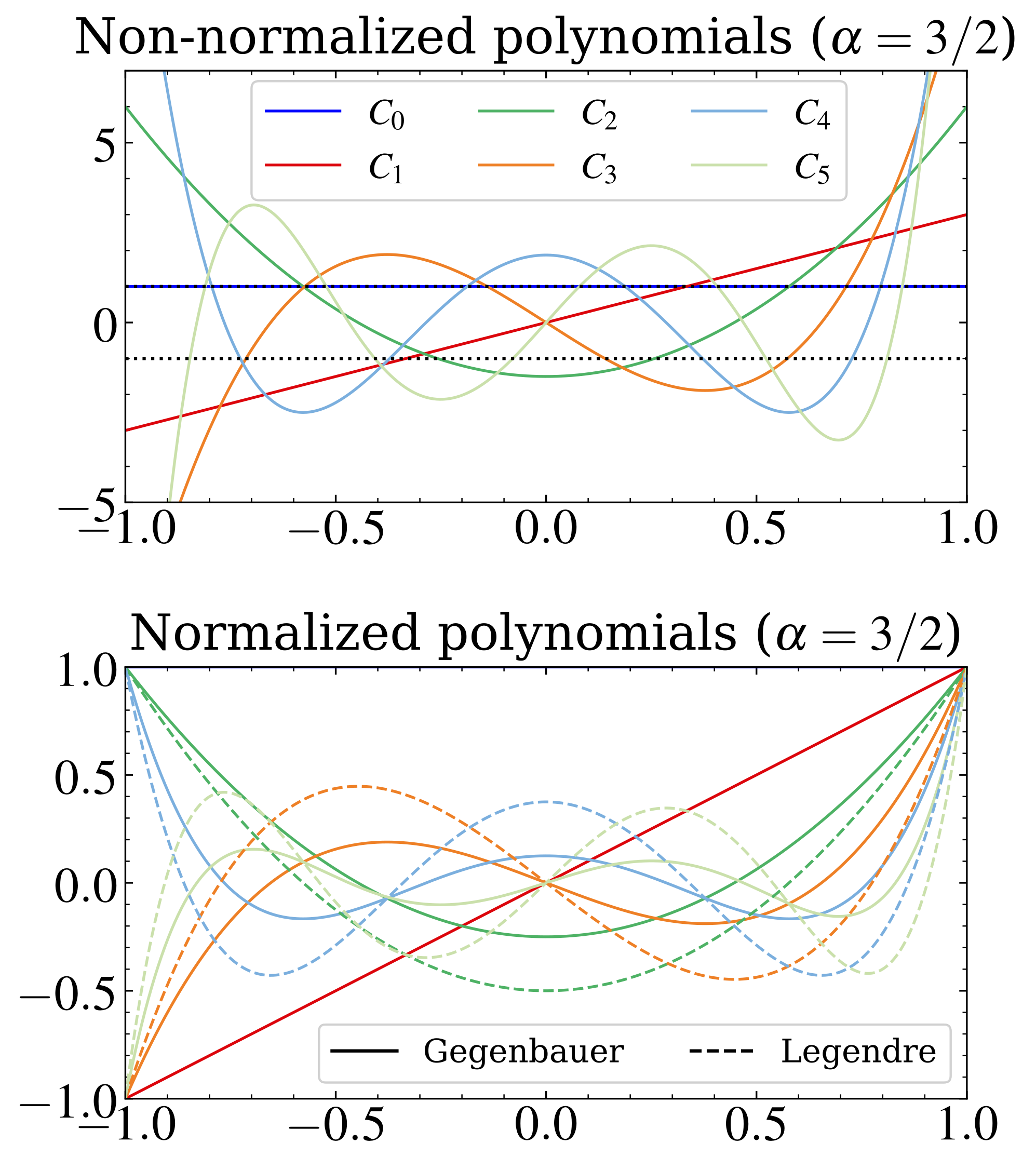}

\caption{Gegenbauer polynomial (with $\alpha = 3/2$ coefficients without (top panel) and with (bottom panel) the normalization employed in \citet{doha1991}. The dotted lines in the top panel represent the threshold $\pm1$, while the dashed lines in the bottom panel represent the corresponding Legendre polynomials.}
\label{Fig::RKGcoefficients}%
\end{figure}

In this work, we do not adopt the usual standardization for the Gegenbauer polynomial (see, e.g., \citealt{erdelyi1953}), but instead the one reported in \citet{doha1991}, requiring that polynomials satisfy the normalization condition:
\begin{equation}
C_s^{(\alpha)}(\pm 1) = \pm 1,
\qquad \forall s \ge 0 ,
\end{equation}
which ensures a bounded amplification factor at the endpoints of the stability interval and simplifies the derivation of the RKG stability polynomial.
This yields the explicit form:
\begin{equation}
C_s^{(\alpha)}(x) =  \left(-\DS\frac{1}{2}\right)^s\DS\frac{\Gamma(\alpha + 
1/2)}{\Gamma(k)}f(x)^{1/2 - \alpha}\der{^s}{x^s}f(x)^{k-1},
\end{equation}
where $\Gamma(\cdot)$ denotes the Gamma function, $k = s + \alpha + 1/2$, $f(x) = 1 - x^2$ and $\alpha > -1/2$.
With this normalization, the first polynomials read
\begin{equation}
\begin{array}{rcl}
C_0^{(\alpha)}(x) & = & 1, \\[6pt]
C_1^{(\alpha)}(x) & = & x,
\end{array}
\end{equation}
and higher orders can be computed through the recurrence relation
\begin{equation}
C_{s+1}^{(\alpha)}(x)=2x\,\frac{\alpha + s}{2\alpha + s} C_s^{(\alpha)}(x)-\frac{s}{2\alpha + s} C_{s-1}^{(\alpha)}(x),\qquad s \ge 1.
\end{equation}
Evaluating \eqref{eq:GegenbauerODE} at $x=1$ yields the useful relations:
\begin{equation}
\begin{array}{ccl}
C_s'^{(\alpha)}(1) & = & \DS\frac{s(s + 2\alpha)}{2\alpha + 1}C_s^{(\alpha)}(1),
 \\ \noalign{\medskip}
C_s^{\prime\prime(\alpha)}(1) & = & \DS\frac{s(s + 2\alpha) - 2\alpha - 1}{3 +
 2\alpha}C_s^{\prime(\alpha)}(1).
\end{array}
\end{equation}

Several classical orthogonal polynomials are recovered for particular values of $\alpha$:

\begin{itemize}
\item $\alpha =0.5$: Legendre polynomials,
\item $\alpha = 1$: Chebyshev polynomials of the second kind (up to a normalization factor),
\item $\alpha = 0$: Chebyshev polynomials of the first kind (in the degenerate limit).
\end{itemize}

Note that in previous works (e.g., \citealt{skaras2021,caplan2024}) the choice $\alpha = 3/2$ is commonly adopted.
For the sake of consistency with that literature, Fig.~\ref{Fig::RKGcoefficients} shows the different Gegenbauer polynomials (up to $5^{\rm th}$ order) without (top panel) and with (bottom panel) the normalization at unity for $\alpha = 3/2$.
The dashed lines in the top panel represent the reference levels $\pm1$, while the dotted line shows the corresponding Legendre polynomial.
For large values of $\alpha$, the classical (non-normalized) Gegenbauer polynomials reach large values near $x=\pm1$, which motivates both the normalization adopted here and the use of larger values of $\alpha$ in the numerical scheme.

\section{Illustration of the RKG scheme}
\label{app:rkgcoefficients} 

In this Appendix, we summarize the derivation of the RKG stability polynomial coefficients and the corresponding stage coefficients used in the SRK formulation.
The stability polynomial is written as (see also Eq. \ref{eq::stabilitypolynomial})
\begin{equation}
 R_s(z) = a_s + b_sC_s^{(\alpha)}(w_0 + w_1z).
\end{equation}
The parameter $w_0$ controls the amount of numerical damping and the extent of the method's stability region, and can be tuned to improve robustness in substepping schemes \citep{meyer2014,osullivan2019}. Since additional damping is not required in the present application \citep{meyer2014}, we set $w_0 = 1$, so that the stability polynomial is centered at $z = 0$.
The coefficients $a_s$, $b_s$, and $w_1$ are determined by imposing 
\begin{equation}
\label{eq::der_stability}
\der{^n}{z^n}R_s(0) = 1,
\end{equation}
where $n = 0,1$ for the RKG1 and $n = 0,1,2$ for the RKG2 algorithms \citet{meyer2014,osullivan2019}.
From $n = 0$ we obtain:
\begin{equation}
a_s + b_s = 1,
\end{equation}
while from $n = 1$ we get
\begin{equation}
b_sw_1 = \DS\frac{1}{C_s'^{(\alpha)}(1)} = \DS\frac{2\alpha + 1}{s(2\alpha + s)}.
\end{equation}

The $1^\mathrm{st}$-order consistency conditions leave one degree of freedom in the coefficients; setting $a_s = 0$ closes the system and yields the classical maximal stability construction \citep{meyer2014},
\begin{equation}
\begin{array}{lcl}
b_s & = & 1, \\ \noalign{\medskip}
w_1 & = & \DS\frac{1 + 2\alpha}{s(s + 2\alpha)}\,.
\end{array}
\end{equation}
For $2^\mathrm{nd}$-order algorithms, the closure relation is derived by evaluating Eq. \ref{eq::der_stability} at $n = 2$, from which we obtain
\begin{equation}
\label{eq::w1_RKG}
w_1 = \DS\frac{C_s^{\prime(\alpha)}(1)}{C_s^{\prime\prime(\alpha)}(1)} = \DS\frac{3 + 2\alpha}{(s + 2\alpha + 1)(s - 1)},
\end{equation}
and, as a consequence,
\begin{equation}
\label{eq::a_b_2nd_RKG}
\begin{array}{lcl}
a_s & = & 1 - \DS\frac{(2\alpha + 1)(s + 2\alpha + 1)(s - 1)}{s(s + 2\alpha)(3 + 2\alpha)},\\ \noalign{\medskip}
b_s & = & \DS\frac{(2\alpha + 1)}{s(s + 2\alpha)}\DS\frac{(s + 2\alpha + 1)(s - 1)}{(3 + 2\alpha)}\,.
\end{array}\
\end{equation}
Note that Eq. \ref{eq::w1_RKG} and \ref{eq::a_b_2nd_RKG} contain factors
that set the denominator to 0 for $s = 0$ and $s = 1$, making these formulas singular.
As a result, the coefficients $b_0$ and $b_1$ cannot be uniquely determined from the closed-form formulas and can thus be specified arbitrarily.
For consistency, we follow \citet{meyer2014} and set $b_0 = b_1 = 1/3$.

By exploiting the three-term recurrence relation of the Gegenbauer polynomials and matching coefficients of the stability polynomial, one obtains the identity
\begin{equation}
a_j - \mu_ja_{j - 1} - \nu_ja_{j - 2} = 1 - \mu_j - \nu_j,
\end{equation}
from which the stage coefficients of the compact RKG formulation (Eq. \eqref{eq::RKGcompactform}) are obtained as
\begin{equation}
\begin{array}{lcl}
\mu_j &=& \;\;2\DS\frac{\alpha + j - 1}{j + 2\alpha - 1}\DS\frac{b_j}{b_{j-1}},\\ \noalign{\medskip}
\nu_j &=& -\DS\frac{j - 1}{j + 2\alpha - 1}\DS\frac{b_j}{b_{j - 2}}, \\ \noalign{\medskip}
\tilde{\mu}_j &=& \;\;\mu_jw_1,\\ \noalign{\medskip} 
\tilde{\gamma}_j &=& -\tilde{\mu}_j a_{j - 1}.
\end{array}
\end{equation}
The coefficients $a_j$ and $b_j$ required in the stage recursion are obtained by evaluating the same closed-form expressions at degree $j$, while keeping $w_1$ fixed to the value determined for the chosen number of stages $s$ \citep{meyer2014,osullivan2019}.
Finally, the number of substages $s$ required for a given super-step follows from the relation between $\tau$ (i.e., the ratio between the parabolic and the hyperbolic timestep) and the number of stages \citep{meyer2014, caplan2024}:
\begin{equation}
\label{eq::substeps inverted}
\tau  = \left\{
\begin{array}{lcl}
\DS\frac{s(s + 2\alpha)}{1 + 2\alpha} & \qquad& 1\mathrm{st}\text{ order}, \\ \noalign{\medskip}
\DS\frac{(s + 2\alpha + 1)(s - 1)}{3 + 2\alpha} & \qquad& 2\mathrm{nd}\text{ order}.
\end{array}\right.
\end{equation}
By inverting Eq. \ref{eq::substeps inverted}, we recover the expression for the minimum number of substeps required as a function os $\tau$ reported in Eq. \ref{eq::substeps}.

For the sake of clarity, all the relevant coefficients for some key values of $\alpha$ (i.e., 0.5, which corresponds to the RKL scheme, 1.5, which corresponds to the value used in \citealt{skaras2021}, and 10, which corresponds to the value used in this paper) are reported in Table \ref{tab:coefficients}.

\begin{table*}
    \centering
    \caption{List of coefficients of the RKG scheme for different values of $\alpha$ as functions of the total number of substeps $s$ and the current substep number $j$.
    }
    \label{tab:coefficients}
    \begin{tabular}{c | c c c c c c}
       $\alpha$ (order) & $b_0$ & $b_1$ & $b_j$ & $w_1$ & $\mu_j$ & $\nu_j$ \\ \hline 
       
     0.5 (1) \rule{0pt}{4ex} & 1   & 1   & 1                                    &  $\DS\frac{2}{s(s + 1)}$ & \multirow{2}{*}[-0.8em]{$\DS\frac{2j-1}{j}\DS\frac{b_j}{b_{j-1}}$} & \multirow{2}{*}[-0.8em]{$-\DS\frac{j-1}{j}\DS\frac{b_j}{b_{j-2}}$} \\ 
     0.5 (2) \rule{0pt}{4ex} & 1/3 & 1/3 & $\DS\frac{(j+2)(j-1)}{2j(j+1)}$      &  $\DS\frac{4}{(s + 2)(s - 1)}$ & \\[2.5ex] \hline
     
     1.5 (1) \rule{0pt}{4ex} & 1   & 1   & 1                                    &  $\DS\frac{4}{s(s+3)}$ & \multirow{2}{*}[-0.8em]{$\DS\frac{2j+1}{j+2}\DS\frac{b_j}{b_{j-1}}$} & \multirow{2}{*}[-0.8em]{$-\DS\frac{j-1}{j+2}\DS\frac{b_j}{b_{j-2}}$} \\  
     1.5 (2) \rule{0pt}{4ex} & 1/3 & 1/3 & $\DS\frac{2(j+4)(j-1)}{3j(j+3)}$     &  $\DS\frac{6}{(s +4)(s-1)}$ & \\[2.5ex] \hline
     
     10  (1) \rule{0pt}{4ex} & 1   & 1   & 1                                    &  $\DS\frac{21}{s(s+20)}$ & \multirow{2}{*}[-0.8em]{$\DS\frac{2j+18}{j+19}\DS\frac{b_j}{b_{j-1}}$} & \multirow{2}{*}[-0.8em]{$-\DS\frac{j-1}{j+19}\DS\frac{b_j}{b_{j-2}}$} \\  
     10  (2) \rule{0pt}{4ex} & 1/3 & 1/3 & $\DS\frac{21(j+21)(j-1)}{23j(j+20)}$ &  $\DS\frac{23}{(s+21)(s-1)}$ & 

    \end{tabular}
\end{table*}

\section{Benchmark: multidimensional resistive field diffusion}
\label{app::field_diff}

Here, we present a multidimensional decay test of a magnetic field due to magnetic resistivity.
As an initial condition, we set a static (i.e., zero initial velocity) magnetized fluid with $\rho = 10^9$.
Such a condition is enforced at every step throughout the domain (through internal boundaries) to ensure that only the magnetic field evolves.
We initialize the magnetic field as a Gaussian starting from the analytical solution of the induction equation with zero velocity and assuming a scalar Ohmic resistivity:
\begin{equation}
\begin{array}{lcl}
B_x(y,t)   & = & f(y)f(z), \\ 
B_y(x,t)   & = & f(x)f(z), \\ 
B_z(x,y,t) & = & f(y)f(x),
\end{array}
\label{eq:field_diff}
\end{equation}
where
\begin{equation}
f(k) = \left\{
\begin{array}{lcl}
\DS\frac{\exp(-k^2/4\eta t)}{\sqrt{t}} & \qquad\qquad & \text{if $k$ is included},  \\ \noalign{\medskip}
1 && \text{otherwise},
\end{array}\right.
\end{equation}
and by setting $t = 1$ in the initial conditions.
We assess the scheme's behavior in both Cartesian and spherical geometries to disentangle purely algorithmic effects from those introduced by the curvilinear metric terms.
For the RKG scheme, we set $\alpha = 10$.

\subsection{Cartesian field diffusion}

We first performed numerical simulations on a two-dimensional Cartesian mesh with $x,y\in[-5,5]$ until $ t=2$.
In this test, we assess the performance of the super–time–stepping scheme when coupled to two different divergence-control strategies: the generalized Lagrange multiplier (GLM) method \citep{dedner2002} and the constrained transport (CT) scheme \citep{mignone2021}.
For this test, we adopt linear reconstruction \citep{toro2009}, a second-order Runge–Kutta integrator \citep{gottlieb2001}, an ideal equation of state, a CFL number of 0.4, and the HLL Riemann solver \citep{harten1983}.
We choose two resistivity regimes with scalar diffusion coefficient: in the low regime, we set $\eta = 0.001$, while we set $\eta = 0.1$ for the high resistivity regime.

\begin{figure}
\centering
\includegraphics[width=0.49\textwidth]{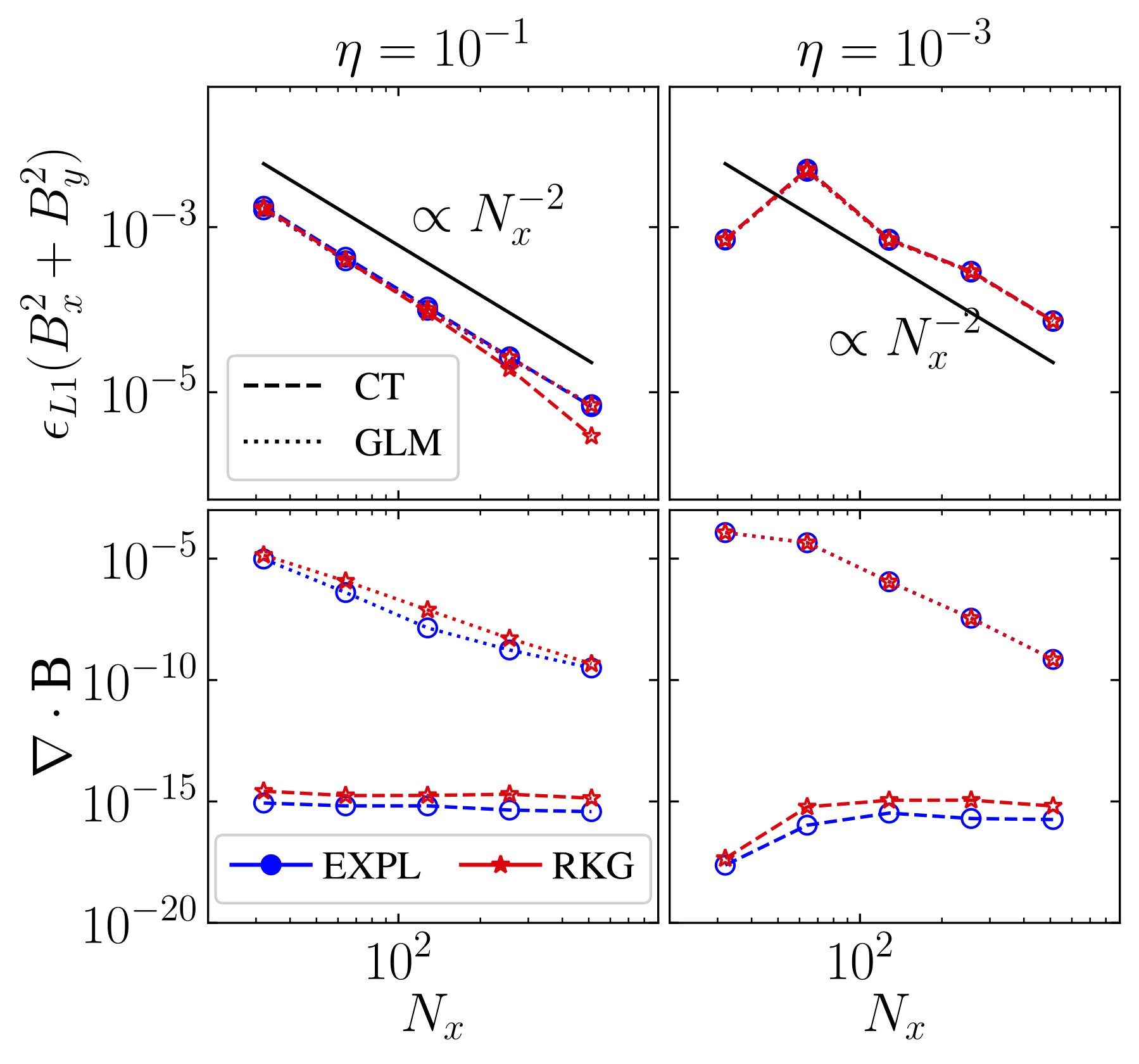}
\caption{Convergence and error on $\nabla\cdot\vect{B}$ for the cartesian field diffusion test. On the top row, the relative error on the 2-dimensional magnetic energy is shown for high (left column) and low resistivities. The bottom panels show the averaged magnetic-field divergence. The solid black lines on the top panels indicate the desired convergence order.}
\label{Fig::res_diffusion}%
\end{figure}

The convergence and accuracy of the RKG scheme compared to an explicit scheme are shown in Fig.\ref{Fig::res_diffusion}.
As a first consideration, we note that both GLM and CT algorithms are able to achieve the expected $2^{\mathrm {nd}} $-order convergence from the full MHD advection-diffusion solver.

\begin{figure}
\centering
\includegraphics[width=0.49\textwidth]{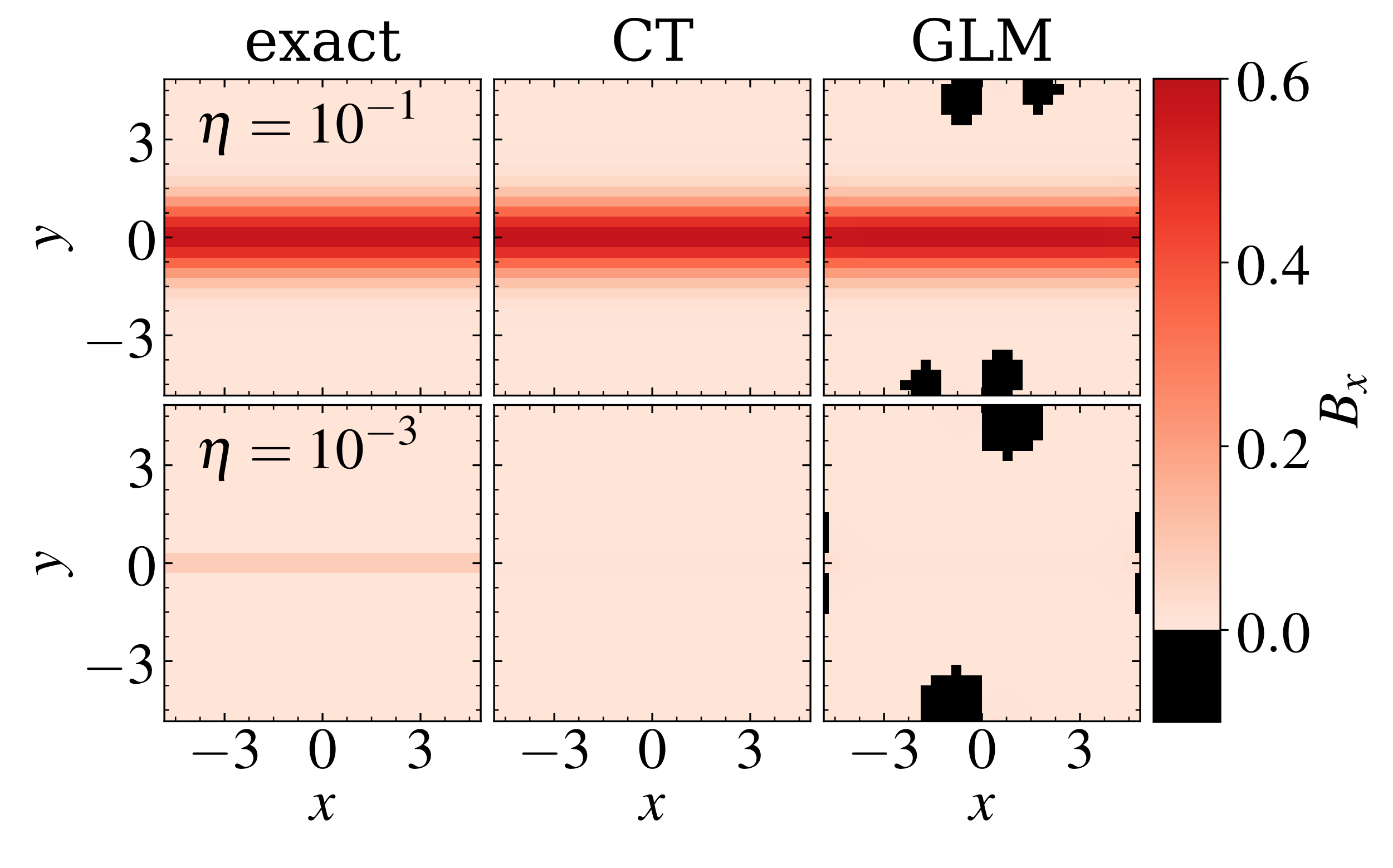}
\caption{Analytical solution (left panel) of the magnetic field $x-$component computed at the final time with the explicit scheme for both CT (middle panels) and GLM (right panels), in the high- (top panels) and low-resistivity (bottom panels) regimes.}
\label{Fig::field_loweta}%
\end{figure}

At the coarsest resolution ($N_x = N_y = 32$), both schemes exhibit a smaller relative error in the low-resistivity case, where numerical diffusion still dominates over the physical dissipation.
However, this apparent improvement does not imply higher physical accuracy. 
This is shown in Fig. \ref{Fig::field_loweta}, which reports the $B_x$ component at the final time for the different $\nabla\cdot\vec{B}$ methods (with resistivity integrated explicitly) at $N_x = N_y = 32$, compared to the analytical solution.
While the high-resistivity case with CT, shown in the top panels, exhibits excellent agreement with the analytical solution, the low-resistivity, low-resolution cases display significant differences. 
More specifically, both CT and GLM simulations exhibit excessive diffusion due to the numerical resistivity, which is inherent to any numerical code and strongly depends on the grid resolution and numerical algorithms employed.
Additionally, the resistive GLM simulations yield few regions near the boundaries where the magnetic field polarity along the $x$-axis reverses.
This unphysical behavior results from numerical diffusion locally exceeding the prescribed physical resistivity and strongly indicates that relying solely on numerical dissipation can lead to spurious features, as its magnitude depends on resolution and the numerical scheme rather than on a controlled physical prescription.

On the other hand, a more monotonous convergence is noticeable for high $\eta$, since the physical resistivity is the dominant process even at low resolution.
Finally, \citet{lesur2023} found that the RKL scheme leads to higher magnetic-field divergence than an explicit method.
We observe the same qualitative behavior with RKG, with divergence levels up to a factor $\lesssim 4$ larger than in the explicit case.
However, the divergence of the magnetic field remains comparable to the reference literature values (e.g., \citealt{toth2000}) and exhibits no qualitative differences when compared to the explicit scheme.

\subsection{Spherical field diffusion}

To check convergence in non-Cartesian geometries, which are a key ingredient in several astrophysically relevant numerical simulations, we perform the same field diffusion test on a spherical mesh ($r$,$\theta$,$\phi$), using Eq. \ref{eq:field_diff} where
\begin{equation}
\begin{array}{lcl}
x & = & r\cos(\phi)\sin(\theta) - 5, \\ \noalign{\medskip}
y & = & r\sin(\phi)\sin(\theta) - 5, \\ \noalign{\medskip}
z & = & r\cos(\theta) - 5.
\end{array}
\end{equation}
The magnetic field, in spherical components, becomes:
\begin{equation}
\begin{array}{lcl}
B_r  & = & B_x(t)\cos\phi\sin\theta + B_y(t)\sin\phi\sin\theta  + B_z(t)\cos\theta, \\ \noalign{\medskip}
B_\theta & = & B_x(t)\cos\phi\cos\theta + B_y(t)\sin\phi\cos\theta - B_z(t)\sin\theta, \\ \noalign{\medskip}
B_r & = & -B_x(t)\sin\phi + B_y(t)\cos\phi. \\ \noalign{\medskip}
\end{array}
\end{equation}

\begin{figure}
\centering
\includegraphics[width=0.49\textwidth]{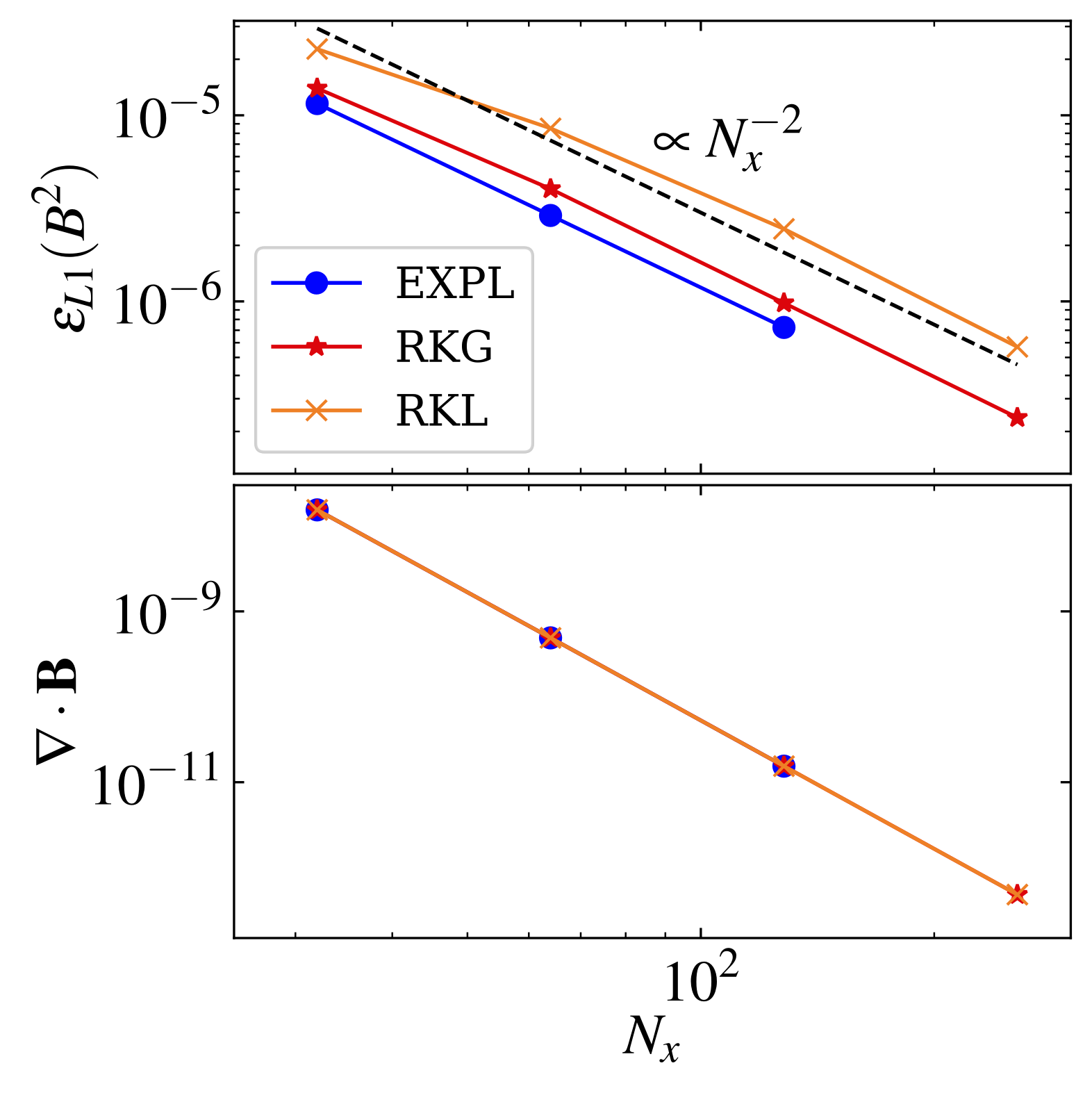}
\caption{Top: $L_1$ error of $B^2$ as a function of the grid resolution for the explicit scheme (EXPL), RKG super–time–stepping, and the RKL variant in the spherical diffusion test. The dashed line shows the expected second–order scaling. Bottom: magnitude of the numerical divergence error of the magnetic field as a function of the grid resolution.}
\label{Fig::sph_diffusion}%
\end{figure}

For this test, we adopt linear reconstruction \citep{toro2009}, a second-order Runge–Kutta integrator \citep{gottlieb2001}, an isothermal equation of state with $c_s = 1$, a CFL number of 0.3,  the UCT-HLL upwind constrained transport method \citep{mignone2021}, and the Roe Riemann solver \citep{roe1981}, and we set the resistivity to $\eta = 1$.
We show the convergence, compared to the expected $2^{\mathrm nd}$ order, in the top panel of Fig. \ref{Fig::sph_diffusion}.
Both the explicit and RKG curves follow the expected second-order slope over the explored resolution range, with only mild pre-asymptotic deviations at the coarsest grid.
To highlight the importance of the $\alpha$ parameter, we also report results obtained with an RKL scheme (a particular case of the RKG scheme when $\alpha = 0.5$).
The comparison in the top panel shows that changing $\alpha$ does not modify the asymptotic convergence rate: RKG and RKL recover the same second–order slope at sufficiently high resolution. 
What $\alpha$ does affect is the pre-asymptotic behavior and the error constant. In particular, the RKL run consistently lies above the RKG run at all resolutions, indicating that: 1) $\alpha$ controls how quickly the method enters its asymptotic regime and 2) how large the errors are at a given grid size.
This means $\alpha$ is a genuine tuning parameter of the STS integrator: different choices lead to significantly different accuracy, even though the formal order is unchanged.

The divergence of the magnetic field is reported in the bottom panel of Fig. \ref{Fig::sph_diffusion}.
Unlike the Cartesian case, here we observe a decrease in $\nabla\cdot\vec {B}$ with resolution.
This behavior is consistent with truncation errors in the discrete divergence operator, which in spherical geometry include metric factors that vanish in the Cartesian limit.
The near-overlap of the RKG and explicit results suggests that divergence control is primarily determined by the spatial discretization rather than by the time integrator, so improvements to $\nabla\cdot\vec{B}$ would likely require higher-order geometric reconstructions.

\section{Benchmark: multidimensional ambipolar Barenblatt solution}
\label{app::barenblatt}

The benchmark performed in this section is the decay of a magnetic field due to ambipolar diffusion.
We start with a non-vanishing magnetic field only in the $z-$component, i.e.:
\begin{equation}
    \vec{B}_0 = (0,0,B_{z,0}),
\end{equation}
where
\begin{equation}
    B_{z,0}(x,y) = \left\{\begin{array}{lll}
    1 &  \qquad & |x - 0.5|^2 + |y - 0.5|^2 = r^2 < (0.9\Delta x)^2, \\ \noalign{\medskip}
    0 &         & \text{otherwise},
    \end{array}\right.
\end{equation}
with $r = \sqrt{(x-0.5)^2 + (y-0.5)^2}$ the radial distance from the center of the domain and 
$\Delta x = \Delta y = 1/N_d$ the uniform grid spacing of a Cartesian grid with $N_d$ cells per dimension, over $x,y \in [0,1]$.
The system of MHD equations then reduces to
\begin{equation}
    \pd{B_z}{t} = \nabla\cdot\left(\DS\frac{v_A^2}{\gamma_{AD}\rho_i}\nabla B_z\right)
\end{equation}
which is a particular case of the Barenblatt-Paddle equation \citep{barenblatt1952}
\begin{equation}
    \pd{B_z}{t} = \nabla\cdot(B_z^\beta\nabla B_z)
\end{equation}
with $\beta = 2$, and \citep{grundy1982, masson2012} has an analytical solution
\begin{equation}
    B_z = \left\{\begin{array}{lcll}
A\tau^\alpha\left[1 - \left(\DS\frac{r}{\eta_0\tau^\delta}\right)^2\right]^{1/\beta} & \qquad\quad & r\leq\eta_0\tau^\delta \\ \noalign{\medskip}
0 & &  r > \eta_0\tau^\delta
    \end{array}\right.
    \label{Eq:bar_solution}
\end{equation}
where $\tau =t/(\gamma_\ad\rho_i)$ (with $t$ final time), and
\begin{equation}
    \left\{\begin{array}{lcl}
    A & = & \left(\DS\frac{\delta\beta\eta_0^2}{2}\right)^{1/2} \\ \noalign{\medskip}
    \alpha & = & \DS\frac{-\mu}{2 + \mu\beta} \\ \noalign{\medskip}
    \delta & = & \DS\frac{1}{2 + \mu\beta} \\ \noalign{\medskip}
    \end{array}\right.
\end{equation}
with $\mu = 2$ (in 2D).
Finally, we compute $\eta_0$ by inverting the relation
\begin{equation}
    M \equiv \int B_{z,0}\,dx\,dy
    =
    \eta_0^{\mu+2/\beta}
    \left(\frac{\delta\beta}{2}\right)^{1/\beta}
    \pi
    \frac{\Gamma(\mu/2)\Gamma(1/\beta+1)}
         {\Gamma(1/\beta+1+\mu/2)} ,
\end{equation}
which yields
\begin{equation}
    \eta_0
    =
    \left[
    \DS\frac{M}{
    \left(\DS\frac{\delta\beta}{2}\right)^{1/\beta}
    \pi
    \DS\frac{\Gamma(\mu/2)\Gamma(1/\beta+1)}
         {\Gamma(1/\beta+1+\mu/2)}
    }
    \right]^{\DS\frac{\beta}{\mu\beta+2}} .
\end{equation}

\begin{figure}
\centering
\includegraphics[width=0.49\textwidth]{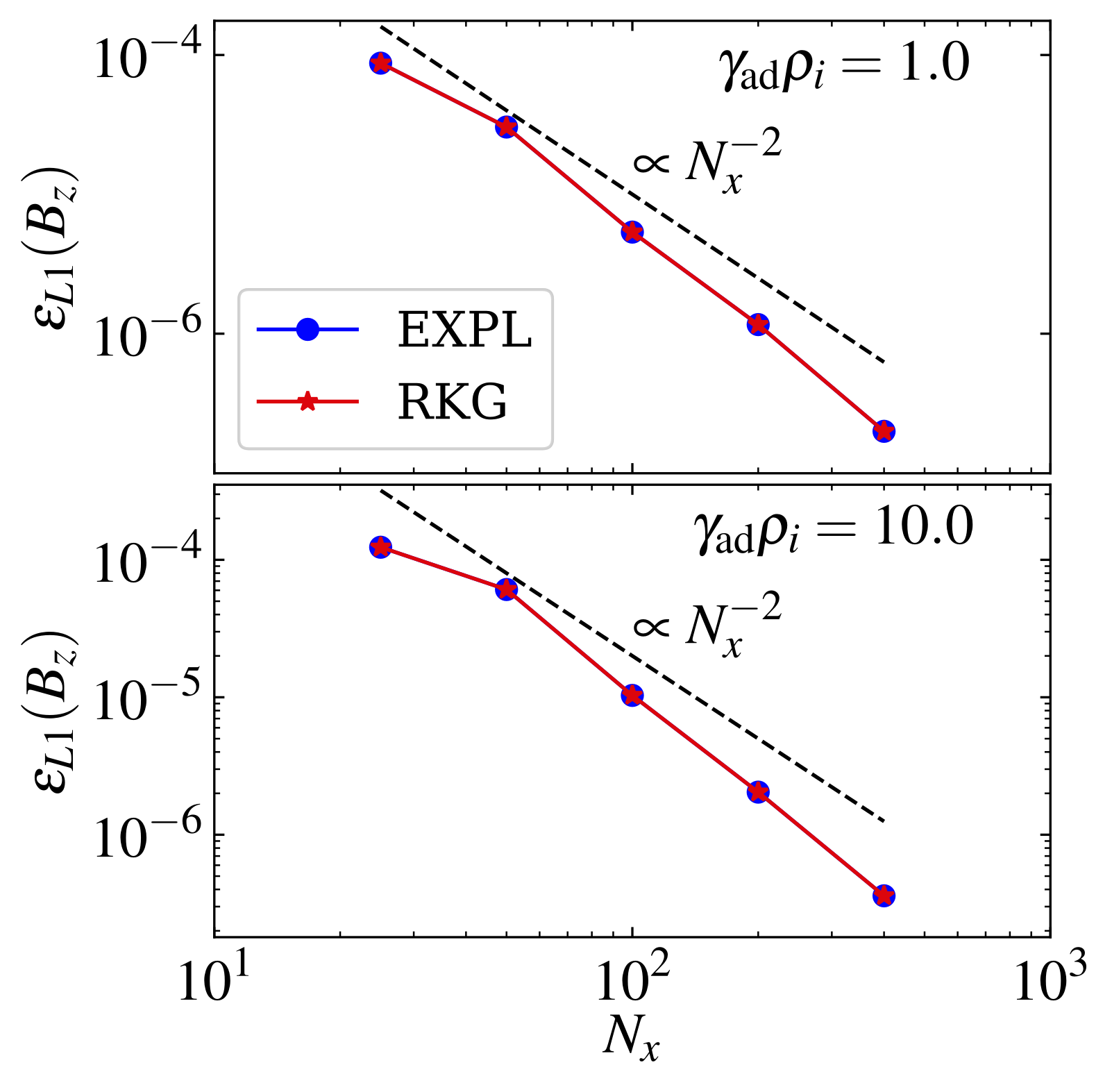}
\caption{L1-error of the numerical ambipolar baremblatt test with explicit (blue) and RKG (red) schemes in the high- (top panel) and low-ambipolar diffusion (bottom panel) regimes.
The dashed line visually highlights the expected $2^{nd}-$order accuracy.}
\label{Fig::cart_barenblatt}%
\end{figure}

For this test, we adopt linear reconstruction \citep{toro2009}, a second-order Runge–Kutta integrator \citep{gottlieb2001}, an isothermal equation of state with $c_s = 1$, a CFL number of 0.45,  the UCT-HLL upwind constrained transport method \citep{mignone2021}, and the Roe Riemann solver \citep{roe1981}.
We show the convergence for a high ($\gamma_\ad = 1$) and low ($\gamma_\ad = 10$) ambipolar coefficient, respectively, in the top and bottom panels of Fig. \ref{Fig::cart_barenblatt}, for both the explicit and the RKG numerical schemes at the final time $t = 20$.
After a weaker convergence order at the lowest resolution, the measured order is $\sim2.5$, which is broadly consistent with a $2^{nd} $- order scheme.
Note that the sharp front, located at $r = \eta_0\tau^\delta$, introduces a non-smooth interface (more specifically, a discontinuity in the gradient of $\vec{B}$) which limits the attainable accuracy and can bias the fitted slope in the pre-asymptotic regime.
Nonetheless, the convergence and error of the RKG scheme are comparable to those of the standard explicit scheme (as shown in Figure \ref{Fig::cart_barenblatt}, see also \citealt{masson2012}).

\section{Resistivity components in purely 2D-MHD}
\label{app::resistive_2D}

We consider a purely two-dimensional MHD configuration in which
\begin{equation}
    v_z = 0,
    \qquad
    B_z = 0,
    \qquad
    \partial_z(\cdot) = 0 .
\end{equation}
The current density is
\begin{equation}
    \vec{J} = \nabla \times \vec{B}
    =
    \left(
    0,\,
    0,\,
    \partial_x B_y - \partial_y B_x
    \right),
\end{equation}
so that only the $z$-component of the current is non–vanishing.
Assuming a diagonal resistivity tensor,
\begin{equation}
    \vec{\mathcal{E}}_{\rm res}
    =
    (\eta_x J_x,\eta_y J_y,\eta_z J_z),
\end{equation}
the resistive electric field reduces to
\begin{equation}
    \vec{\mathcal{E}}_{\rm res}
    =
    (0,0,\eta_z J_z).
\end{equation}
The induction equation,
\begin{equation}
    \partial_t \vec{B}
    =
    -\nabla \times \vec{\mathcal{E}}_{\rm res},
\end{equation}
then becomes
\begin{equation}
    \partial_t \vec{B}
    =
    -\nabla \times (0,0,\eta_z J_z),
\end{equation}
which only depends on $\eta_z$.
Therefore, in a strictly two-dimensional MHD flow with $v_z = 0$ and $B_z = 0$, the evolution of the magnetic field is affected exclusively by the $ z$-component of the resistivity tensor.
The in–plane components $\eta_x$ and $\eta_y$ do not contribute, since the corresponding current components vanish identically.

\end{document}